\documentclass[prd,nofootinbib,float,superscriptaddress,eqsecnum,11pt,reprint,longbibliography]{revtex4-1}
\usepackage{hyperref}
\usepackage{graphicx}
\usepackage{amsmath,amssymb,amsfonts,amsthm,latexsym,stmaryrd,physics}
\usepackage{marginnote}
\usepackage{color}
\usepackage{bm}
\usepackage{subfigure}
\usepackage{verbatim}
\usepackage{lipsum}
\usepackage[utf8]{inputenc}
\usepackage[T1]{fontenc}
\hypersetup{
	colorlinks=true,
	urlcolor=blue,
	citecolor=blue,
	linkcolor=blue
}

{{{ \interfootnotelinepenalty=10000 }}} 

\def\beq{\begin{equation}}
\def\eeq{\end{equation}}
\newcommand{\bea}{\begin{eqnarray}}
\newcommand{\eea}{\end{eqnarray}}
\def\bi{\begin{itemize}}
\def\ei{\end{itemize}}
\def\ba{\begin{array}}
\def\ea{\end{array}}
\def\bfig{\begin{figure}}
\def\efig{\end{figure}}

\def\SL{\mathbb{SL}}
\def\C{\mathbb{C}}
\def\R{\mathbb{R}}
\def\ui{\mathrm{i}}
\def\ue{\mathrm{e}}
\def\ud{\mathrm{d}}

\def\H{\mathcal{H}}
\def\sgn{\text{sgn}}

\newcommand{\hl}[1]{{\color{red} {#1}}}
\newtheorem{theorem}{Theorem}[section]

\newtheorem{lemma}[theorem]{Lemma}

\begin{document}

\title{Asymptotic analysis of spin foam amplitude with timelike triangles}
\author{Hongguang Liu}
\email{liu.hongguang@cpt.univ-mrs.fr}
\affiliation{Aix Marseille Univ, Universit\'e de Toulon, CNRS, CPT, Marseille, France}
\affiliation{Laboratoire de Math\'ematiques et Physique Th\'eorique (UMR CNRS 7350), Universit\'e Fran\c cois Rabelais, Parc de Grandmont, 37200 Tours, France}
\affiliation{Laboratoire Astroparticule et Cosmologie, Universit\'e Denis Diderot Paris 7, 75013 Paris, France}
\author{Muxin Han}
\email{hanm@fau.edu}
\affiliation{Department of Physics, Florida Atlantic University, FL 33431, USA}
\affiliation{Institut f\"ur Quantengravitation, Universit\"at Erlangen-N\"urnberg, Staudtstr. 7/B2, 91058 Erlangen, Germany}

\date{\today}

\begin{abstract}
	The large $j$ asymptotic behavior of $4$-dimensional spin foam amplitude is investigated for the extended spin foam model (Conrady-Hnybida extension) on a simplicial complex. We study the most general situation in which timelike tetrahedra with timelike triangles are taken into account. The large $j$ asymptotic behavior is determined by critical configurations of the amplitude. We identify the critical configurations that correspond to the Lorentzian simplicial geometries with timelike tetrahedra and triangles. Their contributions to the amplitude are phases asymptotically, whose exponents equal to Regge action of gravity. The amplitude may also contains critical configurations corresponding to non-degenerate split signature $4$-simplices and degenerate vector geometries. But for vertex amplitudes containing at least one timelike tetrahedron and one spacelike tetrahedron, critical configurations only give Lorentzian $4$-simplices, while the split signature and degenerate $4$-simplices do not appear.
\end{abstract}

\maketitle

\section{Introduction}
Spin foam models arise as a covariant formulation of Loop Quantum Gravity (LQG), for a review, see \cite{Thiemann:2007zz,Rovelli:2014ssa,Ashtekar:2017yom,Han:2005km, Perez:2012wv}. A spin foam can be regraded as a Feynmann diagram with 5-valent vertices, corresponding to quantum $4$-simplices, as building blocks of the discrete quantum spacetime. The boundary of a $4$-simplex contains $5$ tetrahedra. As one of the popular spin foam models, the Lorentzian Engle-Pereira-Rovelli-Livine/Freidel-Krasnov (EPRL/FK) model comes with a gauge-fixing within each tetrahedron such that in the local frame the timelike normal vector of the tetrahedron reads $u=(1,0,0,0)$ {in a 4D Minkowski spacetime with signature $(-1,1,1,1)$}, known as the "time-gauge". As a result, this model is subject to the restriction that tetrahedra and triangles are all spacelike \cite{Engle:2007wy}, {such that, the tetrahedra lives in a Euclidean subspace. As a result, such spin foam models only correspond to a special class of 4D Lorentzian triangulations}. However, in the extended spin foam model by Conrady and Hnybida, some tetrahedron normal vectors are chosen to be spacelike $u=(0,0,0,1)$. As a result, the model contains timelike tetrahedra and triangles {which live in 3D Minkowski subspaces} \cite{Conrady:2010kc,Conrady:2010vx,Rennert:2016rfp}.

The semiclassical behavior of spin foam model is determined by its large-$j$ asymptotics. Recently there have been many investigations of large-j spin foams, in particular the asymptotics of EPRL/FK model \cite{Conrady:2008mk,Christensen:2009bi,Barrett:2009gg, Barrett:2009mw,Han:2011re,Han:2011rf,Han:2013hna,Han:2013ina,Han:2017xwo}, and models with cosmological constant \cite{Haggard:2014xoa,Haggard:2015kew}. It has been shown that, in large-$j$ asymptotics, the spin foam amplitude is dominated by the contributions from critical configurations, which gives the simplicial geometries and discrete Regge action on a simplicial complex. The resulting geometries from the above analysis only have spacelike tetrahedra and spacelike triangles. Recently, the asymptotics of the Hnybida-Conrady extended model with timelike tetrahedron was investigated in \cite{Kaminski:2017eew}. The critical configurations of the extended model give simplicial geometries containing timelike tetrahedra. But the limitation is that all the triangles are still spacelike within each timelike tetrahedron.

In this paper we extend the semiclassical analysis of extended model to general situations, in which we take into account both timelike tetrahedra and timelike triangles. Our work is motivated by the examples of geometries in classical Lorentzian Regge calculus, and their convergence to smooth geometries \cite{Barrett:1994ks,Gentle:1997df,Gentle:2012tc}. In all examples the Regge geometries contain timelike triangles. In order to have the Regge geometries emerge as critical configurations from spin foam model, we have to extend the semiclassical analysis to contain timelike triangles.

In our analysis, we first derive the large-$j$ integral form of the extended spin foam model with coherent states for timelike triangles. The large-$j$ asymptotic analysis is based on the stationary phase approximation of the integral. The asymptotics of the integral is a sum of contributions from critical configurations.  

Before coming to our main result, we would like to mention some key assumptions for the validity of the result: The following results are valid when we assume every timelike tetrahedron containing at least one spacelike and one timelike triangle. It is the case in all Regge geometry examples mentioned above. Our results also apply to some special cases when all triangles in a tetrahedron are timelike. Moreover all tetrahedra in our discussion are assumed to be nondegenerate. Here we don't consider the critical configurations with a degenerate tetrahedron. Finally, the Hessian evaluated at every critical configuration is assumed to be a non-degenerate matrix.

The main result is summarized as follows: Firstly for a single 4-simplex and its vertex amplitude, it is important to have boundary data satisfy the length matching condition and orientation matching condition. Namely, (1) among the 5 tetrahedra reconstructed by the boundary data (by Minkowski Theorem), each pair of them are glued with their common triangles matching in shape (match their 3 edge lengths), and (2) all tetrahedra have the same orientation. The amplitude has critical configurations only if these 2 conditions are satisfied, otherwise the amplitude is suppressed asymptotically, The critical configurations have geometrical interpretations as geometrical 4-simplices, which may generally have one of three possible signatures: Lorentzian, split, or degnerate. 

\begin{itemize}
    \item When the 4-simplex has Lorentzian signatures: The contribution at the critical configuration is given by a phase, whose exponent is Regge action with a sign related to orientations, i.e. the vertex amplitude gives asymptotically
    \begin{equation}
        A_{v} \sim  N_{+} \ue^{\ui S_{\Delta}} + N_{-} \ue^{- \ui S_{\Delta}}   
	\end{equation}
	up to an overall phase depending on the boundary coherent state.
	The Regge action in the 4-simplex reads $S_{\Delta} = \sum_{f} A_f \theta_f$ with $A_f$ the area of triangle $f$. $\theta_f$ relates to the dihedral angle $\Theta_f$ by $\theta_f=\pi-\Theta_f$. The area spectrum is different between timelike and spacelike triangles in a timelike tetrahedron.
	\begin{equation}
	A_f = \left\{ \begin{array}{ll} \frac{n_f}{2} & \text{timelike triangle} \\ \gamma {j_f} & \text{spacelike triangle} \end{array} \right.
	\end{equation}
	$n_f \in \mathbb{Z}_{+}$ satisfies the simplicity constraint $n_f = \gamma s_f$ where $s_f \in \mathbb{R}_{+}$ labels the continuous series irreps of $\text{SU}(1,1)$. $j_f \in \mathbb{Z}_{+}/2$ labels the  discrete series irreps of $\text{SU}(1,1)$.
	 $N_{\pm}$ are geometric factors depend on the lengths and orientations of the reconstructed $4$ simplex.
    \item The reconstructed 4-simplices have split signatures: The vertex amplitude gives asymptotically
    \begin{equation}
        A_v \sim N_{+} \ue^{\ui \gamma^{-1} S_{\Delta}} + N_{-} \ue^{- \ui \gamma^{-1} S_{\Delta}} 
	\end{equation}
	up an overall phase. Here $S_{\Delta} = \sum_{f} A_f \theta_f$ where $\theta_f$ is a boost dihedral angle.
	
	\item The reconstructed 4-simplices are degenerate (vector geometry) and there is a single critical point. The asymptotical vertex amplitude is given by a phase depending on the boundary coherent states.
\end{itemize}

It is important to remark that for a vertex amplitude containing at least one timelike and one spacelike tetrahedron, critical configurations only give Lorentzian 4-simplices, while the split signature and degenerate 4-simplex do not appear. The last 2 cases only appear when all tetrahedra are timelike in a vertex amplitude. The situation is similar to Lorentzian EPRL/FK model, where the Euclidean signature and degenerate 4-simplex appear because all tetrahedra are spacelike.

Our analysis is generalized to the spin foam amplitude on a simplicial complex $\mathcal{K}$ with many $4$-simplices. We identify the critical configurations corresponding to simplicial geometries with all $4$-simplices being Lorentzian and globally oriented. The configurations come in pairs, corresponding to opposite global orientations. Each pair gives the following asymptotic contribution to the spin foam amplitude (up to an overall phase)
\begin{equation}\label{NSNS}
  N_{+} \ue^{\ui S_{\mathcal{K}}} + N_{-} \ue^{- \ui S_{\mathcal{K}}}   
\end{equation}
where 
\begin{equation}
	S_{\mathcal{K}} = \sum_{f\ \text{bulk}} A_f \varepsilon_f + \sum_{f\ \text{boundary}} A_f \left(\theta_f+p_f\pi\right)
\end{equation}
is the Regge action on the simplicial complex, up to a boundary term with $p_f\in\mathbb{Z}$ ($p_f$ is the number of 4-simplices sharing $f$ minus 1). The additional boundary term $p_f A_f\pi$ doesn't affect the Regge equation of motion. Here the simplicial geometries and Regge action generally contain timelike tetrahedra and timelike triangles. $\varepsilon_f$ is the deficit angle. $\varepsilon_f$ and $\theta_f$ at timelike triangles are given by 
\begin{eqnarray}
\varepsilon_f=2 \pi - \sum_f \Theta_f(v),\quad \theta_f=\pi - \sum_f \Theta_f(v)
\end{eqnarray}
$\Theta_f(v)$ is the dihedral angle within the $4$-simplex at $v$. It is a rotation angle between spacelike normals of tetrahedra, because the tetrahedra sharing a timelike triangle are all timelike.

To obtain (\ref{NSNS}), we have assumed each bulk triangle is shared by an even number of $4$-simplices.  This assumption is true in many important examples of classical Regge calculus.

This paper is organized as follows. In section \ref{sec2}, we write the coherent states for timelike triangles in large $j$ approximation and express the spin foam amplitude in terms of the coherent states. In section \ref{sec3}, we derive and analyze the critical equations. The critical equations are reformulated in geometrical form for a timelike tetrahedron containing both spacelike and timelike triangles. Then in section \ref{sec4}, we reconstruct nondegenerate simplicial geometries from critical configurations. In section \ref{sec5}, the critical configurations for degenerate geometries are analyzed. Finally in section \ref{sec6}, we derive the difference between phases evaluated at pairs of critical configurations corresponding to opposite orientated simplicial geometries.



\section{Spinfoam amplitude in terms of SU(1,1) continuous coherent states}\label{sec2}

{
The spin foam models are defined as a state sum model over simplicial manifold $\mathcal{K}$ and it's dual, which  consists of simplicies $\sigma_{v}$, tetrahedra $\tau_{e}$, triangles $f$, edges and vertices ($v$,$e$ and $f$ are labels for vertices, edges and faces on the dual graph respectively). A triangulation is obtained by gluing simplicies $\sigma$ with pairs of their boundaries (tetrahedrons $\tau$). The phase space associated with manifold $\mathcal{K}$ are
\begin{equation}
	P_{\mathcal{K}} = T^{*}\text{SL}(2,\C)^L, \qquad (\Sigma^{IJ}_f, h_f ) \in T^*\text{SL}(2,\C)
\end{equation}
for a Lorentzian model, where $L$ is the number of triangles, $h_f \in \text{SL}(2,\C)$ is the holonomy along the edges and $\Sigma^{IJ}_f \in \mathfrak{sl}(2,\C)$ is its conjugate momenta. $h_f$ can be decomposed as 
\begin{equation}
	h_f = \prod_{v \subset \partial f} g_{ev}g_{ve'}
\end{equation}
where $g_{ve} \in \text{SL}(2,\C)$ and $g_{ev} = {g_{ve}}^{-1}$. $\Sigma^{IJ}_f$ is subject to the simplicity constraint 
\begin{equation}\label{eq:simp_cons_1}
	\frac{\gamma}{1+\gamma^2} (u_e)^I ((1 - \gamma *) {\Sigma_f }_{IJ}) =0
\end{equation}
where $u_e$ is a $4$ normal vector associated to each tetrahedron $t_e$, $\gamma$ is a real number known as the Immirizi parameter, and $*$ is the Hodge dual operator. Geometrically, the simplicity constraint implies that, each triangle $f$ in tetrahedron $t_e$ is associated with a simple bivector 
\begin{equation}
	B_f = \frac{\gamma}{1+\gamma^2} (1 - \gamma *) {\Sigma_f }
\end{equation}

The state sum is defined over all quantum states of the physical Hilbert space on a given $\mathcal{K}$, given as
\begin{equation}
	Z(\mathcal{K}) = \sum_{J} \prod_{f} \mu_f (J_f) \prod_{v} A_v(J_f, i_{e})
\end{equation}
Here $J=\vec{j}_f$ represents the combination of labels of the $\text{SL}(2,\C)$ irreps associated to each triangle. $i_e$ is the intertwiner associated with each tetrahedron
\begin{equation}
	i_e \in \text{Inv}_{G} [ V_{J_1} \otimes \dots \otimes V_{J_4}  ]
\end{equation}
which impose the gauge invariance. The vertex amplitude $ A_v(J_f, i_{e})$ associated with each $4$ simplex $\sigma_v$ captures the dynamics of the model, while the face amplitude $ \mu_f (J_f)$ is a weight for the $J$ sum.

Usually a partial gauge fixing is taken to the above models, which corresponding to pick a special normal $u$ for all of the tetrahedra $\forall_e, \; u_e =u$ . As a result, the intertwiners associated with each tetrahedron defined above is replaced by the intertwiners of the the stabilizer group $H \in G$. There are two different gauge fixing: 
\begin{itemize}
		\item $u=(1,0,0,0)$, $H = SU(2)$ , {EPRL/FK models}
		\item $u=(0,0,0,1)$, $H = SU(1,1)$, {Conrady-Hnybida Extension}
\end{itemize}
which, after impose the quantum simplicity constraint (\ref{eq:simp_cons_1}) lead to the following conditions \cite{Engle:2007wy,Freidel:2007py,Conrady:2010kc}
\begin{itemize}
	\item $u=(1,0,0,0)$, spacelike triangles
	\begin{equation}\label{eq:simp1}
		\rho =  \gamma n , \qquad n = j
	\end{equation} 
	\item $u=(0,0,0,1)$, spacelike triangles
	 \begin{equation}\label{eq:simp1}
		\rho =  \gamma n , \qquad n = j
	\end{equation} 
	\item $u=(0,0,0,1)$, timelike triangles
	    \begin{equation}\label{eq:simplicity}
			\rho = - n / \gamma, \qquad s = \frac{1}{2}\sqrt{n^2/\gamma^2 -1}
		\end{equation} 
\end{itemize}
Here $(\rho \in \R , \;n \in \mathbb{Z}/2)$ are labels of $\text{SL}(2,\C)$ irreps, $j \in \mathbb{N}/2$ is the label of $\text{SU}(2)$ irreps or $\text{SU}(1,1)$ discrete series and $s \in \R$ is the label of $\text{SU}(1,1)$ continous series, we will give a brief introduction of $\text{SU}(1,1)$ and  $\text{SL}(2,\C)$ representation theory later.
As a result, the area spectrum is given by
\begin{equation}
A_f = \left\{ \begin{array}{ll} \frac{n_f}{2} & \text{timelike triangle} \\ \gamma {j_f} & \text{spacelike triangle} \end{array} \right.
\end{equation}

}

The spin foam vertex amplitude can be expressed in the coherent state representation:
\begin{widetext}
\begin{equation}\label{eq:amplitude}
	A_v(K)=\sum_{j_f} \prod_{f} \mu(j_f) \int_{\SL (2, \C)} \prod_{e} dg_{\nu e} \prod_{(e,f)} \int_{S^2} d N_{ef} \bra{ \Psi_{\rho_f n_f}(N_{ef})} D^{(\rho_f,n_f)}(g_{ev}g_{ve'}) \ket{ \Psi_{\rho_f n_{f}} (N_{e'f})}
\end{equation}
\end{widetext}
Here $N$ is the unit vector in a sphere or hyperbolid which labels the  coherent states $\ket{ \Psi_{\rho n}}$ of $\text{SL}(2, \C)$ in the unitary irrep $\H_{(\rho,n)}$. 
By $\text{SU}(1,1)$ decomposition of $\text{SL}(2, \C)$ unitary irrep, $\text{SL}(2, \C)$ irrep is isomorphic to a direct sum of irreps of $\text{SU}(1,1)$. The area of timelike triangles is related to $\text{SU}(1,1)$ spin $s$ and the Immirzi parameter $\gamma$ by $A_f = \gamma \sqrt{s^2+1/4}$ which is consistent with the spectrum from canonical approach \cite{Conrady:2010kc, Liu:2017bfk}. 
However, the solution of quantum simplicity constraint (\ref{eq:simp_cons_1} on timelike triangles induced a $Y$-map where the physical Hilbert space $\H \in \H_{(\rho,n)}$ is isomorphic to continuous series of $\text{SU}(1,1)$ with spin $s$ fixed by (\ref{eq:simplicity}). As a result, the area spectrum is now given by
\begin{equation}
    A_f = \gamma \sqrt{s^2+1/4} = \frac{n_f}{2} 
\end{equation}
which is quantized.

In the following, we first give a brief introduction of the $\text{SU}(1,1)$ and $\text{SL}(2,\C)$ representation theory. Then we write the $\text{SL}(2,\C)$ states explicitly using continuous $\text{SU}(1,1)$ coherent states in terms of spinor variables. Finally we derive the integral from of spin foam amplitude on timelike triangles with a spin foam action.

\subsection{Representation theory of $\text{SL}(2,\C)$ and $\text{SU}(1,1)$ group}
$\text{SL}(2,\C)$ group has $6$ generators $J^i$ and $K^i$ with commutation relation
\begin{equation}
	\begin{split}
		& [J^i, J^i]= \epsilon^{ij}_k J^k, \;\;[J^i, K^j]= \epsilon^{ij}_k K^k, \\
		& \qquad \qquad [K^i, K^j]= - \epsilon^{ij}_k J^k
	\end{split}
\end{equation}
The unitary representations of the group are labelled by pairs of numbers $(\rho \in \R, n \in \mathbb{Z}_{+})$ from the two Casimirs 
\begin{equation}
\begin{split}
	&C_1= 2 (\vec{J}^2 - \vec{K}^2) = \frac{1}{2}(n^2-\rho^2-4)\\
	&C_2= -4 \vec{J} \cdot \vec{K} = n \rho
\end{split}
\end{equation}
The Hilbert space $\H_{(\rho,n)}$ of unitary irrep of $\text{SL}(2,\C)$ can be represented as a space of homogeneous functions $F: \C^2 \backslash \{0\} \to \C $ with the homogeneity
property
\begin{equation}\label{eq:rep_F}
	F(\beta z_1, \beta z_2)=\beta^{i \rho/2 +n/2-1}\beta^{*i \rho/2-n/2-1} F(z_1,z_2)
\end{equation}
The inner product in $\H_{(\rho,n)}$ is given by
\begin{equation}\label{eq:sl2c_inner}
	\braket{F_1}{F_2} = \int_{\mathbb{CP}_1} \pi((F_1)^* F_2 \omega)
\end{equation}
where $\pi: \mathbb{C}^2\setminus \{0\} \to \mathbb{CP}_1$. $\omega$ is the $\text{SL}(2,\C)$ invariant 2-form defined by
\begin{equation}\label{eq:omega}
	\omega = \frac{\ui}{2} (z_2 \ud z_1 - z_1 \ud z_2) \wedge (\bar{z}_2 \ud \bar{z}_1 - \bar{z}_1 \ud \bar{z}_2)
\end{equation}

$\text{SU}(1,1)$ group is a subgroup of $\text{SL}(2,\C)$ with generators $\vec{F} = (J^3,K^1,K^2)$. $\vec{F}$ and $\vec{G} = \ui \vec{F} = (K^3,-J^1,-J^2)$ transform as Minkowski vectors under $\text{SU}(1,1)$. The Casimir reads $Q=(J^3)^2-(K^1)^2-(K_2)^2$.
The unitary representation of $\text{SU}(1,1)$ group is usually built from the eigenstates of $J^3$ which is labelled by $j,m$:
\begin{equation}\label{eq:basis_su1,1_j3}
	\braket{jm}{jm'} = \delta_{mm'}
\end{equation}
where $m$ is the eigenvalue of $J^3$ and $j$ related to the eigenvalues of the Casimir $Q$.

The unitary irrep of SU(1,1) contains two series: the discrete series and continuous series. For the discrete series, one has
\begin{equation}
	Q \ket{jm} = j (j+1) \ket{jm}, \quad \text{with}  \;\; j = -\frac{1}{2},-1,-\frac{3}{2}, ...
\end{equation}
The eigenvalue $m$ of $J^3$ takes the values
\begin{equation}
	m = -j, -j+1, -j+2 .... \quad \text{or} \quad m = j, j-1, j-2....
\end{equation}
The Hilbert spaces of spin $j$ are denoted by $\mathcal{D}_j^{\pm}$ with $m^ >_< 0$. For the continuous series, $Q$ takes continuous value
\begin{equation}
	Q \ket{jm} = j (j+1) \ket{jm}
\end{equation}
where $j=-1/2+\ui s$ and $s$ is a real number $s \in \R_{+}$. Thus in continuous case, we can use $s$ instead of $j$ to represent the spin. The eigenvalues $m$ takes the values
\begin{equation}
	m = 0 , \pm 1, \pm 2, ... \qquad \text{or} \qquad m = \pm \frac{1}{2}, \pm \frac{3}{2}, ...
\end{equation}
The irreps of this series are denoted by $C_s^{\epsilon}$ where $\epsilon = 0, 1/2$ corresponding to the integer $m$ and half-integer $m$ respectively. 

Instead of $\ket{jm}$, one may also choose the generalized continuous eigenstates $\ket{j \lambda \sigma}$ of $K^1$ as the basis of the irrep Hilbert space \cite{lindblad1970continuous}:
\begin{equation}\label{eq:c_basis}
	\braket{j \lambda' \sigma'}{j \lambda \sigma}=\delta(\lambda -\lambda')\delta_{\sigma \sigma'}
\end{equation}
where $\sigma=0,1$ distinguish the two-fold degeneracy of the spectrum and $\lambda$ here is a real number. For continuous series irreps, Casimir $Q$ takes 
\begin{equation}
	 Q \ket{j \lambda \sigma}= j(j+1) \ket{j \lambda \sigma} = - \left(s^2 + \frac{1}{4}\right)\ket{j \lambda \sigma}.
\end{equation}

\subsection{Unitary irreps of $\text{SL}(2,\C)$ and the decomposition into $\text{SU}(1,1)$ continuous state}
The Hilbert space $\mathcal{H}_{(\rho,n)}$ can be decomposed as a direct sum of irreps of $\text{SU}(1,1)$. The decomposition can be derived from the homogeneity property and the Plancherel decomposition of $\text{SU}(1,1)$.
As shown in \cite{Conrady:2010sx}, the functions $F$ in the $\text{SL}(2,\C)$ Hilbert space satisfying (\ref{eq:rep_F}) can be described by pairs of functions $f^{\alpha}: \text{SU}(1,1) \to \C, \alpha=\pm 1$ via
\begin{equation}
	F(z_1,z_2)=\sqrt{\pi} (\alpha \langle z,z \rangle)^{i \rho/2-1} f^{\alpha}( v^{\alpha}(z_1,z_2)), \; \; 
\end{equation}
where $v^{\alpha}$ is the induced $\text{SU}(1,1)$ matrix 
\begin{equation}\label{eq:valpha_z}
 	v^{\alpha}= \left\{ \begin{array}{l}\frac{1}{\sqrt{\langle z, z \rangle}} \left( \begin{array}{ll} z_1 &z_2\\ \bar{z}_2 &\bar{z}_1 \end{array} \right), \quad \alpha = 1 \\ \frac{1}{\sqrt{- \langle z, z \rangle}} \left( \begin{array}{ll}  \bar{z}_2 &\bar{z}_1 \\z_1 &z_2 \end{array} \right), \quad \alpha=-1 \end{array} \right.
 \end{equation}
 with $\langle z,z \rangle=z^{\dagger} \sigma_3 z=\bar{z}_1 z_1-\bar{z}_2 z_2$ being $\text{SU}(1,1)$ invariant inner product. Here $\alpha$ is a signature
\begin{equation}
	\alpha = \left\{ \begin{array}{ll}
		1, & |z_1|>|z_2| \\
	   -1, & |z_1|<|z_2| \end{array} \right.
\end{equation}
Then $\H_{(\rho,n)}$ is isomorphic to the Hilbert space $L^2(\text{SU}(1,1)) \oplus L^2(\text{SU}(1,1))$ with inner product
\begin{equation}
	\braket{(f_1^{+}, f_1^{-})}{(f_2^{+}, f_2^{-})} = \sum_{\alpha} \int dv (f_1^{\alpha}(v))^* f_2^{\alpha}(v)
\end{equation}
where $dv$ is the $\text{SU(1,1)}$ measure.

  The function $f$ in $\text{SU}(1,1)$ continuous series representations with continuous basis reads
\begin{equation}\label{eq:su11_f}
	f^{\alpha}_{j \lambda}(z)= \left\{ \begin{array}{ll}
		\sqrt{2j+1} (D^{j}_{n/2, \lambda}(v(z)),0), & \alpha=1\\
		\sqrt{2j+1} (0,D^{j}_{-n/2, \lambda}(v(z))),& \alpha=-1
		\end{array} \right.
\end{equation}
Noticed that here we assume $s \neq 0$. $D^j_{m \lambda}$ is the Wigner matrix with mixed basis (\ref{eq:basis_su1,1_j3}) and (\ref{eq:c_basis})
\begin{equation}
	\begin{split}
		&D^j_{m \lambda \sigma}(v)=\bra{j,m} v(z) \ket{j,\lambda, \sigma}
	\end{split}
\end{equation}
Recall the quantum simplicity constraint (\ref{eq:simplicity}),
\begin{equation}
 	\rho = - n / \gamma, \qquad s = \frac{1}{2}\sqrt{n^2/\gamma^2 -1}
 \end{equation} 

Asymptotically, when $s \gg 1$, we have
\begin{equation}
	\rho \sim -2 s \sim -\frac{n}{\gamma}
\end{equation}
Since $n$ is discrete, $s$ and $\rho$ are also discrete.
Using the representation matrix of continuous series of $\text{SU}(1,1)$, and some transformations of hypergeometric function and asymptotic analysis, we prove that when $n \gg 1$ and $\lambda = -s$ (the detailed derivation is shown in Appendix \ref{app_f}),
\begin{equation}
	\begin{split}
    &D^{j}_{{\frac{n}{2}}, -s}(v)=\frac{1}{\sqrt{s{|\gamma + \Im(\bar{v}_1 v_2)|}}} \cross\\
    & \left( \tilde{T}^{j}_{+\sigma } \Big(\frac{v_1-v_2}{\sqrt{2}}\Big)^{{\frac{n}{2}}-\ui s} \Big(\frac{\overline{v_1-v_2}}{\sqrt{2}}\Big)^{-{\frac{n}{2}}-\ui s} \right.\\
    &\qquad \left. -\tilde{T}^{j}_{-\sigma } \Big(\frac{v_1+v_2}{\sqrt{2}}\Big)^{{\frac{n}{2}}+\ui s} \Big(\frac{\overline{v_1+v_2}}{\sqrt{2}}\Big)^{-{\frac{n}{2}}+\ui s}\right)
	\end{split}
\end{equation}
where $\sqrt{2}\tilde{T}^j_{\pm \sigma } =\sqrt{2} S^j_{n/2,-s, \sigma}/T^j_{\pm \sigma}$ are some phases: $\tilde{T}_{\pm}\overline{\tilde{T}_{\pm}}=1/2$ \footnote{ {Here we ignore the regulator in (\ref{eq:final_D_matrix}) for the zero points of $|\gamma + \Im(\bar{v}_1 v_2)|$ since it will appear naturally as the integration contribution from this $1/2$ singularity in the inner product. One can check Appendix \ref{app_f} for details.}}. The detailed definition of $S^j_{n/2,-s, \sigma}$ and $T^j_{\pm\sigma}$ are given in (\ref{eq:s_j_m}) and (\ref{eq:def_t}).

The $m=-n/2$ case  in (\ref{eq:su11_f}) can be obtained by the relation
\begin{equation}
	D^{\sigma j}_{-m, \lambda}(v)=-(-1)^{\sigma}\ue^{-\ui \pi m}D^{\sigma j}_{m, \lambda}(\bar{v})
\end{equation}

When $\alpha=1$, we would like to write elements of $v^{\alpha} \in \text{SU}(1,1)$ {{introduced in (\ref{eq:valpha_z})}} as
\begin{align}
	\frac{v_1-v_2}{\sqrt{2}}=\frac{\langle \bar{z},l^{+}_0 \rangle}{\sqrt{\langle z, z \rangle}}, \;\;
	\frac{v_1+v_2}{\sqrt{2}}=\frac{\langle \bar{z},l^{-}_0 \rangle }{\sqrt{\langle z, z \rangle}}.
\end{align}
where
\begin{equation}\label{eq:def_l+-}
 l^{\pm}_0=\frac{1}{\sqrt{2}}(n_1 \pm n_2)=\frac{1}{\sqrt{2}} \left( \begin{array}{l} 1\\ \pm 1 \end{array} \right)
\end{equation}
Notice that, $\langle l^{+}_0, l^{+}_0 \rangle=\langle l^{-}_0, l^{-}_0 \rangle=0$, $\langle l^{-}_0, l^{+}_0 \rangle=1$, thus they form a null basis in $\C^2$.
Similarly, for $\alpha=-1$, we have
\begin{align}
    &\frac{v_1-v_2}{\sqrt{2}}= - \frac{\langle l^{+}_0, \bar{z} \rangle}{\sqrt{-\langle z, z \rangle}}, &\frac{v_1+v_2}{\sqrt{2}}=\frac{\langle l^{-}_0, \bar{z} \rangle}{\sqrt{-\langle z, z \rangle}}
\end{align}

With this notation, we finally obtain
\begin{widetext}
\begin{equation}
	\begin{split}
	&F^{(\rho,n)}_{-s, \sigma, \alpha }(z)=\frac{ \sqrt{\pi} \alpha^{n/2+\sigma+1}  }{\sqrt{s} \sqrt{\alpha \langle z,z \rangle}\sqrt{|\alpha (\gamma - \ui) \langle z,z \rangle + 2 \ui \alpha {\langle \bar{z},l^{-}_0 \rangle \langle l^{+}_0,\bar{z} \rangle}}|} \\
	& \cross \left( \tilde{T}^{j}_{+\sigma}  {(\alpha \langle z,z \rangle)}^{i \rho/2+\ui s} {( \langle l^{+}_0, \bar{z} \rangle \langle \bar{z} , l^{+}_0 \rangle)}^{ -\ui s} \left(\frac{{\langle \bar{z}, l^{+}_0 \rangle }}{\langle l^{+}_0, \bar{z} \rangle } \right)^{\frac{n}{2}}  - \tilde{T}^{j}_{- \sigma} {(\alpha \langle z,z \rangle)}^{i \rho/2-\ui s} {(\langle l^{-}_0, \bar{z} \rangle \langle \bar{z} , l^{-}_0 \rangle)}^{\ui s} \left(\frac{{\langle \bar{z}, l^{-}_0 \rangle }}{\langle l^{-}_0, \bar{z} \rangle } \right)^{\frac{n}{2}} \right)
	\end{split}
\end{equation}
\end{widetext}
One can check the homogeneity property (\ref{eq:rep_F}):
\begin{equation}
 	F(\lambda z)= \lambda^{m+\ui \rho/2 -1} \bar{\lambda}^{-m+\ui \rho/2 -1} F(z)
 \end{equation} 

The coherent state is built from the reference state $\lambda = - s$, and we choose $\sigma=1$, according to \cite{Conrady:2010vx},
\begin{widetext}
\begin{equation}\label{eq:coherent_states}
	\begin{split}
	&\Psi^{(\rho,n)}_{\tilde{g}, \alpha}(z)= D^{(\rho,n)}(\tilde{g})F^{(\rho,n)}_{-s,1,\alpha }(z) =\frac{ \sqrt{\ui \pi}\tilde{S}^j_{m,-s,\sigma} \alpha^{-2 \ui s+m}  }{\sqrt{|\langle z,z \rangle|}\sqrt{|(\gamma - \ui) \langle z,z \rangle + 2 \ui {\langle \bar{z},l^{-} \rangle \langle l^{+},\bar{z} \rangle|}}} \cross \\
	& \left( \tilde{T}^{j}_{+1}  {\langle z,z \rangle}^{i \rho/2+\ui s} {(\langle l^{+}, \bar{z} \rangle \langle \bar{z} , l^{+} \rangle)}^{ -\ui s} \left(\frac{{\langle \bar{z}, l^{+} \rangle }}{\langle l^{+}, \bar{z} \rangle } \right)^{\frac{n}{2}} -\tilde{T}^{j}_{-1} { \langle z,z \rangle}^{i \rho/2-\ui s} {(\langle l^{-}, \bar{z} \rangle \langle \bar{z} , l^{-} \rangle)}^{\ui s} \left(\frac{{\langle \bar{z}, l^{-} \rangle }}{\langle l^{-}, \bar{z} \rangle } \right)^{\frac{n}{2}} \right)
	\end{split}
\end{equation}
\end{widetext}
where $\tilde{g} \in SU(1,1)$, and $l^{\pm}=\tilde{g}^{-1 \dagger} l^{\pm}_0$ is defined though
\begin{equation}
	\langle l^{\pm}_0, \overline{\tilde{g}^t z} \rangle=\langle \tilde{g}^{-1 \dagger}l^{\pm}_0, \bar{z} \rangle={\langle l^{\pm} ,\bar{z}\rangle }
\end{equation}

\subsection{Spinform amplitude}
Now we can write down explicitly the inner product between the coherent states appearing in the amplitude (\ref{eq:amplitude}) { by inserting (\ref{eq:coherent_states}) and using (\ref{eq:sl2c_inner})}:
\begin{widetext}
\begin{equation}\label{eq:amp}
\begin{split}
&\bra{ \Psi_{\tilde{g}_{e'f} \delta}^{(\rho_{f},n_f)}} D^{(\rho_f,n_f)}(g_{ve'}g_{ev}) \ket{ \Psi_{\tilde{g}_{ef}\delta}^{(\rho_{f},n_f)}}=\sum_{\alpha} \int_{CP_1} \omega_{z_{vf}} {\Psi^{(\rho_f,n_f)}_{\tilde{g}_{e'f} \delta \alpha} \left(g^t_{ve'} z_{vf} \right)} \overline{\Psi^{(\rho_f,n_f)}_{\tilde{g}_{ef} \delta \alpha}}\left(g^t_{ev} z_{vf} \right)\\
=&\int_{CP_1 /\langle Z, Z \rangle = 0} \frac{\omega_{z_{vf}}}{h_{vef} h_{ve'f}} \left( N_{f+} \ue^{S_{vf+}}+  N_{f-}\ue^{S_{vf-}}+  N_{fx+} \ue^{S_{vfx+}}+  N_{fx-}\ue^{{S_{vfx-}}} \right)\\
\end{split}
\end{equation}
where $N$ are some normalization factors, {$\omega$ is the $\text{SL}(2,\mathbb{C})$ invariant measure defined in (\ref{eq:omega}).
}
The exponents read
\begin{equation}\label{eq:action}
	\begin{split}
	S_{vf \pm}=S_{ve'f \pm} - S_{vef \pm}, \quad S_{vfx \pm} = S_{ve'f \pm} - S_{vef \mp}
	\end{split}
\end{equation}
with
\begin{eqnarray}\label{action1}
S_{vef \pm} = s_f \left[ \gamma \ln \frac{{\langle Z_{vef},l^{\pm}_{ef} \rangle } }{{\langle l^{\pm}_{ef}, Z_{vef} \rangle } } \mp \ui \ln {{\langle Z_{vef},l^{\pm}_{ef} \rangle }{\langle l^{\pm}_{ef}, Z_{vef} \rangle }} +\ui(-1 \pm 1) \ln {\langle Z_{vef},Z_{vef} \rangle } \right]
\end{eqnarray}
\end{widetext}
where $Z_{vef}=g^{\dagger}_{ve} \bar{z}_{vf}$.{ $l^{\pm}_{ef}$ here is defined as $l^{\pm}=v(N_{ef})^{-1 \dagger} l^{\pm}_0$ with $l^{\pm}_0$ defined in (\ref{eq:def_l+-}), and $v(N_{ef}) \in \text{SU}(1,1)$ which encoding the unit normal.} $\langle Z_{ve'f},Z_{ve'f} \rangle$ has the same sign as $\langle Z_{vef},Z_{vef} \rangle$. The integrand is invariant under the following gauge transformations:
\begin{eqnarray}
	&g_{ve} \to g_{v} g_{ve}, \qquad z_{vf} \to \lambda_{vf} (g_{v}^{T})^{-1} z_{vf} \\ 
	&g_{ve} \to s_{ve} g_{ve}, \qquad s_{ve} = \pm 1\\
	&g_{ve} \to g_{ve}v_{e} , \qquad l^{\pm}_{ef} \to  v_{e} l^{\pm}_{ef}, 
\end{eqnarray}
where $g_v\in \text{SL}(2,\C), v_{e} \in \text{SU}(1,1)$, and $\lambda_{vf}\in\C\setminus\{0\}$.

It's worth to point out that both $S_{vf \pm}$ and $S_{vfx \pm}$ are purely imaginary, {and they are all proportional to $s_f$ which will be uniform scaled later to derive the asymptotics.} The real valued function $h$ is given by
\begin{equation}\label{eq:h_singular}
\begin{split}
	h_{vef}&={|\langle Z_{vef},Z_{vef} \rangle|} \sqrt{\left|\gamma - \ui + \frac{2 {\langle l^{-}_{ef}, Z_{vef} \rangle \langle Z_{vef}, l^{+}_{ef} \rangle}}{{\langle Z_{vef},Z_{vef} \rangle}} \right|}\\
\end{split}
\end{equation}
$h_{vef}$ can be $0$ when we integrate over $z$ on $\mathbb{CP}_1$ and $\text{SL}(2,\C)$ group elements $g$ in (\ref{eq:amplitude}), and the zeros of $h$ are exactly the points where we define the principle value, i.e. at $\langle Z, Z \rangle = 0$. However, as shown in Appendix \ref{ana_singu}, the singularities due to $h$ are of half order thus the final integral is remain finite at these points.

\section{Analysis of critical points}\label{sec3}
{
As we shown above, the actions $S_{vf \pm}$ and $S_{vfx \pm}$ are pure imaginary, and they are proportional to $s_f$. Thus we can use stationary phase approximation to evaluate the amplitude in the semi-classical limit where $s$ is uniformly scaled by a factor $\Lambda\to\infty$. Note that the denominator $h$ defined by (\ref{eq:h_singular}) in (\ref{eq:amp}) contains $1/2$ order singular point at $\langle Z,Z \rangle=0$, as shown in Appendix \ref{ana_singu}. Then the integral is of the following type
\begin{equation}
    I = \int \ud x \frac{1}{\sqrt{x-x_0}} g(x) \ue^{\Lambda S(x)} 
\end{equation}
Here $g$ is an analytic function which does not scale with $\Lambda$. There are two different asymptotic equations for such type integral according to the critical point $x_c$ located exactly at the branch point $x_0$ or away from it. According to \cite{bleistein1966uniform}, if $x_c$ located exactly at $x_0$, the leading order contribution will locate at the critical points (which is also the branch points), and the asymptotic expansion is given by
\begin{equation}\label{eq:asym_I}
	I \sim g(x_c) \frac{\pi \ue^{\ui \pi (\mu-2)/8}}{\Gamma(3/4)}\left(\frac{2}{\Lambda|\det H(x_c)|}\right)^{1/4} \ue^{\Lambda S(x_c)} 
\end{equation}
where $H(x_c)$ is the Hessian matrix at $x_c$, and $\mu = \text{sgn} \det H(x_c)$.

As we explain in the following sections, the critical points of Eq.(\ref{eq:amp}) are always located at the branch points, when every tetrahedron containing the timelike triangle $f$ also contain at least one spacelike triangle. It is quite generic to have every tetrahedron contain both timelike and spacelike triangles in a simplicial geometry. In addition, in case that we consider tetrahedra with all triangles timelike, for a single vertex amplitude, the critical point is again located at the branch points, when the boundary data give the closed geometrical boundary of a 4-simplex (i.e. the tetrahedra at the boundary are glued with shape matching). We don't consider the possibility other than (\ref{eq:asym_I}). 

}
\subsection{Equation of Motion}

Since both $S_{vf \pm}$ and $S_{vfx \pm}$ are purely imaginary, their critical points, or namely critical configurations, are solutions of equations of motion. The equations of motion are given by variations of $S$'s respects to spinors $z$, $\text{SU}(1,1)$ group elements $v$ and $\text{SL}(2,\C)$ group elements $g$.

Before calculating the variation, we would like to introduce a decomposition of spinor $Z$. We first introduce following lemmas:
\begin{lemma}\label{lemma:l+o}
Given a specific $l^{+}$ satisfying $\langle l^{+},l^{+} \rangle=0$, there exist $\tilde{l}^{-}$, s.t. $\langle l^{+}, \tilde{l}^{-} \rangle=1, \langle \tilde{l}^{-}, \tilde{l}^{-} \rangle=0$. For two elements $\tilde{l}_{1}^{-}$ and $\tilde{l}_{2}^{-}$ satisfying the condition, they are related by
\begin{equation}
	\tilde{l}_{1}^{-} = \tilde{l}_{2}^{-}+ \ui \eta l^{+}, \qquad \qquad \eta \in R
\end{equation}
\end{lemma}
This is easy to proof since $\langle \tilde{l}^{-}+ \ui \eta l^{+}, \tilde{l}^{-}+ \ui \eta l^{+} \rangle =\eta^2 \langle l^{+}, l^{+}\rangle +  \langle \tilde{l}_{2}^{-}, \tilde{l}_{2}^{-} \rangle - \ui \eta \langle l^{+}, \tilde{l}^{-} \rangle + \ui \eta \langle \tilde{l}^{-}, l^{+} \rangle$ and $\langle l^{+}, \tilde{l}^{-} + \ui \eta l^{+} \rangle = \langle l^{+}, \tilde{l}^{-} \rangle + \ui \eta \langle l^{+}, l^{+}\rangle$.

\begin{lemma}\label{lemma:basis}
For a given $l^{+}$ and $\tilde{l}^{-}$ defined by Lemma \ref{lemma:l+o}, $l^{+}$ and $\tilde{l}^{-}$ form a null basis in two dimensional spinors space.
\end{lemma}
This lemma is proved by using the fact that given $l^{+}$ and $\tilde{l}^{-}$, there exists a $\text{SU}(1,1)$ element $\tilde{g}$, such that $l^{+}=\tilde{g} l_{0}^{+}$ and $\tilde{l}^{-}=\tilde{g} l_{0}^{-}$, and the fact that $l_{0}^{+}$ and $l_{0}^{-}$ forms a null basis.

With Lemma \ref{lemma:basis}, for a given $l^{+}$ or $l^{-}$, we have
\begin{theorem} For given $l^{+}$ and $\tilde{l}^{-}$ defined by Lemma \ref{lemma:l+o}, spinor $Z_{vef}$ always can be decomposed as 
\end{theorem}
\begin{equation}\label{eq:Z_com}
	Z_{vef} = \zeta_{vef} (\tilde{l}^{\mp }_{ef} + \alpha_{vef} l^{\pm}_{ef})
\end{equation}
where $\zeta_{vef} \in \C$ and  $\alpha_{vef} \in \C$.

At the vertex $v$, from the action $S_{vef+}$ ($S_{vef-}$), we only have $l^{+}$ ($l^{-}$) enters the action, thus we can choose arbitrarily $\tilde{l}^{\mp}_{vef}$ to form a basis. By Lemma \ref{lemma:l+o}, we can always write $\tilde{l}'^{\mp}_{vef} = \tilde{l}'^{\mp}_{vef} + \ui \Im(\alpha_{vef}) l^{\pm}_{ef}$ s.t.,
\begin{equation}\label{eq:Z_com_spe}
	Z_{vef} = \zeta_{vef} (\tilde{l}^{\mp }_{vef} + \Re(\alpha_{vef}) l^{\pm}_{ef}).
\end{equation}
$\Im(\alpha)$ is basis dependent. It is easy to check that if we replace $Z$ inside the action (\ref{eq:action}) by the decomposition (\ref{eq:Z_com}), the action is independent of $\Im(\alpha)$, which means that $\Im(\alpha)$ is a gauge freedom. 

We will drop the tilde on $\tilde{l}$ in the following. One should keep in mind that we have the freedom to choose the $l^{-}$ ($l^{+}$) such that for some vertices $v$, $\Im(\alpha_{vef})=0$.

From the decomposition of $Z_{vef}$, there is naturally a constraint. By the fact $Z_{vef}=g_{ve}^{\dagger} \bar{z}_{vf}$, we have
\begin{equation}\label{eq:z_Z}
	\bar{z}_{vf}=g_{ve}^{-1 \dagger} \; Z_{vef}=g_{ve'}^{-1 \dagger}\; Z_{ve'f}.
\end{equation}
In terms of decomposition of $Z_{vef}$
\begin{equation}
	g_{ve}^{-1 \dagger} \; (l^{\pm}_{ef} + \alpha_{vef} l^{\mp}_{ef} )=\frac{\zeta_{ve'f}}{\zeta_{vef}} g_{ve'}^{-1 \dagger} \; (l^{\pm}_{e'f} + \alpha_{ve'f} l^{\mp}_{vef})
\end{equation}
This can be written as
\begin{equation}\label{eq:sol_parall}
	g_{ve} \; J \; (l^{\pm}_{ef} + \alpha_{vef} l^{\mp}_{ef}) = \frac{\bar{\zeta}_{ve'f}}{\bar{\zeta}_{vef}} g_{ve'} \; J \; (l^{\pm}_{e'f} + \alpha_{ve'f} l^{\mp}_{ef})\\
\end{equation}
where we used the anti-linear map $J$:
\begin{equation}
	J (a,b)^{T}=(-\bar{b},\bar{a}), \;\;\; J g J^{-1}=-J g J =g^{-1 \dagger}
\end{equation}

\subsubsection{variation respect to $z$}
From the definition of $\text{SU}(1,1)$ inner product, for arbitrary spinor $u$ we have
\begin{equation}
	\begin{split}
	&\delta_{\bar{z}} \langle u, Z \rangle=\delta_{\bar{z}} (u^{\dagger} \eta g^{\dagger} \bar{z})=  (g \eta u)^{\dagger} \delta \bar{z},\\
	&\delta_{{z}} \langle Z, u \rangle=\delta_{{z}} ((g^{\dagger} \bar{z})^{\dagger} \eta u)= (\delta {{z}})^{T} (g \eta u)
	\end{split}
\end{equation}
Then it is straight forward to see the variation of $S_{vef}$ leading to
\begin{equation}
	\begin{split}
	\delta_{\bar{z}} S_{vef \pm} =& (\frac{n_f}{2} \pm \ui s_f) \frac{(g_{ve} \eta l^{\pm}_{ef})^{\dagger}} {{\langle l^{\pm}_{ef}, Z_{vef} \rangle}} -\ui(\rho_f \pm s_f) \frac{(g_{ve} \eta Z_{vef})^{\dagger}} {{\langle Z_{vef}, Z_{vef} \rangle}}
	\end{split}
\end{equation}
and
\begin{equation}
	\delta_{{z}} S=-\overline{\delta_{\bar{z}} S}
\end{equation}
which comes from the fact that $S$ is pure imaginary. With the definition of $S_{vf}$ in (\ref{eq:action}), after inserting the decomposition, we obtain the following equations
\begin{widetext}
\begin{align}
	&\delta S_{vf+} = (\gamma - \ui) s_f( \frac{g_{ve} \eta l^{+}_{ef}}{\bar{\zeta}_{vef}} - \frac{ g_{ve'} \eta l^{+}_{e'f} }{\bar{\zeta}_{ve'f} }) = 0 \qquad \text{with} \;\; Z= \zeta (l^{-} + \alpha l^{+} ) \label{eq:va_z_s+}\\
	&\delta S_{vf-} = - \ui s_f (\frac{g_{ve} \eta n_{vef}}{\Re(\alpha_{vef})\bar{\zeta}_{vef}} - \frac{ g_{ve'} \eta n_{ve'f} }{\Re(\alpha_{ve'f})\bar{\zeta}_{ve'f} }) =0 \qquad \text{with} \;\; Z= \zeta (l^{+} + \alpha l^{-} ) \label{eq:va_z_s-}\\
	&\delta S_{vfx + } = -(\gamma - \ui)s_f \frac{g_{ve'} \eta l^{+}_{e'f}}{\bar{\zeta}_{ve'f}} - \ui s_f \frac{ g_{ve} \eta n_{vef} }{\Re(\alpha_{vef})\bar{\zeta}_{vef} } =0 \quad \text{with} \;\; Z_{e'}= \zeta (l^{-} + \alpha l^{+} )  \;\&\; Z_{e}= \zeta (l^{+} + \alpha l^{-} ) \label{eq:va_z_sx}\\
	&\delta S_{vfx - } = (\gamma - \ui)s_f \frac{g_{ve} \eta l^{+}_{vef}}{\bar{\zeta}_{vef}} + \ui s_f \frac{ g_{ve'} \eta n_{ve'f} }{\Re(\alpha_{ve'f})\bar{\zeta}_{ve'f} } = 0 \quad \text{with} \;\; Z_{e}= \zeta (l^{-} + \alpha l^{+} ) \;\&\; Z_{e'}= \zeta (l^{+} + \alpha l^{-} )
\end{align}
\end{widetext}
where
\begin{equation}\label{eq:n_def}
	\begin{split}
	n_{vef}:=l^{+}_{ef}+ \ui \left(\gamma \Re({\alpha_{vef}})+\Im({\alpha_{vef}}) \right)  l^{-}_{ef}\\
	\end{split}
\end{equation}
Note that $n_{vef}$ here satisfies Lemma. \ref{lemma:basis} and can form a basis with $l^{-}_{ef}$ given in $S_{vef-}$.

\subsubsection{variation respect to $\text{SU}(1,1)$ group elements $v_{ef}$}
Since $l^{\pm}= v^{-1 \dagger} l^{\pm}_0$ with $v \in \text{SU}(1,1)$, the variation respect to $l^{\pm}$ is the variation respect to the $\text{SU}(1,1)$ group element $v$. If we considering a small perturbation of $v$ which is given by $v'=v \ue^{-\epsilon_i F^i}$, where $F^i$ are generators of $\text{SU}(1,1)$ group, we have $v'^{-1}= \ue^{ \epsilon_i F^i} v^{-1}$. The variation is then given by
\begin{equation}
 	\delta {v}^{-1}= \epsilon_i F^i v^{-1}, \;\;\; \delta v^{-1 \dagger}= \epsilon_i v^{-1 \dagger} (F^i)^{\dagger} 
 \end{equation} 
Thus for arbitrary spinor $u$, we have
\begin{equation}
\begin{split}
&\delta \langle u, m \rangle =\delta \langle u, v^{-1 \dagger} m_0 \rangle =  \epsilon^i \langle u,  v^{-1 \dagger} F_i^{\dagger}  m_0 \rangle\\
&\delta \langle m, u \rangle =\delta \langle v^{-1 \dagger} m_0, u \rangle = \epsilon^i \langle v^{-1 \dagger} F_i^{\dagger} m_0, u \rangle
\end{split}
\end{equation}
When $S_{ef}=S_{vef \pm }-S_{v'ef \pm}$, the variation reads
\begin{widetext}
\begin{equation}
	\begin{split}
	\delta S= & \epsilon^i  (\frac{n_f}{2} \mp \ui s_f)\left(\frac{\langle Z_{v'ef},  v_{ef}^{-1 \dagger} F_i^{\dagger}  l^{\pm}_0 \rangle}{{\langle Z_{v'ef}, l^{\pm}_{ef} \rangle }}-\frac{\langle Z_{vef},  v_{ef}^{-1 \dagger} F_i^{\dagger}  l^{\pm}_0 \rangle}{{\langle Z_{vef}, l^{\pm}_{ef} \rangle }} \right)\\
	&+ \epsilon^i  (\frac{n_f}{2} \pm \ui s_f) \left( \frac{\langle v_{ef}^{-1 \dagger} F_i^{\dagger} l^{\pm}_0, Z_{vef} \rangle} {{\langle l^{\pm}_{ef}, Z_{vef} \rangle}}-\frac{\langle v_{ef}^{-1 \dagger} F_i^{\dagger} l^{\pm}_0, Z_{v'ef} \rangle} {{\langle l^{\pm}_{ef}, Z_{v'ef} \rangle}} \right)
	\end{split}
\end{equation}
While $S_{ef}=S_{vef \pm }-S_{v'ef \mp}$, we have

\begin{equation}
	\begin{split}
	\delta S= & \epsilon^i  (\frac{n_f}{2})\left(\frac{\langle Z_{vef},  v_{ef}^{-1 \dagger} F_i^{\dagger}  l^{\pm}_0 \rangle}{{\langle Z_{vef}, l^{\pm}_{ef} \rangle }} -\frac{\langle Z_{v'ef},  v_{ef}^{-1 \dagger} F_i^{\dagger}  l^{\mp}_0 \rangle}{{\langle Z_{v'ef}, l^{\mp}_{ef} \rangle }}+ \frac{\langle v_{ef}^{-1 \dagger} F_i^{\dagger} l^{\pm}_0, Z_{vef} \rangle} {{\langle l^{\pm}_{ef}, Z_{vef} \rangle}}-\frac{\langle v_{ef}^{-1 \dagger} F_i^{\dagger} l^{\mp}_0, Z_{v'ef} \rangle} {{\langle l^{\mp}_{ef}, Z_{v'ef} \rangle}}\right)\\
	&+ \epsilon^i  s_f \left( \frac{\langle v_{ef}^{-1 \dagger} F_i^{\dagger} l^{\pm}_0, Z_{vef} \rangle} {{\langle l^{\pm}_{ef}, Z_{vef} \rangle}}+\frac{\langle v_{ef}^{-1 \dagger} F_i^{\dagger} l^{\pm}_0, Z_{v'ef} \rangle} {{\langle l^{\pm}_{ef}, Z_{v'ef} \rangle}} + \frac{\langle Z_{v'ef},  v_{ef}^{-1 \dagger} F_i^{\dagger}  l^{\mp}_0 \rangle}{{\langle Z_{v'ef}, l^{\mp}_{ef} \rangle }} + \frac{\langle Z_{vef},  v_{ef}^{-1 \dagger} F_i^{\dagger}  l^{\pm}_0 \rangle}{{\langle Z_{vef}, l^{\pm}_{ef} \rangle }}  \right)
	\end{split}
\end{equation}

Since $F^i=1/2 (\ui \sigma_3, \sigma_1, \sigma_2, )$ is $\text{SU}(1,1)$ generators, we have
\begin{eqnarray}
&(F^0)^{\dagger} l^{\pm}_0= \frac{\ui}{2 \sqrt{2}} \left(\begin{array}{ll} 1 &0 \\0 & -1 \end{array} \right) \left(\begin{array}{l} 1\\ \pm 1 \end{array} \right) = \frac{\ui}{2 } l^{\mp}_0\\
&(F^1)^{\dagger} l^{\pm}_0= \frac{1 }{2 \sqrt{2}} \left(\begin{array}{ll} 0 &1 \\1 &0 \end{array} \right) \left(\begin{array}{l} 1\\ \pm 1 \end{array} \right) =\pm \frac{1}{2} l^{\pm}_0\\
&(F^2)^{\dagger} l^{\pm}_0= \frac{1}{2 \sqrt{2}} \left(\begin{array}{ll} 0 &-\ui \\ \ui &0 \end{array} \right) \left(\begin{array}{l} 1\\ \pm1 \end{array} \right) =\mp \frac{1}{2} l^{\mp}_{0}
\end{eqnarray}
Then in the first case we only left with one equation, which reads
\begin{equation}
		0= (\frac{n_f}{2} \mp \ui s_f)\left(\frac{\langle Z_{v'ef}, \ui l^{\mp}_{ef} \rangle}{{\langle Z_{v'ef}, l^{\pm}_{ef} \rangle }}-\frac{\langle Z_{vef}, \ui l^{\mp}_{ef} \rangle}{{\langle Z_{vef}, l^{\pm}_{ef} \rangle }} \right)+(\frac{n_f}{2} \pm \ui s_f) \left( \frac{\langle \ui l^{\mp}_{ef}, Z_{vef} \rangle} {{\langle l^{\pm}_{ef}, Z_{vef} \rangle}}-\frac{\langle \ui l^{\mp}_{ef}, Z_{v'ef} \rangle} {{\langle l^{\pm}_{ef}, Z_{v'ef} \rangle}} \right)
\end{equation}
\end{widetext}
 After inserting the decomposition $Z = \zeta (l^{\mp} + \alpha l^{\pm}) $ correspondingly, we get
\begin{equation}
	\begin{split}
	0= &(\frac{n_f}{2} \mp \ui s_f)\left(\bar{\alpha}_{v'ef}-\bar{\alpha}_{vef} \right)+ (\frac{n_f}{2} \pm \ui s_f) \left( \alpha_{v'ef}-\alpha_{vef} \right)\\
	=&2 \ui s_f \gamma  \Re(\alpha_{v'ef}-\alpha_{vef}) \pm 2 \ui s_f \Im(\alpha_{vef}-{\alpha}_{v'ef}) 
	\end{split}
\end{equation}
The solution reads
\begin{equation}\label{eq:trans_v_+}
	\gamma \Re({\alpha_{vef}}) \mp \Im({\alpha_{vef}})=\gamma \Re({\alpha_{v'ef}}) \mp \Im({\alpha_{v'ef}})
\end{equation}
Here $\Im(\alpha)$ is the decomposition of $Z$ respect to $l^{\mp}_{ef}$ specified by $v_{ef}$. Note that in this case, we only have $l^{+}_{ef} (l^{-}_{ef})$ in the action, thus there is an ambiguity of $v_{ef}$. However, changing $v_{ef}$ corresponds to adding the same constant to both $\Im(\alpha_v)$ and $\Im(\alpha_v')$, thus the relation is kept unchange. After absorbing $\Im(\alpha)$ into $\tilde{l}$ by a redefinition, the equation actually tells us that,
\begin{equation}
	\tilde{l}^{\mp}_{vef} - \tilde{l}^{\mp}_{v'ef} = \pm \gamma (\Re({\alpha_{vef}}) - \Re({\alpha_{v'ef}})) l^{\pm}_{ef} 
\end{equation}
which fixes the transformation of $\tilde{l}_{vef}$ between vertices and removes the ambiguity between different vertices $v$ in the bulk. With this redefinition, it is easy to see that $n_{vef}$ defined in (\ref{eq:n_def}) satisfies $n_{vef}=n_{ve'f}$, thus we ignore the $v$ variable and define
\begin{eqnarray}
	n_{ef} := n_{vef} = n_{v'ef}
\end{eqnarray}

In the mixing case there will be two different equations for $F_2$ and $F_3$,  which leads to 
\begin{widetext}
\begin{align}
	0= & \frac{n_f}{2}\left( \Re \frac{\langle Z_{v'ef},  l^{\pm} \rangle}{{\langle Z_{v'ef}, l^{\mp}_{ef} \rangle }}- \Re \frac{\langle Z_{vef}, l^{\mp} \rangle}{{\langle Z_{vef}, l^{\pm}_{ef} \rangle }} \right) \pm \ui s_f \left( \ui \Im \frac{\langle Z_{v'ef},  l^{\pm} \rangle}{{\langle Z_{v'ef}, l^{\mp}_{ef} \rangle }}+ \ui \Im\frac{\langle Z_{vef}, l^{\mp} \rangle}{{\langle Z_{vef}, l^{\pm}_{ef} \rangle }} \right)\\
	0= & \frac{n_f}{2}\left( \Re \frac{\langle Z_{v'ef},  l^{\pm} \rangle}{{\langle Z_{v'ef}, l^{\mp}_{ef} \rangle }}+ \Re \frac{\langle Z_{vef}, l^{\mp} \rangle}{{\langle Z_{vef}, l^{\pm}_{ef} \rangle }} \right) \pm \ui s_f \left( \ui \Im \frac{\langle Z_{v'ef},  l^{\pm} \rangle}{{\langle Z_{v'ef}, l^{\mp}_{ef} \rangle }}- \ui \Im\frac{\langle Z_{vef}, l^{\mp} \rangle}{{\langle Z_{vef}, l^{\pm}_{ef} \rangle }} \right)
\end{align}
The equations give the solution
\begin{eqnarray}
	&\label{eq:trans_v_x+}\gamma \Re(\alpha_{v'ef}) \pm \Im(\alpha_{v'ef}) =0, \qquad \text{with} \qquad \; Z_{v'ef}= \zeta_{v'ef} ( l^{\pm}_{ef} + \alpha_{v'ef} l^{\mp}_{ef}) \notag\\
	&\label{eq:trans_v_x-}\gamma \Re(\alpha_{vef}) \mp \Im(\alpha_{vef}) =0, \qquad \text{with} \qquad  \; Z_{vef}= \zeta_{vef} ( l^{\mp}_{ef} + \alpha_{vef} l^{\pm}_{ef}) \notag
\end{eqnarray}
\end{widetext}
Here $l^{+}$ and $l^{-}$ completely fix the group element $v$. $\alpha$ corresponds to the decomposition of $Z$ with these $l^{+}$ and $l^{-}$. The $n_{vef}$ in this case is simply $n_{vef}= l_{ef}^{+}$.

\subsubsection{variation respect to $\text{SL}(2,\C)$ elements $g$}
With the small perturbation of $g$ which is given by ${g}'={g} \ue^{ L}$, the variation of $\text{SL}(2,\C)$ group element $g$ is given by
\begin{equation}
	\delta g=  g L, \;\;\; \delta g^{\dagger}= -  L^{\dagger} g^{\dagger}
\end{equation}
where $L$ is a linear combination of $\text{SL}(2,\C)$ generators, $L=\epsilon_{i} F^i + \tilde{\epsilon}_i G^i = (\epsilon_{i} + \ui \tilde{\epsilon_i})F^i$. Here $F$s are $\text{SU}(1,1)$ lie algebra generators defined as above, and we use the fact that in spin $1/2$ representation $G=\ui F$.
1
Then for arbitrary $u$, we have
\begin{equation}
\begin{split}
&\delta \langle u,Z \rangle =\delta \langle u, {g}^{\dagger} \bar{z} \rangle = \langle u, L^{\dagger} {g}^{\dagger} \bar{z} \rangle=  \langle u, L^{\dagger} Z \rangle\\
&\delta \langle Z, u \rangle =\delta \langle {g}^{\dagger} \bar{z}, u \rangle =(L^{\dagger} {g}^{\dagger} \bar{z})^{\dagger} \eta u= \langle L^{\dagger} Z, u \rangle.
\end{split}
\end{equation}
The variation leads to
\begin{widetext}
\begin{equation}\label{334}
	\begin{split}
	\delta S
	=& \sum_f \epsilon_{ef}(v) \left(-(\frac{n_f}{2} \mp \ui s_f)\left(\frac{ {\langle L^{\dagger} Z_{vef}, l^{\pm}_{ef} \rangle }}{{\langle Z_{vef}, l^{\pm}_{ef} \rangle }} \right)+(\frac{n_f}{2} \pm \ui s_f) \left( \frac{ {\langle l^{\pm}_{ef}, L^{\dagger} Z_{vef} \rangle}} {{\langle l^{\pm}_{ef}, Z_{vef} \rangle}}\right) \right.\\
	&\left. - \ui (\rho_f \pm s_f) \left(\frac{{\langle L^{\dagger} Z_{vef}, Z_{vef} \rangle}+{\langle Z_{vef}, L^{\dagger} Z_{vef} \rangle}} {{\langle Z_{vef}, Z_{vef} \rangle}} \right) \right)\\
	\end{split}
\end{equation}
\end{widetext}
where $\epsilon_{ef}(v)= \pm 1$ is determined according to the face orientation is consistent to the edge $e$ or opposite (up to a global sign). We have
\begin{equation}
	\epsilon_{ef}(v) = -\epsilon_{e'f}(v), \quad \epsilon_{ef}(v) = -\epsilon_{ef}(v').
\end{equation}
We write $\epsilon_{ef}(v)=+1$ in the following for simplicity, and recover general $\epsilon$ at the end of the derivation.

From the property of $\text{SU}(1,1)$ generator,
\begin{equation}
 	\eta F \eta= - F^{\dagger}
 \end{equation}
we have
\begin{equation}
\begin{split}
	\langle F^{\dagger} Z, u \rangle= - Z^{\dagger} F \eta u=  - Z^{\dagger} \eta F^{\dagger} u= -\langle Z, F^{\dagger} u \rangle\\
\end{split}
\end{equation}
Then (\ref{334}) can be written as
\begin{equation}\label{337}
	\begin{split}
	&\sum_{f} (\frac{n_f}{2} \mp \ui s_f)\left(\frac{ {\langle Z_{vef}, F^{\dagger}  l^{\pm}_{ef} \rangle }}{{\langle Z_{vef}, l^{\pm}_{ef} \rangle }} \right)\\
	& \qquad +(\frac{n_f}{2} \pm \ui s_f) \left( \frac{ {\langle l^{\pm}_{ef}, F^{\dagger} Z_{vef} \rangle}} {{\langle l^{\pm}_{ef}, Z_{vef} \rangle}}\right) = 0
	\end{split}
\end{equation}
and
\begin{equation}\label{338}
	\begin{split}
	&\sum_{f}- (\frac{n_f}{2} \mp \ui s_f)\frac{ {\langle Z_{vef}, F^{\dagger}  l^{\pm}_{ef} \rangle }}{{\langle Z_{vef}, l^{\pm}_{ef} \rangle }}+(\frac{n_f}{2} \pm \ui s_f)\frac{ {\langle l^{\pm}_{ef}, F^{\dagger} Z_{vef} \rangle}} {{\langle l^{\pm}_{ef}, Z_{vef} \rangle}} \\
	&\qquad \qquad - 2 \ui (\rho_f \pm s_f) \frac{{\langle Z_{vef},  F^{\dagger} Z_{vef} \rangle}} {{\langle Z_{vef}, Z_{vef} \rangle}}=0
	\end{split}
\end{equation}
After inserting the decomposition of $Z$ and solution of simplicity constraint, we have the following equations:
For both $S_{\pm}$, (\ref{337}) becomes
\begin{equation}\label{eq:va_e}
	\begin{split}
	&0=\delta_{F} S_{\pm} \\
	&=  \mp 2 \ui  \sum_{f} s_f {\langle l^{\mp}_{ef} \mp \ui (\gamma Re(\alpha_{vef}) \mp \Im(\alpha_{vef})) l^{\pm}_{ef}, F^{\dagger} l^{\pm}_{ef} \rangle}\\
	\end{split}
\end{equation}
(\ref{338}) will leads to different equations for different actions $S_{\pm}$ due to the appearance of $\langle Z_{vef},  F^{\dagger} Z_{vef} \rangle$ term.
The variation of $S_{+}$ reads
\begin{equation}\label{va_e1_+}
	\begin{split}
	&0=\delta_{G} S_{+}\\
	= & {- 2 \gamma} \sum_{f} s_f {\langle l^{-}_{ef} - \ui (\frac{1}{\gamma} \Re(\alpha_{vef})- \Im(\alpha_{vef})) l^{+}_{ef}, F^{\dagger} l^{+}_{ef} \rangle},\\
	\end{split}
\end{equation}
while the variation of $S_{-}$ reads
\begin{equation}\label{va_e1_-}
\begin{split}
	\delta_{G} S_{-} 
	&=2 \ui \sum_{f} s_f \frac{\langle n_{vef}, F^{\dagger} n_{vef} \rangle}{Re(\alpha_{vef})} + 2 \gamma \sum_{f} s_f \langle n_{vef}, F^{\dagger} l^{-}_{ef} \rangle.
\end{split} 
\end{equation}

{
\subsubsection{summary}
As a summary, after we introduce the decomposition of $Z$ as (\ref{eq:Z_com}):
\begin{equation}
	Z_{vef} = \zeta_{vef} (\tilde{l}^{\mp }_{ef} + \alpha_{vef} l^{\pm}_{ef})
\end{equation}
and a spinor $n$ as (\ref{eq:n_def})
\begin{equation}
	\begin{split}
	n_{vef}:=l^{+}_{ef}+ \ui \left(\gamma \Re({\alpha_{vef}})+\Im({\alpha_{vef}}) \right)  l^{-}_{ef}\\
	\end{split}
\end{equation}
the equation of motion is given by the following equations
\begin{itemize}
	\item parallel transport equations
\begin{widetext}
	\begin{align}
		&S_{vf+} :  \frac{g_{ve} \eta l^{+}_{ef}}{\bar{\zeta}_{vef}} = \frac{ g_{ve'} \eta l^{+}_{e'f} }{\bar{\zeta}_{ve'f} }, & g_{ve}^{-1 \dagger} \; (l^{-}_{ef} + \alpha_{vef} l^{+}_{ef} )=\frac{\zeta_{ve'f}}{\zeta_{vef}} g_{ve'}^{-1 \dagger} \; (l^{-}_{e'f} + \alpha_{ve'f} l^{+}_{vef}) \\
		&S_{vf-} : \frac{g_{ve} \eta n_{vef}}{\Re(\alpha_{vef})\bar{\zeta}_{vef}} = \frac{ g_{ve'} \eta n_{ve'f} }{\Re(\alpha_{ve'f})\bar{\zeta}_{ve'f} } , & g_{ve}^{-1 \dagger} \; (l^{+}_{ef} + \alpha_{vef} l^{-}_{ef} )=\frac{\zeta_{ve'f}}{\zeta_{vef}} g_{ve'}^{-1 \dagger} \; (l^{+}_{e'f} + \alpha_{ve'f} l^{-}_{vef})\\
		&S_{vfx + } : \frac{ g_{ve} \eta n_{vef} }{\Re(\alpha_{vef})\bar{\zeta}_{vef} } = -(1+ \ui \gamma )\frac{g_{ve'} \eta l^{+}_{e'f}}{\bar{\zeta}_{ve'f}}  , &g_{ve}^{-1 \dagger} \; (l^{+}_{ef} + \alpha_{vef} l^{-}_{ef} )=\frac{\zeta_{ve'f}}{\zeta_{vef}} g_{ve'}^{-1 \dagger} \; (l^{-}_{e'f} + \alpha_{ve'f} l^{+}_{vef})\\
		&S_{vfx - } : -(1+ \ui \gamma ) \frac{g_{ve} \eta l^{+}_{vef}}{\bar{\zeta}_{vef}} =\frac{ g_{ve'} \eta n_{ve'f} }{\Re(\alpha_{ve'f})\bar{\zeta}_{ve'f} }   & g_{ve}^{-1 \dagger} \; (l^{-}_{ef} + \alpha_{vef} l^{+}_{ef} )=\frac{\zeta_{ve'f}}{\zeta_{vef}} g_{ve'}^{-1 \dagger} \; (l^{+}_{e'f} + \alpha_{ve'f} l^{-}_{vef})
\end{align}
Here $S_{vf\pm} = S_{ve'f\pm}- S_{vef\pm}$, $S_{vfx\pm} = S_{ve'f\pm}- S_{vef\mp}$ with $S_{vef\pm}$ is the action given in (\ref{action1}), the same for $S_{ef \pm}$ and $S_{efx\pm}$.
\item vertcies relations
\begin{align}
	& S_{ef \pm} : & \gamma \Re({\alpha_{vef}}) \mp \Im({\alpha_{vef}})=\gamma \Re({\alpha_{v'ef}}) \mp \Im({\alpha_{v'ef}})\\
	& S_{ef \pm x} : & \gamma \Re({\alpha_{vef}}) \mp \Im({\alpha_{vef}})=\gamma \Re({\alpha_{v'ef}}) \pm \Im({\alpha_{v'ef}} ) = 0
\end{align}
\item closure constraints
	\begin{align}
			&0=  - 2 \ui  \sum_{f\; \text{/w} \;S_{+(x)}} s_f {\langle l^{-}_{ef} - \ui (\gamma \Re(\alpha_{vef}) - \Im(\alpha_{vef})) l^{+}_{ef}, F^{\dagger} l^{+}_{ef} \rangle} + 2 \ui \sum_{f\; \text{/w} \;S_{-(x)}} s_f \langle n_{ef}, {F}^{\dagger} l^{-}_{ef} \rangle \\
			&0= {- 2 \gamma} \sum_{f\; \text{/w} \;S_{+(x)}} s_f {\langle l^{-}_{ef} - \ui (\frac{1}{\gamma} \Re(\alpha_{vef})- \Im(\alpha_{vef})) l^{+}_{ef}, {F}^{\dagger} l^{+}_{ef} \rangle} +2 \sum_{f\; \text{/w} \;S_{-(x)}} \ui s_f \frac{\langle n_{ef}, {F}^{\dagger} n_{ef} \rangle}{\Re(\alpha_{vef})} + \gamma s_f \langle n_{vef}, {F}^{\dagger} l^{-}_{ef} \rangle 
\end{align}
\end{widetext}

\end{itemize}
}

\subsection{Bivector representation}
For given spinors $l^{-}$ and $l^{+}$, there is a 3-vector $v^i$ associated to them
\begin{equation}
    v^i= 2  \langle l^{+}, F^{i} l^{-} \rangle 
\end{equation}
From which we can define a $\text{SU}(1,1)$ valued bivector in spin-$\frac{1}{2}$ representation
\begin{widetext}
\begin{equation}\label{eq:bivector_tensorp}
\begin{split}
      V=2 \langle l^{+}, F^{i} l^{-} \rangle F^{i}=-  2 (l^{+})^{\dagger} (F^{i})^{\dagger} \eta l^{-} F^{i} = - \frac{1}{2} (l^{+})^{\dagger} \sigma_{i} \eta l^{-} \sigma_{i} = -\eta l^{-} \otimes (l^{+})^{\dagger} + \frac{1}{2} \langle l^{+}, l^{-} \rangle I_2
\end{split}
\end{equation}
\end{widetext}
where we use the fact $\eta F \eta=- F^{\dagger}$ and the completeness of pauli matrix.
Since $\langle l^{-}, F l^{+} \rangle = - \langle l^{+}, F l^{-} \rangle$, 
\begin{equation}\label{eq:v_bi_tp}
     V= - 2 \langle l^{-}, F^{i} l^{+} \rangle F_{i}=  \eta l^{+} \otimes (l^{-})^{\dagger} - \frac{1}{2} \langle l^{+}, l^{-} \rangle I_2
\end{equation}
From the fact
\begin{equation}
	K^i=-K_i = J^{0i}, \qquad J^i=J_i=\frac{1}{2}{\epsilon^{0i}}_{jk} J^{jk}
\end{equation}
where $J^{i} =  * K^{i}$. We have in spin $1/2$ representation $* \to \ui$ and $J^{i} =   \ui K^{i}$.
The bivector can be encoded into $\text{SL}(2,\C)$ bivector that in spin-1 representation reads 
\begin{equation}
 V^{IJ}=\left( \begin{array}{llll} 0 & -v^1 & -v^2 & 0\\ v^1 & 0 & v^0 & 0\\ v^2 & -v^0 & 0 & 0\\0 & 0 & 0 & 0 \end{array} \right),
\end{equation}
Then $(* V)^{IJ}$ reads
\begin{equation}
(* V)^{IJ}=\left( \begin{array}{llll} 0 & 0 & 0 & v^0\\ 0 & 0 & 0 & -v^2\\ 0 & 0 & 0 & v^1\\-v^0 & v^2 & -v^1 & 0 \end{array} \right)= (v_{ef}^I \wedge u^J)
\end{equation}
where the encoded 4-vector $v_{ef}^{I}:=(v^0, -v^2, v^1,0)$, $u^I=(0,0,0,1)$. Clearly one can see that
\begin{equation}
	v^I = \ui (\bra{l^{-}} \hat{\sigma}^{I} \ket{l^{+}} + u^I)
\end{equation}
where $\hat{\sigma}=(\sigma_0, -\sigma_1,-\sigma_2,-\sigma_3)$.

Since $\langle l^{+}, F^{i} l^{-} \rangle = \langle l^{+}_0, v^{\dagger} F^{i} v^{-1 \dagger} l^{-}_{0} \rangle$, in this sense, $v_i$ is nothing else but the $\text{SO}(1,2)$ rotation of 3 vector $v_0 = ( 0, 0, 1 )$ with group element $v^{-1 \dagger}$.

Similarly, we can define
\begin{equation}\label{eq:w_define}
    W^{\pm}= 2 \ui \langle l^{\pm}, F^{i} l^{\pm} \rangle F^{i}=   - \ui \eta l^{\pm} \otimes (l^{\pm})^{\dagger} 
\end{equation}
with
\begin{equation}
W^{\pm IJ}= w_{ef}^{\pm I} \wedge u^J, \qquad w^{\pm I}:=\bra{l^{\pm}} \hat{\sigma}^{I} \ket{l^{\pm}}
\end{equation}
Here $w_{ef}^{\pm I}$ is a null vector $w_{ef}^{\pm I} {w_{ef}^{\pm}}_{I}=0$.

We introduce $\text{SO}(1,3)$ group elements $G$ given by
\begin{equation}
	G_{ve} = \pi(g_{ve})
\end{equation}
where $\pi : \text{SL}(2,\C) \to \text{SO}(1,3)$.
 Since the action (\ref{eq:action}) is invariant under the transformation $g_{ve} \to \pm g_{ve}$, two group elements related to $g_{ve}$ are gauge equivalent if they satisfy
\begin{equation}
	\tilde{G}_{ve} = G_{ve} I^{s_{ve}}, \qquad s_{ve}=\{ 0,1\}
\end{equation}
where $I$ is the inversion operator. With this gauge transformation, we can always assume $G_{ve} \in \text{SO}_{+}(1,3)$.

We can write the critical equations in terms of bivectors. The detailed analysis is in Appendix \ref{app:bivector_critical}. Given any solution to the critical equations, we can define a bivector
 \begin{equation}\label{eq:xvef_def}
	\begin{split}
		 X_{vef}&= -2 \ui \langle l^{-}, F^{i} l^{+} \rangle F_{i} -\ui \bar{\alpha}_{vef} \langle l^{+}, F^{ i} l^{+} \rangle F_{i} \\
		 &= V_{ef} - (\Im(\alpha_{vef}) + \Re(\alpha_{vef}) *)W_{ef}^{+}
	\end{split}
\end{equation}
or
 \begin{equation}
	\begin{split}
		 X_{vef}&= - 2 \ui \langle n, F^{\dagger i} l^{-} \rangle F_{i} - \frac{\ui + \gamma}{(1+\gamma^2)\Re({\alpha}_{vef}) } \langle n, F^{\dagger i} n \rangle F_{i}\\
		 & = - V_{ef} - \frac{1 - \gamma *}{(1+\gamma^2)\Re({\alpha}_{vef}) } W^{+}_{ef}
	\end{split}
\end{equation}
corresponding to their action is composited by $S_{vef+}$ or $S_{vef-}$. Here $V_{ef}$ is a spacelike bivector and $W_{ef}$ is a null bivector. In spin-1 representation, we can express the above bivector as
 \begin{equation}
      X_{ef}^{IJ}=(*) (\tilde{v}_{vef}^I \wedge \tilde{u}_{vef}^J)
 \end{equation}
 where
 \begin{align}
	 &\tilde{v}_{vef} = \left\{ \begin{array}{ll} v_{ef} - \Im(\alpha_{vef}) w_{ef}^{+},  & S_{vef+} \\  v_{ef}- \frac{\gamma}{(1+\gamma^2)\Re(\alpha_{vef})} w_{ef}^{+}, & S_{vef-} \end{array} \right. \; \\
	 & \tilde{u}_{vef} = \left\{ \begin{array}{ll} u + \Re(\alpha_{vef}) w_{ef}^{+},  & S_{vef+} \\  u + \frac{1}{(1+\gamma^2)\Re(\alpha_{vef})} w_{ef}^{+}, & S_{vef-} \end{array} \right.
 \end{align}
 with
 \begin{align}
	& v_{ef} = \left\{ \begin{array}{ll} - 2 \ui \langle l^{-}_{ef}, F^{ i} l^{+}_{ef} \rangle,  & S_{vef+} \\ - 2 \ui \langle n_{ef}, F^{ i} l^{-}_{ef} \rangle & S_{vef-} \end{array} \right. , \\
	& w^{+}_{ef} =\left\{ \begin{array}{ll} 2 \langle l^{+}_{ef}, F^{ i} l^{+}_{ef} \rangle,  & S_{vef+} \\ 2 \langle n_{ef}, F^{i} n_{ef} \rangle & S_{vef-} \end{array} \right.
 \end{align}

The bivector $X_{vef}$ satisfies the parallel transport equation:
 \begin{equation}\label{eq:para_trans_t}
     g_{ve} X_{vef} g_{ve}^{-1}=g_{ve'} X_{ve'f} g_{ve'}^{-1}
 \end{equation}
 This corresponds to
 \begin{equation}
     X_{f}(v):=g_{ve} X_{vef} g_{ev} = v^{I}_{ef}(v) \wedge N^I_{e}(v)
 \end{equation}
 where
 \begin{equation}
     v^{I}_{ef}(v) := G_{ve} \tilde{v}_{vef}, \qquad N^I_e(v) = G_{ve} \tilde{u}_{vef}
 \end{equation}
The closure constraint in terms of the bivector variable then reads
 \begin{equation}\label{eq:clo_bivec}
     2 \sum_{f} \gamma \epsilon_{ef}(v) s_f X_{f}(v)=  \sum_{f} \epsilon_{ef}(v) B_{f}(v) =0
 \end{equation}
 where $B_f= 2 \gamma s_f X_f =n_f X_f$ with $B_f^2= - n_f^2$.
 Note that the closure constraint is composed by two independent equations enrolling $\tilde{v}$ and $w^{+}$
 \begin{equation}\label{eq:clo_vec}
	\begin{split}
	 &\sum_{f} \epsilon_{ef}(v) \tilde{v}_{vef} = 0,\\
	 &\left\{ \begin{array}{ll} \sum_{f} \epsilon_{ef}(v) \Re(\alpha_{vef}) w^{+}_{ef} = 0, & S_{vef+}\\ \sum_{f} \epsilon_{ef}(v) (\Re(\alpha_{vef})^{-1} w^{+}_{ef} = 0, & S_{vef-}\end{array} \right.
	 \end{split}
 \end{equation}

\subsection{Timelike tetrahedron containing both spacelike and timelike triangles}

The timelike tetrahedron in a generic simplicial geometry contains both spacelike and timelike triangles. For spacelike triangles, the irreps of $\text{SU}(1,1)$ are in the discrete series, in contrast to the continuous series used in timelike triangles. The simplicity constraint is also different from (\ref{eq:simplicity}). This leads to different face actions on triangles with different signature, and the total action is expressed by the sum of these actions. The action on spacelike triangle and corresponding critical point equations have already been derived in \cite{Kaminski:2017eew}. The results are reviewed in Appendix \ref{app:sp_face}.

The variations with respect to $z_{vf}$ and $v_{ef}$ give equations of motions (\ref{eq:para_trans_t}) for timelike triangles and (\ref{eq:para_trans_sp}) for spacelike triangles respectively. In addition, for timelike triangles, solutions should satisfy (\ref{eq:trans_v_+}), (\ref{eq:trans_v_x+}) or (\ref{eq:trans_v_x-}).

The variation respect to $\text{SL}(2, \C)$ group element $g_{ve}$ involves all faces connected to $e$, which may include both spacelike and timelike triangles. In general, from (\ref{eq:va_e} - \ref{va_e1_-}) and  (\ref{eq:sp_cl_1}-\ref{eq:sp_cl_2}), the action including different types of triangles gives
\begin{widetext}
\begin{equation}\label{mix_clo_1}
	\begin{split}
	\delta_{F} S= &  - 2 \ui  \sum_{f\; \text{/w} \;S_{+(x)}} s_f {\langle l^{-}_{ef} - \ui (\gamma \Re(\alpha_{vef}) - \Im(\alpha_{vef})) l^{+}_{ef}, F^{\dagger} l^{+}_{ef} \rangle}\\
	& + 2 \ui \sum_{f\; \text{/w} \;S_{-(x)}} s_f \langle n_{ef}, {F}^{\dagger} l^{-}_{ef} \rangle -2 \sum_{f\; \text{/w} \;S_{\text{sp}}} j_f{\langle \xi^{\pm}_{ef}, F^{\dagger} \xi^{\pm}_{ef} \rangle}=0\\
	\end{split}
\end{equation}
\begin{equation}\label{mix_clo_2}
	\begin{split}
	& {- 2 \gamma} \sum_{f\; \text{/w} \;S_{+(x)}} s_f {\langle l^{-}_{ef} - \ui (\frac{1}{\gamma} \Re(\alpha_{vef})- \Im(\alpha_{vef})) l^{+}_{ef}, {F}^{\dagger} l^{+}_{ef} \rangle}\\
	& +2 \sum_{f\; \text{/w} \;S_{-(x)}} \ui s_f \frac{\langle n_{ef}, {F}^{\dagger} n_{ef} \rangle}{\Re(\alpha_{vef})} + \gamma s_f \langle n_{vef}, {F}^{\dagger} l^{-}_{ef} \rangle + 2 \ui \gamma \sum_{f\; \text{/w} \;S_{\text{sp}}} j_f{\langle \xi^{\pm}_{ef}, {F}^{\dagger} \xi^{\pm}_{ef} \rangle} = 0
	\end{split}
\end{equation}
Summation of the two equations leads to
\begin{equation}\label{eq:null_clo_mix}
 	(1 + \gamma^2) \sum_{f\; \text{/w} \;S_{+(x)}} s_f \Re(\alpha_{vef}) \langle l^{+}_{ef}, F^{i} l^{+}_{ef} \rangle + \sum_{f\; \text{/w} \;S_{-(x)}} s_f \frac{\langle n_{ef}, {F}^{i} n_{ef} \rangle}{\Re(\alpha_{vef})} =0
 \end{equation} 
\end{widetext}
 This equation only involves timelike triangles. Since ${w^{+ i}}_{ef}=\langle l^{+}_{ef}, {F^i}l^{+}_{ef} \rangle$ (or $w^{+i}_{ef}=\langle n_{ef}, F^i n_{ef} \rangle$ in $S_{-(x)}$ case) are null vectors, the above equation implies summing over null vectors equal to $0$. In a tetrahedron contains both timelike and spacelike triangles, the number of timelike triangles, which is also the number of null vectors here, is less than $4$. If one has less than 4 null vectors sum to $0$ in $4$-dimensional Minkowski space, then they are either trivial or colinear. The only possibility to have a nondegenerate tetrahdron from (\ref{eq:null_clo_mix}) is that all the timelike faces are in the action $S_{+}$ and set $\Re(\alpha)=0$. The solution reads
 \begin{equation}\label{eq:alpha_zero}
  	\Re(\alpha_{vef})=0 \qquad \& \qquad  \forall_{f \in t_e} , \; S_f = S_{+(x)}.
 \end{equation}
{It means that in order to have critical point, the action associated to each triangle $f$ of the tetrahedron $t_e$ must be $S_+$ or $S_{+x}$, other actions do not have stationary point.
 The closure constraint is now given by (\ref{mix_clo_1}) minus (\ref{mix_clo_2}) 
 }
 \begin{equation}\label{eq:clo_mix}
	\begin{split}
		 &{- 2 \ui} \sum_{f\; \text{/w} \;S_{+(x)}} s_f {\langle l^{-}_{ef} + \ui \Im(\alpha_{vef}) l^{+}_{ef}, {F}^{i} l^{+}_{ef} \rangle} = 0\\
		 &-2 \sum_{f\; \text{/w} \;S_{sp}} j_f{\langle \xi^{\pm}_{ef}, F^{i} \xi^{\pm}_{ef} \rangle}=0
	\end{split}
 \end{equation}

The parallel transport equations for timelike triangles still keep the same form as (\ref{eq:va_z_s+}-\ref{eq:va_z_sx}). After we impose condition (\ref{eq:alpha_zero}), the parallel transport equation becomes
\begin{equation}\label{eq:pa_mix}
	\begin{split}
			&g_{ve} l^{+}_{ef} \otimes (l^{-}_{ef} + \ui \Im(\alpha_{vef}) l^{+}_{ef})^{\dagger} g_{ev} \\
			&\qquad = g_{ve'} l^{+}_{e'f} \otimes (l^{-}_{e'f} + \ui \Im(\alpha_{ve'f}) l^{+}_{e'f})^{\dagger} g_{e'v} 
	\end{split}
\end{equation}
One recognize the same composition of spinors $l^{-}_{ef} + \ui \Im(\alpha_{vef}) l^{+}_{ef}$ in (\ref{eq:clo_mix}) and (\ref{eq:pa_mix}). This is exactly the spinor satisfying Lemma (\ref{lemma:l+o}). Recall (\ref{eq:trans_v_+}), coming from the variation respect to $SU(1,1)$ group elements $v_{ef}$, we have
\begin{equation}\label{eq:trans_v_mix}
	\Im(\alpha_{vef})=\Im(\alpha_{v'ef})
\end{equation} 
in $S_{+}$ case or $\Im(\alpha_{vef})=0$ in $S_{x+}$ case respectively. However, recall for $S_{+}$ case, there is an ambiguity in defining $\tilde{l}^{-}$ and $\Im(\alpha)$ from lemma \ref{lemma:l+o}. This ambiguity does not change the action, and gives the same vector $v^i = \langle \tilde{l}_{ef}^{-}, F^{i} l^{+}_{ef} \rangle$. Thus we can always remove the $\Im(\alpha_{vef})$ by a redefinition of $l^{-}_{ef}$, which does not change the geometric form of the critical equations. With (\ref{eq:trans_v_mix}), this redefinition will extended to both end points of the edge $e$. Thus we always make the choice that $\Im(\alpha_{vef})=0$ and drop all $\Im(\alpha_{vef})$ terms in (\ref{eq:clo_mix}) and (\ref{eq:pa_mix})

In bivector representation, we can build bivectors for timelike triangles,
\begin{equation}
 	 X_{ef}= * ( {v}_{ef} \wedge u), 
 \end{equation} 
 with $v_{ef}$ a normalized vector defined by $v_{ef}^I = \ui (\bra{l_{ef}^{+}} \hat{\sigma}^{I} \ket{l_{ef}^{-}}- u^{I})$. The parallel transportation equation implies we can define a bivector $X_f(v)$ independent of $e$
 \begin{equation}\label{eq:361}
 	X_{f}(v)=G_{ve} X_{ef} G_{ev}
 \end{equation}
 Clearly in this case we have
 \begin{equation}
 	N_e \cdot X_{f}(v) = 0, \qquad \text{with} \;\; N_e=G_{ve} u
 \end{equation}
 For spacelike triangles, the bivector is defined in (\ref{eq:bi_sp}). One see they have exactly the same form as in the timelike case and follow the same condition, except now $v_{ef}^{I}=\bra{\xi^{\pm}_{ef}} \hat{\sigma}^{I} \ket{\xi^{\pm}_{ef}}-\bra{\xi^{\pm}_{ef}}\ket{\xi^{\pm}_{ef}} u^I$ instead.
 With bivectors $X_{ef}$ and $X_f$, (\ref{eq:clo_mix}) becomes (after recover the sign factor $\epsilon_{ef}(v)$)
 \begin{equation}{}
 	\sum_{f\; \text{/w} \;S_{+(x)}} \epsilon_{ef}(v) s_f  X_f(v) - \sum_{f\; \text{/w} \;S_{sp}} \epsilon_{ef}(v) j_f X_f(v) = 0
 \end{equation}

In summary, the critical equations for a timelike tetrahedron with both timelike and spacelike triangles imply a nondegenerate tetrahedron geometry only when timelike triangles have action $S_{+(x)}$. Suppose we have a solution $(j_f, g_{ve},z_{vf})$, one can define bivectors
\begin{equation}\label{eq:b_def}
	B_{ef} = 2 A_f X_{ef} = 2 A_f * ( v_{ef} \wedge u)
\end{equation}
where 
\begin{equation}
	v_{ef}^I = \left\{ \begin{array}{ll} - \ui (\bra{l^{+}_{ef}} \sigma^I \ket{l^{-}_{ef}} - u^I ) & \text{for timelike triangle} \\\bra{\xi^{\pm}_{ef}} \sigma^I \ket{\xi^{\pm}_{ef}} - \langle \xi^{\pm},\xi^{\pm} \rangle u^I & \text{for spacelike case} \end{array} \right., 
\end{equation}
and 
\begin{equation}
A_f = \left\{ \begin{array}{ll} \gamma s_f = n_f/2 & \text{for timelike triangle} \\ \gamma j_f = \gamma n_f/2  & \text{for spacelike triangle} \end{array}  \right.
\end{equation}
We define $B_{ef}(v)$ as
\begin{equation}
	B_{f}(v) :=  G_{ve} B_{ef} G_{ev}
\end{equation}
The critical point equations imply
\begin{align}\label{eq:cri_eq_final_mix}
	&B_{ef}(v)=B_{e'f}(v)=B_{f}(v)\\
	& N_e \cdot B_f(v)=0\\
	&\sum_{f \in t_e} \epsilon_{ef}(v) B_{f}(v)=0  
\end{align}
where $N_e^I = G_{ve} u^I$, $\epsilon_{ef}(v)=\pm1$ and changes it's sign when exchanging vertex and edge variables. 

\subsection{Tetrahedron containing only timelike triangles}

Starting from the critical equations derived above, we can see what happens when all faces appear inside the closure constrain is timelike. For simplicity, we will use $S_{+}$ action as an example, the other cases will follow similar properties as they can be written in similar forms as $S_{+}$.

Suppose we have a solution to critical equations with all the face actions being $S_{+}$. As we have shown above, the solution satisfies two closure constraints,
\begin{align}
	&\sum_f s_f (v_{ef} + \Im(\alpha_{vef})w^{+}_{ef}) = 0,\\
	&\sum_f s_f \Re(\alpha_{vef}) w^{+}_{ef} = 0
\end{align}
Clearly here we have family of solutions generated by the continuous transformations
\begin{equation}\label{eq:c_ambi_clo}
	\begin{split}
		&\Re(\alpha_{vef}) \to  \tilde{C}_{ve} \Re(\alpha_{vef}), \\
		&\Im(\alpha_{vef}) \to \Im(\alpha_{vef}) + {C}_{ve} \Re(\alpha_{vef})
	\end{split}
\end{equation}
In other words, the closure constraint only fixes $\alpha$ up to $C_{ve}$ and $\tilde{C}_{ve}$. 

 Back to the bivectors inside the parallel transportation equation, it is easy to see, the bivector can be rewritten as
\begin{widetext}
\begin{equation}
	X= V + (\Im(\alpha) + \Re(\alpha) * ) W^{+} = X^0 + \Re(\alpha) ( C + \tilde{C} * ) W^{+}
\end{equation}
where $X_0 = V +  \Im(\alpha_{vef}^0)$ for some given $\Im(\alpha_{vef}^0)$.
Suppose we have a solution to some fixed $C$ and $\tilde{C}$, the parallel transported bivector then reads
\begin{equation}\label{eq:para_trans_t_1}
	G_{ve} X_{ef} G_{ev} = G_{ve} X_{ef}^0 G_{ev} + \Re(\alpha) ( C + \tilde{C} * ) G_{ve} W^{+}_{ef} G_{ev} = *( (G_{ve}\tilde{v}_{vef}) \wedge (G_{ve}\tilde{u}_{vef}) )
\end{equation}
\end{widetext}
From the fact that in spin-$1/2$ representation $* \to \ui$, we define $c : = C + \ui \tilde{C}$.

From the parallel transported vector $\tilde{v}_f := G_{ve}\tilde{v}_{vef}$ and $\tilde{u}_f := G_{ve} \tilde{u}_{vef}$, one can determine a null vector $\tilde{w}_f$ related to face $f=(e,e')$ uniquely up to a scale by
\begin{equation}\label{wvwu}
 \tilde{w}_f . \tilde{v}_f= \tilde{w}_f.\tilde{u}_f= 0
\end{equation}
From the definition of $\tilde{v}$ and $\tilde{u}$, we see that $w_{ef}.\tilde{u}_{vef}= w_{ef} . \tilde{v}_{vef} = 0$ and the same relation for $e'$. Since $G \in \text{SO}_{+}(1,3)$ which preserves the inner product, we then have
\begin{equation}\label{eq:null_equal}
	\tilde{w}_f \propto G_{ve} w_{ef} \propto G_{ve'} w_{e'f}
\end{equation}

Suppose a solution to critical equations determines a geometrical $4$-simplex up to scaling and reflection with normals $N_{e}(v)=G_{ve} u$ (Appendix \ref{app:geo} for the geometrical interpretation of the critical solution. We suppose the solution is non-degenerate here. The degenerate case will be discussed in Sec. \ref{sec5}). From this $4$-simplex, we can get its boundary tetrahedron with faces normals $v^g_{ef}(v) = G_{ve} v^s_{ef}$.  For two edges $e$ and $e'$ belong to the same face $f$, $N_e$ and $N_{e'}$ determine uniquely a null vector (up to scaling), which is perpendicular to $N_e$ and $N_{e'}$. Then from (\ref{wvwu}) and (\ref{eq:null_equal}), the vector is proportional to $\tilde{w}_{f}$. Then it implies that,
\begin{equation}
	 v^s_{ef} = \tilde{v}_{ef}+d_{ef} w_{ef}
\end{equation}
The tetrahedra determined by $ v^s_{ef}$ (by Minkowski Theorem) satisfy the length matching condition, which further constrain $d_{ef}$. 10 $d_{ef}$'s are over-constrained by 20 length matching conditions. $d_{ef}=0$ corresponds to a solution if the boundary data (relating to $\tilde{v}_{ef}$) also satisfy the length matching condition. We have the parallel transportation equation:
\begin{widetext}
\begin{equation}
  	g_{ve} X^0_{ef} g_{ev} + d_{ef} g_{ve} W^{+}_{ef} g_{ev} = g_{ve'} X^0_{e'f} g_{e'v} + d_{e'f} g_{ve'} W^{+}_{e'f} g_{e'v}
\end{equation}  
However, from (\ref{eq:para_trans_t_1}) we know that
\begin{equation}\label{eq:para_new}
	g_{ve} X^0_{ef} g_{ev} + \Re(\alpha_{ef}) c_{ve} g_{ve} W^{+}_{ef} g_{ev} = g_{ve'} X^0_{e'f} g_{e'v} + \Re(\alpha_{e'f}) c_{ve'} g_{ve'} W^{+}_{e'f} g_{e'v} 
\end{equation}
which means
\begin{equation}\label{eq:c_i_condition}
	(\Re(\alpha_{vef}) c_{ve} - d_{ef}) g_{ve} W^{+}_{ef} g_{ev}  = (\Re(\alpha_{ve'f}) c_{ve'} - d_{e'f})  g_{ve'} W^{+}_{e'f} g_{e'v}
\end{equation}
\end{widetext}
They are $10$ complex equations, with $5$ complex $c_{ve}$, thus again give an over-constrained system. 

A special case is that the boundary data itself satisfy the length matching condition. In this case, $d_{ef}=0$ correspond to a critical solution. It can be further proved that (\ref{eq:c_i_condition}) with $d_{ef}=0$ implies
\begin{equation}
	\forall_{e} \; c_{ve} = 0\label{special}
\end{equation}
The condition is nothing else but (\ref{eq:alpha_zero}), and it is easy to see that in this case the critical equations reduce to (\ref{eq:b_def} - \ref{eq:cri_eq_final_mix}).

\section{Geometric Interpetation and Reconstruction}\label{sec4}

The critical solutions of spinfoam action are shown to satisfy certain geometrical bivector equations, we would like to compare them with a discrete Lorentzian geometry. The general construction of a discrete Lorentzian geometry and the relation with critical solutions for spacelike triangles were discussed in detail in \cite{Han:2011re} and \cite{Kaminski:2017eew}. We will see that our solutions, which include timelike triangles, can be applied to a similar reconstruction procedure. We demonstrate the detailed analysis in Appendix \ref{app:geo}. The main result is summarized here. The result is valid when every timelike tetrahedron contains both spacelike and timelike triangles. It is also valid for tetrahedra containing only timelike triangles in the special case with Eq.(\ref{special}).

The following condition at a vertex $v$ implies the nondegenerate 4-simplex geometry:
\begin{equation}
	\prod_{e1,e2,e3,e4=1}^5 \det(N_{e1},N_{e2},N_{e3},N_{e4}) \neq 0\label{nondegcond}
\end{equation}
which means any $4$ out of $5$ normals are linearly independent. Since $N_{e}=G_{ve}u$, the above non-degeneracy condition is a constraint on $G_{ve}$. Here $u=(0,0,0,1)$ or $u=(1,0,0,0)$ for a timelike or spacelike tetrahedron.

Then we can prove that satisfying the nondegeneracy condition, each solution $B_{ef}(v)$ at a vertex $v$ determines a geometrical $4$-simplex uniquely up to shift and inversion. The bivectors $B^{\Delta}_{ef}(v)$ of the reconstructed $4$-simplex satisfy
\begin{equation}
	B^{\Delta}_{ef}(v) = r(v) B_{ef}(v) 
\end{equation}
where $r(v)=\pm 1$ relates to the 4-simplex (topological) orientation defined by an ordering of tetrahedra. The reconstructed normals are determined up to a sign
\begin{equation}
	N^{\Delta}_{ve} = (-1)^{s_{ve}} N_{ve}
\end{equation}
We can prove that for a vertex amplitude, the solution exists only when the boundary data determines tetrahedra that are glued with length-matching (the pair of glued triangles have their edge-lengths matched). 

Given the boundary data, we can determines geometric group elements $G^{\Delta} \in O(1,3)$ from reconstructed normals $N^{\Delta}$. Then it can be shown that, after one choose $s_{v}$ and $s_{ve}$, such that
\begin{equation}\label{eq:ori_match}
	\forall_{e} \; \det \; G_{ve}^{\Delta} = (-1)^{s_v} = r(v).
\end{equation}
$G_{ve}^{\Delta}$ relates to $G_{ve}$ by
\begin{equation}
	G_{ve} = G_{ve}^{\Delta} I^{s_{ve}} (I R_{u})^{s_{v}}
\end{equation}
where $R_N$ is the reflection respecting to normalized vector $N$ defined as
\begin{equation}
	(R_{N})^I_J=\mathbb{I}^I_J - \frac{2 N^I N_J}{N \cdot N}
\end{equation}
The choice of $s_{ve}=\pm1$ corresponds to a gauge freedom and is arbitrary here. Condition \ref{eq:ori_match} is called the orientation matching condition, which essentially means that the orientations of 5 boundary tetrahedra determined by the boundary condition are required to be the same. 

For a vertex amplitude, the non-degenerate geometric critical solutions exist if and only if the length matching condition and orientation matching condition are satisfied. Up to gauge transformations, there are two gauge inequivalent solutions which are related to each other by a reflection respect to any normalized 4 vector $e_{\alpha}$ (this reflection is referred to as the parity transformation in e.g. \cite{Han:2011re,Han:2011rf,Barrett:2009mw,Barrett:2009gg})
\begin{equation}
	\tilde{B}_{ef}(v) = R_{e_{\alpha}}(B_{ef}(v)), \qquad \tilde{s}_{v} =s_v +1
\end{equation}
which means
\begin{equation}
	\tilde{G}_{ve} = R_{e_{\alpha}} G_{ve} (I R_{N})
\end{equation}

Geometrically the second one corresponds to the reflected simplex. These two critical solutions correspond to the same 4-simplex geometry, but associates to different sign of the oriented 4-simplex volume $V(v)$. $\sgn(V(v))$ is referred to as the (geometrical) orientation of the 4-simplex\footnote{$\sgn(V(v))$ is a discrete analog of the volume element compatible to the metric in smooth pseudo-Riemannian geometry.}, which shouldn't be confused with $r(v)$. This result generalizes \cite{Kaminski:2017eew} to the spin foam vertex amplitude containing timelike triangles. 

The reconstruction can be extended to simplicial complex $\mathcal{K}$ with many $4$-simplices, in which some critical solutions of the full amplitude correspond to nondegenerate Lorentzian simplicial geometries on $\mathcal{K}$ (see Appendix \ref{app:geo}). But similar to the situation in \cite{Han:2011re,Han:2011rf}, 4-simplices in $\mathcal{K}$ may have different $\sgn(V(v))$. We may divide the complex $\mathcal{K}$ into sub-complexes, such that each sub-complex is globally orientated, i.e. the sign of the orientated volume $\sgn(V)$ is a constant. Then we have the following result

	For critical solutions corresponding to simplicial geometries with all $4$-simplices globally oriented, 
	picking up a pair of them corresponding to opposite global orientations, they satisfy
	\begin{equation}\label{eq:g_f_tilde_o}
		\tilde{G}_f = \left\{\begin{array}{l} R_{u_e} G_f(e) R_{u_e} \quad \text{internal faces} \\ I^{r_{e1}+r_{e0}} R_{u_{e1}} G_f(e_1,e_0) R_{u_{e0}} \quad \text{boundary faces} \end{array} \right.
	\end{equation} 
	where $G_f=  \prod_{v \subset \partial f} G_{e'v}G_{ve}$ is the face holonomy. We will use this result to derive the phase difference of their asymptotical contributions to the spin foam amplitude. Note that, the asymptotic formula of the spinfoam amplitude is given by summing over all possible configuration of orientations.

\section{Split signature and Degenerate $4$ simplex}\label{sec:de_sol}\label{sec5}

This section discusses the critical solutions that violate the non-degeneracy condition (\ref{nondegcond}). We refer to these solutions as degenerate solutions. If the non-degeneracy condition is violated, then in each 4-simplex, all five normals $N_e$ of tetrahedra $t_e$ are parallel, since we only consider nondegenerate tetrahedra \cite{Kaminski:2017eew}. When it happens with all $t_e$ timelike (or spacelike), with the help of gauge transformation $G_{ve} \to G G_{ve}$, we can write $N_e(v)=G_{ve}u,\ u=(0,0,0,1)$, where all the group variables $G_{ve} \in \text{SO}_{+}(1,2)$. However, when the vertex amplitude contains at least one timelike and one spacelike tetrahedron, the non-degeneracy condition (\ref{nondegcond}) cannot be violated since timelike and spacelike normals certainly cannot be parallel. Therefore the solutions discussed in this section only appear in the vertex amplitude with all tetrahedra timelike. Moreover, these degenerate solutions appears when the boundary data are special, i.e. correspond to the boundary of a split signature 4-simplex or a degenerate 4-simplex, as we see in a moment.

When the tetrahedron contains both timelike and spacelike triangles, the closure constraint (\ref{eq:null_clo_mix}) concerning $w$ involves at most $3$ null vectors, which directly leads to $\Re(\alpha_{vef})=0$ as the only solution. For degenerate solutions, the bivector $X_{f}(v)= g_{ve} X_{ef} g_{ev}$ in (\ref{eq:361}) becomes
\begin{equation}
	X_{f}(v) = * G_{ve} (v_{ef} \wedge u) G_{ev} = G_{ve} v_{ef} \wedge u = v^{g}_{vef} \wedge u
\end{equation}
The parallel transportation equation (\ref{eq:cri_eq_final_mix}) becomes
\begin{equation}\label{eq:vec_pa}
	v^g_{f}(v) = v^g_{ve} =v^g_{ve'} = 2 A_f G_{ve} v_{ef}.
\end{equation}  
Thus, the degenerate critical solutions satisfy
\begin{equation}\label{eq:vec_all}
	v^g_{f}(v) = v^g_{ve} =v^g_{ve'}, \qquad \sum_{f} \epsilon_{ef}(v) v^g_f(v) =0
\end{equation}
and the collection of vectors $v^g_f(v)$ is referred to as a vector geometry in \cite{Barrett:2009gg}.

In the case that all triangles in a tetrahedron are timelike, we use $S_{vf+}$ as an example. The degeneracy implies $G_{ve} u = G_{ve'} u =u$,. The parallel transportation equation (\ref{eq:para_trans_t_1}) becomes
\begin{widetext}
\begin{equation}\label{7.1}
	( G_{ve} \tilde{v}_{vef} - G_{ve'} \tilde{v}_{ve'f} ) \wedge u = c_{ve} \Re(\alpha_{vef}) G_{ve} w^{+}_{ef} \wedge u - c_{ve'} \Re(\alpha_{ve'f}) G_{ve'} w^{+}_{e'f} \wedge u.
\end{equation}
\end{widetext}
$c_{ve} = C_{ve} + \ui \tilde{C}_{ve}$ is the factor which solves the closure constrain with a given normalization of $\Re(\alpha_{vef})$, e.g. $\sum_f \Re(\alpha_{vef}) =1$ as shown in (\ref{eq:c_ambi_clo}). (\ref{7.1}) directly leads to
\begin{widetext}
\begin{align}
	\label{eq:deg_pa_1}& G_{ve} (\tilde{v}_{vef}+ {C}_{ve}\Re(\alpha_{vef})  {w}_{ef}) =  G_{ve'} (\tilde{v}_{ve'f}+ {C}_{ve}\Re(\alpha_{vef})  {w}_{ef}) \\
	\label{eq:deg_pa_2} & \tilde{C}_{ve}\Re(\alpha_{vef}) G_{ve} {w}_{ef} =  \tilde{C}_{ve'} \Re(\alpha_{ve'f}) G_{ve'} w_{e'f} 
\end{align}
\end{widetext}
Notice that from (\ref{eq:deg_pa_1}), since $w_{ef}$ is null and $w_{ef} \cdot v_{ef}=0$, we have
\begin{equation}
	G_{ve} {w}_{ef} \propto G_{ve'} w_{e'f}.
\end{equation}
It implies that (\ref{eq:deg_pa_2}) is only a function of $\tilde{C}$. However, at a vertex $v$, there are only $5$ independent $\tilde{C}$ variables out of $10$ equations. Thus (\ref{eq:deg_pa_2}) are over constrained equations and give $5$ consistency condition for $G_{ve}$ unless $\tilde{C}=0$. 

 Actually one can show that, there is no solution when $\tilde{C} \neq 0 $. We give the proof here. For simplicity, we only focus on a single $4$-simplex.
 
 Suppose we have solutions to above equations with $\tilde{C} \neq 0 $, then the following equations hold according to (\ref{eq:deg_pa_1}), (\ref{eq:deg_pa_2}) and the closure constraint (\ref{eq:clo_+})
\begin{equation}
	\begin{split}
	&	{v}^g_{f}(v) = v_{ef}^g(v) = v_{e'f}^g(v), \quad \sum_{f \subset t_e} \epsilon_{ef}(v) v_{ef}^g(v) = 0 \;,\\
	 & {w}^g_{f}(v) = w_{ef}^g(v) = w_{e'f}^g(v), \quad \sum_{f \subset t_e} \epsilon_{ef}(v) w_{ef}^g(v) = 0 ,
	\end{split}
\end{equation}
where
\begin{equation}
	\begin{split}
		& v_{ef}^g(v)=G_{ve}\tilde{v}_{ef} + {C_i} \Re(\alpha_{vef}) G_{ve}{w}_{ef}\\
		& w_{ef}^g(v)=\tilde{C_i}  \Re(\alpha_{vef}) G_{ve} {w}_{ef}
\end{split}
\end{equation}
Suppose $v^g$ satisfy the length matching condition. From above equations, $\tilde{v}^g_{ef}={v}^g_{ef} + a w^g_{ef}$ with arbitrary real number $a$ are also solutions. This means $\tilde{v}^g$ should also satisfy the length matching condition. However the transformation from $v$ to ${v}+ a w$ changes the edge lengths of the tetrahedron, and the length matching condition gives constraint to $a$. This conflict with the fact that $a$ is arbitrary to form the solution. It means that we can not have a solution with $\tilde{C} \neq 0 $ and length matching condition satisfied.

Thus, when boundary data satisfies the length matching condition, the only possible solution of (\ref{eq:deg_pa_2}) is $\tilde{C}_{ve}=0$. This corresponds to $\Re(\alpha)=0$ thus only possible with action $S_{+}$. One recognizes that this is the same condition as in the case of tetrahedron with both timelike and spacelike triangles, e.g. (\ref{eq:alpha_zero}). In this case $C_{ve}$ thus $\Im(\alpha)$ can be uniquely determined by the closure and length matching condition. The critical point equations again becomes (\ref{eq:vec_pa}) and (\ref{eq:vec_all})

In the end of this section, we introduce some relations between the vector geometry and non-degenerate split signature 4-simplex. As shown in Appendix \ref{app:flip_sig}, the vector geometries in $3$ dimensional subspace $V$ can be map to the split signature space $M'$ with signature $(-,+,+,-)$ (flip the signature of $u=(0,0,0,1)$), with the map $\Phi^{\pm}: \wedge^2 {M^4}' \to V$ for bivectors $B$,
\begin{equation}\label{def_phi_flip}
	\Phi^{\pm}(B) =  (\mp B - *' B) \cdot' u.
\end{equation}
$\Phi^{\pm}$ naturally induced a map from $g \in \text{SO}(2,2)$ to the subgroup $h \in \text{SO}(1,2)$, defined by
\begin{equation}
	\Phi^{\pm}(g B g^{-1}) = \Phi^{\pm}(g) \Phi^{\pm}(B)
\end{equation}
If the vertex amplitude has the critical solutions being a pair of non-gauge-equivalent vector geometries $\{ G_{ve}^{\pm} \}$,  they are equivalent to a pair of non-gauge-equivalent $\{G_{ve}\in SO(M') \}$ satisfying the nondegenerate condition. One of the non-degenerate $\{G_{ve} \}$ satisfies $G^{\pm}_{ve}=\Phi^{\pm}(G_{ve})$, while the other $\{\tilde{G}_{ve} \}$ satisfies
\begin{equation}
		\Phi^{\pm}(\tilde{G})= \Phi^{\pm}(R_{u} G R_{u}) = \Phi^{\mp}(G)
\end{equation}
When the vector geometries are gauge equivalent, the corresponding geometric $SO(M')$ solution is degenerate. In this case the reconstructed $4$ simplex is degenerate and the $4$ volume is $0$.

\section{Summary of Geometries}

We summarize all possible reconstructed geometries corresponding to critical configurations of Conrady-Hnybida extended spin foam model (include EPRL model) here. We first introduce the length matching condition and orientation matching condition for the boundary data. Namely, (1) among the 5 tetrahedra reconstructed by the boundary data (by Minkowski Theorem), each pair of them are glued with their common triangles matching in shape (match their 3 edge lengths), and (2) all tetrahedra have the same orientation. The amplitude will be suppressed asymptotically if orientation matching condition is not satisfied. 

For given boundary data satisfies length matching condition and orientation matching condition, we may have the following reconstructed $4$ simplex geometries corresponding to critical configurations of Conrady-Hynbida model:
\begin{itemize}
    \item Lorentzian $(-+++)$ $4$ simplex geometry: reconstructed by boundary data which may contains
    \begin{itemize}
        \item both timelike and spacelike tetrahedra,
        \item all tetrahedra being timelike.
        \item all tetrahedra being spacelike.
    \end{itemize}
    \item Split signature $(-++-)$ $4$ simplex geometry: This case is only possible when every boundary tetrahedron are timelike.
    \item Euclidean signature $(++++)$ $4$ simplex geometry: This case is only possible when every boundary tetrahedron are spacelike.
	\item Degenerate $4$ simplex geometry: This case is only possible when all boundary tetrahedron are timelike or all of them are spacelike.
\end{itemize}

When length matching condition is not satisfied, we might still have one gauge equivalence class of solutions which determines a single vector geometry. This solution exists again only when all boundary tetrahedron are timelike or all of them are spacelike. 

Our analysis is generalized to a simplicial complex $\mathcal{K}$ with many $4$-simplices. A most general critical configuration of Conrady-Hnybida model may mix all the types of geometries on the entire $\mathcal{K}$. One can always make a partition of $\mathcal{K}$ into sub-regions such that in each region we have a single type of reconstructed geometry with boundary. However, this may introduce nontrivial transitions between different types of geometries through boundary shared by them as suggested in \cite{Han:2011re}. It is important to remark that, if we take the boundary data of each $4$ simplex to contain at least one timelike and one spacelike tetrahedron, critical configurations will only give Lorentzian 4-simplices.

\section{Phase difference}\label{sec6}
In this section, we compare the difference of the phases given by a pair of critical solutions with opposite (global) $\sgn(V)$ orientations on a simplical complex $\mathcal{K}$. Recall that the amplitude is defined with SU(1,1) and SU(2) coherent states at the timelike and spacelike boundary. When we define the coherent state, we have a phase ambiguity from $K_1$ direction in SU(1,1) (or $J_3$ direction in SU(2)), thus the action is determined up to this phase. Thus the phase difference $\Delta S$ is the essential result in the asymptotic analysis of spin foam vertex amplitude. The phase difference at a spacelike triangle has already been discussed in \cite{Kaminski:2017eew}, we only focus on timelike triangles here. 

Given a timelike triangle $f$, in Lorentzian signature, the normals $N_e$ and $N_{e'}$ are spacelike and span a spacelike plane, while in split signature they form a timelike surface. The dihedral angles ${\Theta}_f$ at $f$ are defined as follows: In Lorentzian signature, the dihedral angle is ${\Theta}_f = \pi - \theta_f$ where
\begin{equation}\label{eq:def_dihedral_1}
	\cos \theta_f = N^{\Delta}_{e} \cdot N^{\Delta}_{e'}, \qquad \theta_f \in (0, \pi)
\end{equation}
While in split signature, the boost dihedral angle $ \theta_f$ is defined by 
\begin{equation}\label{eq:def_dihedral_2}
	\cosh \theta_f = | N^{\Delta}_{e} \cdot' N^{\Delta}_{e'} |, \qquad \theta_f \gtrless 0 \;\; \text{while} \;\; N^{\Delta}_{e} {\cdot'} N^{\Delta}_{e'} \gtrless 0; 
\end{equation}

\subsection{Lorentzian signature solutions}

As we shown before, when every tetrahedron has both timelike and spacelike triangles, the critical solutions only comes from $S_{+}$. So we focus on $S_{+}$ action.

From the action (\ref{eq:action}), after inserting the decomposition (\ref{eq:Z_com}), we find
\begin{widetext}
	\begin{equation}\label{eq:phase_onev}
	\begin{split}
	S_{vf+}
	=& \frac{n_f}{2} \ln \frac{\zeta_{vef} \bar{\zeta}_{ve'f}}{\bar{\zeta}_{vef} \zeta_{ve'f}} - \ui s_f \ln \frac{\zeta_{ve'f} \bar{\zeta}_{ve'f}}{\bar{\zeta}_{vef} \zeta_{vef}}= -2 \ui \gamma s_f (\arg( \zeta_{ve'f})-\arg( \zeta_{vef})  -2 \ui s \ln \frac{|\zeta_{ve'f}|}{|\zeta_{vef}|}\\
	=&-2 \ui s_f ( \theta_{e'vef} + \gamma \phi_{e'vef})
	\end{split}
	\end{equation}
\end{widetext}
where $\theta$ and $\phi$ are defined by
\begin{equation}
\begin{split}
&	\theta_{e'vef}:=\ln \frac{|\zeta_{ve'f}|}{|\zeta_{vef}|}, \\
& \phi_{e'vef}:=\arg( \zeta_{ve'f})-\arg( \zeta_{vef} ) \\
\end{split}
\end{equation}
The face action at a triangle dual to a face $f$ then reads
\begin{equation}\label{eq:phase_all}
	S_f=\sum_{v \in \partial f} S_{vf}=  -2 \ui s_f \left( \sum_{v \in \partial f} \theta_{e'vef} +  \gamma  \sum_{v \in \partial f} \phi_{e'vef}\right)
\end{equation}
We start the analysis from faces dual to boundary triangles (boundary faces) and then going to internal faces.

\subsubsection{Boundary faces}
For critical configurations solving critical equations (we keep $\Im(\alpha) = 0$ by redefinition of $l^{-}_{ef}$), they satisfy
\begin{align}
	&{\label{eq:pa_l}}g_{ve} \eta l^{+}_{ef}=\frac{\bar{\zeta}_{vef}}{\bar{\zeta}_{ve'f}}g_{ve'} \eta l^{+}_{e'f}
	\\
	&g_{ve} J l^{-}_{ef}=\frac{\bar{\zeta}_{ve'f}}{\bar{\zeta}_{vef}}g_{ve'}^{-1 \dagger} J l^{-}_{e'f}.
\end{align}
We then have
\begin{align}
& G_f (e_1,e_0) \;\; \eta l^{+}_{e_0 f} \\
&\qquad  = \ue^{ -\sum_{v \in p_{e_1 e_0}} \theta_{e'vef} + \ui \sum_{v \in p_{e_1 e_0}} \phi_{e'vef}} \;\;  \eta l^{+}_{e_1 f} \notag\\
& G_f (e_1,e_0) \;\; Jl^{-}_{e_0 f}\\
&\qquad = \ue^{ \sum_{v \in p_{e_1 e_0}} \theta_{e'vef} - \ui  \sum_{v \in p_{e_1 e_0}} \phi_{e'vef}} \;\;  Jl^{-}_{e_1 f} \notag
\end{align}
where $G_f (e_1,e_0)$ is the product of edge holonomy along the path $p_{e_0 e_1}$
\begin{equation}
	G_f (e_1,e_0):=g_{e_1 v_1} ... g_{e'v_0} g_{v_0 e_0}
\end{equation}
Suppose we have holonomies $G$ and $\tilde{G}$ from the pair of critical solutions with global $\sgn(V)$ orientation, then one can see
\begin{align}
	&\tilde{G}^{-1} G \; \eta l^{+}_{e_0 f} \\
	&\qquad=\ue^{ -\sum_{v \in p_{e_1 e_0}} \Delta \theta_{e'vef} + \ui \sum_{v \in p_{e_1 e_0}} \Delta \phi_{e'vef}}\; \eta l^{+}_{e_0 f} \notag\\
	&\tilde{G}^{-1} G \; Jl^{-}_{e_0 f}\\
	&\qquad = \ue^{ \sum_{v \in p_{e_1 e_0}} \Delta \theta_{e'vef} - \ui  \sum_{v \in p_{e_1 e_0}} \Delta \phi_{e'vef}} \; J l^{-}_{e_0 f} \notag
\end{align}
For a single 4-simplex, the above equations read
\begin{align}
	&(\tilde{g}_{e'v} \tilde{g} _{ve})^{-1}(g_{e'v} g _{ve}) \;\; \eta l^{+}_{e_0 f} =\frac{\bar{\zeta'}_{vef}}{\bar{\zeta'}_{ve'f}} \frac{\bar{\zeta}_{vef}}{\bar{\zeta}_{ve'f}} \;\; \eta l^{+}_{ef}\\
	&\qquad =\ue^{-\Delta \theta_{e'vef} +\ui \Delta \phi_{e'vef}} \;\; \eta l^{+}_{ef} \notag\\
	&(\tilde{g}_{e'v} \tilde{g} _{ve})^{-1}(g_{e'v} g _{ve}) \;\; Jl^{-}_{e_0 f} =\frac{\bar{\zeta'}_{ve'f}}{\bar{\zeta'}_{vef}} \frac{\bar{\zeta}_{ve'f}}{\bar{\zeta}_{vef}} \;\; Jl^{-}_{e_0 f}\\
	&\qquad =\ue^{ \Delta \theta_{e'vef} - \ui \Delta \phi_{e'vef}} \;\; Jl^{-}_{e_0 f} \notag
\end{align}
which lead to
\begin{align}
	&g_{ve} (\tilde{g}_{e'v} \tilde{g} _{ve})^{-1}g_{e'v}\;\;\; g _{ve} \; \eta l^{+}_{e_0 f} \label{eq:615}\\
	&\qquad  =\ue^{-\Delta \theta_{e'vef} + \ui \Delta \phi_{e'vef}} \;\;\; g _{ve} \; \eta l^{+}_{e_0 f} \notag\\
	&g_{ve} (\tilde{g}_{e'v} \tilde{g} _{ve})^{-1}g_{e'v} \;\;\; g _{ve}\; J l_{e_0 f} \label{eq:616} \\
	&\qquad =\ue^{\Delta \theta_{e'vef} - \ui  \Delta \phi_{e'vef}} \;\;\;  g _{ve} \;Jl^{-}_{e_0 f} \notag
\end{align}
We can define an operator $T_{ef}$ by
\begin{equation}
	T_{ef}:=\eta l^{+}_{ef}  \otimes (l^{-}_{ef})^{\dagger}=\ket{\eta l^{+}_{ef}} \bra{l^{-}_{ef}}
\end{equation}
From the facts $\braket{l^{-}_{ef}}{\eta l^{+}_{ef}} = \langle l^{-}_{ef}, l^{+}_{ef} \rangle=1, \braket{l^{-}_{ef}}{J l^{-}_{ef}}=0$, the action of this operator leads to
\begin{equation}
\begin{split}
&T_{ef} \ket{\eta l^{+}_{ef}}= \ket{\eta l^{+}_{ef}} \braket{l^{-}_{ef}}{\eta l^{+}_{ef}}=\ket{\eta l^{+}_{ef}}\\
&T_{ef} \ket{J l^{-}_{ef}}= 0
\end{split}
\end{equation}
From the definition of (\ref{eq:xvef_def}) (with $\alpha=0$), by using (\ref{eq:v_bi_tp}) and (\ref{eq:w_define}), one then see 
\begin{equation}\label{eq:x_action}
	X_{ef} \ket{\eta l^{+}_{ef}}=\frac{1}{2}\ket{\eta l^{+}_{ef}}, \;\;\; X_{ef} \ket{J l^{-}_{ef}}= -\frac{1}{2} \ket{J l^{-}_{ef}}
\end{equation}
Then we have
\begin{widetext}
\begin{align}
	&2 X_f \;\; g_{ve} \ket{\eta l^{+}_{ef}}=2 g_{ve} X_{ef} g_{ev} g_{ve}  \ket{\eta l^{+}_{ef}} =g_{ve} 2X_{ef} \ket{\eta l^{+}_{ef}}=g_{ve} \ket{\eta l^{+}_{ef}}\\
	&2 X_f \;\; g_{ve} \ket{Jl^{-}_{ef}}= 2 g_{ve} X_{ef} g_{ev} g_{ve}  \ket{Jl^{-}_{ef}} =g_{ve} 2X_{ef} \ket{Jl^{-}_{ef}}=-g_{ve} \ket{J l^{-}_{ef}}.
\end{align}
\end{widetext}
From (\ref{eq:615}) and (\ref{eq:616}), it is easy to see
\begin{equation}\label{eq:g_diff_one}
	g_{ve} (\tilde{g}_{e'v} \tilde{g} _{ve})^{-1}g_{e'v} =\ue^{-2 \Delta \theta_{e'vef} X_f + 2 \ui \Delta \phi_{e'vef} X_f}
\end{equation}

For a general simplicial complex with boundary, given a boundary face $f$ with two edges $e_0$ and $e_1$ connecting to the boundary, and $v$ is the bulk end-point of $e_0$ if we define
\begin{equation}
	G_f(e_1,e_0)= G_f(v, e_1)^{-1} g_{ve_0}
\end{equation}
It can be proved that
\begin{equation}
	G_f(v, e_1) X_{e_1 f} G_f(v,e_1)^{-1} = g_{v e_0} X_{e_0 f} g_{e_0 v}
\end{equation}
which is the generalization of the parallel transportation equation within a single 4-simplex. Then we can apply the same derivation as the single-simplex case by replacing $g_{ve'} \to G(v, e_1)$, which leads to
\begin{widetext}
\begin{equation}\label{eq:g_diff_boundary}
	g_{ve} \tilde{G}_f(e_1,e_0)^{-1}{G}_f(e_1,e_0) g_{ev}=\ue^{-2 \sum_{v \in \partial f} \Delta \theta_{e'vef} X_f + 2 \ui \sum_{v \in \partial f} \Delta \phi_{e'vef} X_f}.
\end{equation}
\end{widetext}

\subsubsection{Internal faces}

The discussion of internal face $f$ is similar to the boundary case, we have
\begin{align}
\label{eq:holo_action_l+} G_f \; \eta l^{+}_{ef} = \ue^{ -\sum_{v \in \partial f} \theta_{e'vef} +  \ui \sum_{v \in \partial f} \phi_{e'vef}} \;\;  \eta l^{+}_{ef}\\
\label{eq:holo_action_l-}G_f \; J l^{-}_{ef} = \ue^{ \sum_{v \in \partial f} \theta_{e'vef} - \ui  \sum_{v \in \partial f} \phi_{e'vef}} \;\;  J l^{-}_{ef}
\end{align}
where $G_f$ is the face holonomy
\begin{equation}
G_f := \prod_{v \in \partial f}^{\leftarrow} g_{e'v} g_{ve}.
\end{equation}
By the action of bivector $X_{ef}$ in (\ref{eq:x_action}),
\begin{align}
 &\ue^{ -\sum_{v \in \partial f} \theta_{e'vef} 2 X_{ef} + \ui \sum_{v \in \partial f} \phi_{e'vef} 2 X_{ef}} \;\;  \ket{\eta l^{+}_{ef}}\\
 &\qquad =\ue^{ -\sum_{v \in \partial f} \theta_{e'vef} + \ui \sum_{v \in \partial f} \phi_{e'vef}} \;\;  \ket{\eta l^{+}_{ef}} \notag \\
 &\ue^{ -\sum_{v \in \partial f} \theta_{e'vef} 2 X_{ef} + \ui \sum_{v \in \partial f} \phi_{e'vef} 2 X_{ef}} \;\;  \ket{J \; l^{-}_{ef}}\\
 &\qquad =\ue^{ \sum_{v \in \partial f} \theta_{e'vef} -  \ui  \sum_{v \in \partial f} \phi_{e'vef}} \;\;  \ket{J \; l^{-}_{ef}} \notag
\end{align}
Compare to (\ref{eq:holo_action_l+}) and (\ref{eq:holo_action_l-}), we see that
\begin{equation}\label{eq:631}
	G_f=\ue^{ -\sum_{v \in \partial f} \theta_{e'vef} 2 X_{ef} + \ui \sum_{v \in \partial f} \phi_{e'vef} 2 X_{ef}}
\end{equation}
Given $G_f$ and $\tilde{G}_f$ from a pair of critical solutions with opposite $\sgn(V)$ orientation, we find
\begin{equation}\label{eq:g_diff_internal}
	g_{ve} \tilde{G}_f^{-1}{G}_f g_{ev}=\ue^{-2 \sum_{v \in \partial f} \Delta \theta_{e'vef} X_f + 2 \ui  \sum_{v \in \partial f} \Delta \phi_{e'vef} X_f}
\end{equation}

\subsubsection{Phase difference}

For a pair of globally orientated (constant $\sgn(V)$) critical solutions with opposite orientation, from (\ref{eq:phase_all}) we have
\begin{equation}
	\Delta S_f=  -2 \ui s_f \left( \sum_{v \in \partial f} \Delta \theta_{e'vef} + \gamma  \sum_{v \in \partial f} \Delta \phi_{e'vef}\right)
\end{equation}
where $\Delta \theta$ and $\Delta \phi$ are determined by
\begin{equation}
	g_{ve} \tilde{G}_f^{-1} {G}_f g_{ev}=\ue^{-2 \sum_{v \in \partial f} \Delta \theta_{e'vef} X_f + 2 \ui \sum_{v \in \partial f} \Delta \phi_{e'vef} X_f}
\end{equation}
$G_f\equiv{G}_f(e_1,e_0)$ if $f$ is a boundary face. Since $\gamma s_f = n_f/2\in \mathbb{Z}/2$, we may restrict
\begin{equation}
\sum_{v \in \partial f} \Delta \phi_{e'vef}\in [-\pi,\pi].
\end{equation}
because $\Delta S_f$ is an exponent. 

After projecting to $\text{SO}_{+}(1,3)$, 
\begin{equation}
	g_{ve} \tilde{G}^{-1}_f G_f g_{ev} \to G_{ve} \tilde{G}_f^{-1} G_f G_{ev}, \qquad  \ui \to *
\end{equation}
For spacelike normal vector $u=(0,0,0,1)$, from 
it is easy to see $G$ and $\tilde{G}$ are related by
\begin{equation}
	\tilde{G}=R_{e_{0}} G R_{u} I \in SO_{+}(1,3)
\end{equation}
and 
\begin{equation}\label{eq:g_f_o_sp}
	\tilde{G}_f = R_{u_e} G_f R_{u_e} 
\end{equation}
for both internal and boundary triangles $f$.
The equation then leads to
\begin{widetext}
\begin{equation}
	G_{ve} \tilde{G}_{f}^{-1} G_f G_{ev}= G_{ve} R_{u} G^{-1}_f R_{u} G_f G_{ev} = R_{N_e} R_{N_{e'}}
\end{equation}
\end{widetext}
for both internal and boundary triangles $f$.
$N_{e}$ and $N_{e'}$ here are given by
\begin{equation}
	N_{e} = G_{ve} u, \qquad N_{e'} = G_{ve} (G_f^{-1} u),
\end{equation}
thus $N_{e'}$ is the parallel transported vector along the face.

Therefore in both internal case and boundary case, we have
\begin{equation}
	R_{N_{e}} R_{N_{e'}} = \ue^{-2 \sum_{v \in \partial f} \Delta \theta_{e'vef} X_f + 2 * \sum_{v \in \partial f} \Delta \phi_{e'vef} X_f}
\end{equation}

On the other hand, from the fact that, $R_{N} = G R_u G$, and the fact that $G^{\Delta}_{ve} = G I^{s_{ve}} (I R_u)^s_{v}$, we have
\begin{equation}
	R_{N_e} R_{N_e'} = R_{N_e^{\Delta}} R_{N_{e'}^{\Delta}}
\end{equation} 
Since $R_{N^{\Delta}}$ is a reflection respect to spacelike normal $N^{\Delta}$, we have (see Appendix \ref{app_rr_d})
\begin{equation}
 	R_{N_{e}^{\Delta}} R_{N_{e'}^{\Delta}} = \ue^{2 \theta_{f} \frac{N_{e}^{\Delta} \wedge N_{e'}^{\Delta}}{|N_{e}^{\Delta} \wedge N_{e'}^{\Delta}|}}
 \end{equation} 
where $f$ is the triangle dual to the face determined by edges $e$ and $e'$. $\theta_{f}\in[0,\pi]$ satisfies $N_e^{\Delta} \cdot N_{e'}^{\Delta} = \cos(\theta_f)$. From the geometric reconstruction,
\begin{equation}
	B_f= n_f X_f = -\frac{1}{Vol^{\Delta}} r W_{e}^{\Delta} W_{e'}^{\Delta} * (N_{e'}^{\Delta} \wedge N_{e}^{\Delta}), 
\end{equation}
Since $|B_f|^2= - n_f^2$, we have
\begin{equation}
	\left| \frac{1}{Vol^{\Delta}} r W_e^{\Delta} W_{e'}^{\Delta} \right| |N_{e'}^{\Delta} \wedge N_e^{\Delta}|= n_f
\end{equation}
Thus
\begin{equation}\label{eq:x_nn}
		X_f = \frac{B_f}{n_f}= \sigma_{f} \frac{* (N_{e'}^{\Delta} \wedge N_e^{\Delta})}{| (N_{e'}^{\Delta} \wedge N_e^{\Delta})|}
	\end{equation}
where $\sigma_{f}=- r \text{sign}(W_{e'}^{\Delta} W_e^{\Delta})$. Since $N_{e}$ and $N_{e'}$ are both spacelike, we have $\sigma_{f}=-r$. Keep in mind that $r$ is the orientation and is a constant sign on the (sub-)triangulation. Therefore
\begin{widetext}
\begin{equation}
	\ue^{2 r \sum_{v \in \partial f} \Delta \theta_{e'vef} \frac{* (N_{e'}^{\Delta} \wedge N_e^{\Delta})}{|N_{e'}^{\Delta} \wedge N_e^{\Delta}|} + 2 r \sum_{v \in \partial f} \Delta \phi_{e'vef} \frac{ N_{e'}^{\Delta} \wedge N_e^{\Delta} }{|N_{e'}^{\Delta} \wedge N_e^{\Delta}|}} =\ue^{2 \theta_{f} \frac{N_{e}^{\Delta} \wedge N_{e'}^{\Delta}}{|N_{e}^{\Delta} \wedge N_{e'}^{\Delta}|}}
\end{equation}
\end{widetext}
which implies
\begin{equation}\label{eq:corres_phi}
	\begin{split}
		& \sum_{v \in \partial f} \Delta \theta_{e'vef}=0, \\
		&-  r \sum_{v \in \partial f} \Delta \phi_{e'vef}=  \theta_{f}  \mod \;\;  \pi
	\end{split}
\end{equation}
The phase difference is then 
\begin{equation}
	\Delta S_f = 2 \ui r A_f \theta_f \mod \; \ui \pi 
\end{equation}
where $A_f = \gamma s_f = n_f/2\in \mathbb{Z}/2$ is the area spectrum of the timelike triangle.

The $i\pi$ ambiguity relates to the lift ambiguity from $G_f\in \mathrm{SO}^+(1,3)$ to $\mathrm{SL}(2,\mathbb{C})$. Some ambiguities may be absorbed into gauge transformations $g_{ve}\to -g_{ve}$. Firstly we consider a single 4-simplex, (\ref{eq:corres_phi}) reduces to $\Delta \theta_{e'vef}=0$ and $\Delta \phi_{e'vef}= - \theta_{f}$ mod $\pi$ ( Here we use the notation that we move the orientation $r$ from $ \Delta \phi$ in (\ref{eq:corres_phi}) to the definition of $\Delta S$. Keep in mind $\Delta S$ always depends on the orientation $r$). However it is shown in Appendix \ref{app:deformation_phase} that this ambiguity can indeed be absorbed into the gauge transformation of $g_{ve}$, i.e. if we fix the gauge,
\begin{equation}
\Delta \phi_{e'vef}= - \theta_{f}(v) \mod 2\pi, \label{singlephi}
\end{equation}
where $\theta_{f}(v)$ is the angle between tetrahedron normals in the 4-simplex at $v$. Although this fixing of lift ambiguity only applies to a single 4-simplex, it is sufficient for us to obtain $\Delta S_f^{\Delta}$ unambiguously. Applying (\ref{singlephi}) to the case with many 4-simplices 
\begin{equation}
\sum_{v \in \partial f} \Delta \phi_{e'vef}=-\sum_{v \in \partial f}\theta_{f}(v)\ \mod 2\pi
\end{equation}
Since $\theta_{f}(v)$ relates to the dihedral angle $\Theta_f(v)$ by $\theta_{f}(v)=\pi-\Theta_f(v)$, for an internal $f$, $\sum_{v \in \partial f} \Delta \phi_{e'vef}$ relates to the deficit angle $\varepsilon_f=2\pi-\sum_{v \in \partial f}\Theta_{f}(v)$ by 
\begin{equation}\label{eq:delta_phi}
\sum_{v \in \partial f} \Delta \phi_{e'vef}= (2-m_f)\pi-\varepsilon_f \ \mod 2\pi
\end{equation}
where $m_f$ is the number of $v\in\partial f$. Similarly, for a boundary $f$, $\sum_{v \in \partial f} \Delta \phi_{e'vef}$ relates to the deficit angle $\theta_f=\pi-\sum_{v \in \partial f}\Theta_{f}(v)$ by
\begin{equation}
\sum_{v \in \partial f} \Delta \phi_{e'vef}= (1-m_f)\pi-\theta_f \ \mod 2\pi
\end{equation}

As a result, the total phase difference is  
\begin{eqnarray}
	&\exp(\Delta S_f) = \exp\Big\{2 \ui r \sum_{f\ \mathrm{bulk}} A_f \left[(2-m_f)\pi-\varepsilon_f\right] \nonumber\\
	&+  2 \ui r\sum_{f\ \text{boundary}} A_f \left[(1-m_f)\pi-\theta_f\right]\Big\}
\end{eqnarray}
The exponent is a Regge action when all bulk $m_f$ are even, i.e. every internal $f$ has even number of vertices. Obtaining Regge calculus only requires all bulk $m_f$'s to be even, while boundary $m_f$'s can be arbitrary, since the boundary terms $A_f (1-m_f)\pi$ doesn't affect the Regge equation of motion. 

The above phase difference is for a general simplicial complex, the result for a single 4-simplex is simply given by removing the bulk terms and letting all boundary $m_f=1$.

\subsubsection{Determine the phase for bulk triangles}

For the internal faces in the bulk, we can determine the phase at critical point uniquely. 

Recall (\ref{eq:631}, the holonomy $G_f(v)= g_{ve} G_f(e) g_{ev}$ at vertex $v$ reads
\begin{equation}
	G_f(v)=\ue^{ -\sum_{v \in \partial f} \theta_{e'vef} 2 X_{f}(v) + \ui \sum_{v \in \partial f} \phi_{e'vef} 2 X_{f}(v)}
\end{equation}
Recall (\ref{eq:parallel_e}) as we shown in Appendix \ref{app:geo}, for edges $E_{l1}(v)$ and  $E_{l1}(v)$ of the triangle $f$ in the frame of vertex $v$,
\begin{equation}\label{eq:635}
	\begin{split}
	&G_f(v) E_{l1}(v) = \mu E_{l1}(v), \\
	&G_f(v) E_{l2}(v) = \mu E_{l2}(v)
	\end{split}
\end{equation}
where $\mu = (-1)^{\sum_{e \subset \partial f} s_e }= \pm 1$. Here $s_e$ is defined as $s_e = s_{ve}+s_{v'e}+1 $ for edge $e=(v,v')$ with $s_{ve} \in \{0,1\}$.
With edges $E_{l1}(v)$ and  $E_{l1}(v)$, the bivector $X_f(v)$ at vertex $v$ can be expressed as
\begin{equation}\label{eq:636}
	X_{f}(v) = \frac{* (N_{e'}(v) \wedge N_{e}(v) )}{|N_{e'}(v) \wedge N_{e}(v) |} = \frac{ E_{l1}(v) \wedge E_{l2}(v) }{| E_{l1}(v) \wedge E_{l2}(v) |}
\end{equation}
From (\ref{eq:635}) and (\ref{eq:636}), with the fact that $\ue^{X_f(v)}$ is a boost, one immediately see $\mu_e =1$ and
\begin{equation}
	G_f(v)=\ue^{\ui \sum_{v \in \partial f} \phi_{e'vef} 2 X_{f}(v)}= \ue^{ 2 r \sum_{v \in \partial f} \phi_{e'vef} \frac{ N_{e'}^{\Delta} \wedge N_e^{\Delta} }{|N_{e'}^{\Delta} \wedge N_e^{\Delta}|}}
\end{equation}
where we use (\ref{eq:x_nn}). As we proved in Appendix \ref{app_rr_d}, there exists spacelike normalized vector $\tilde{N}$ in the plane span by $N_{e}$ and $N_{e'}$ such that  
\begin{equation}
	G_f(v) = R_{N} R_{\tilde{N}}
\end{equation}

From (\ref{eq:g_f_o_sp}), 
\begin{equation}
	\begin{split}
		G_{ve} \tilde{G}_f(e) G_f(e) G_{ev} &= G_{ve} R_u G_f(e) R_u G_f(e) G_{ev}\\
		&= R_{N} G_{f}(v) R_{N} G_{f}(v)
	\end{split}
\end{equation}
Then it is straightforward to show
\begin{equation}
	\begin{split}
		&G_{ve} \tilde{G}_f(e) G_f(e) G_{ev}=	R_{N} G_{f}(v) R_{N} G_{f}(v) \\
	=& R_{N}  R_{N} R_{\tilde{N}} R_{N}  R_{N} R_{\tilde{N}} = R_{\tilde{N}}R_{\tilde{N}} = 1
	\end{split}
\end{equation}
Thus
\begin{equation}
	\ue^{2 \sum_{v \in \partial f} ( \tilde{\phi}_{e'vef}+\phi_{e'vef}) * X_f} =1
\end{equation}
which leads to
\begin{equation}
	\sum_{v \in \partial f} ( \tilde{\phi}_{e'vef}+\phi_{e'vef}) = 0 \mod \pi
\end{equation}
The $\pi$ ambiguity here relates to the lift ambiguity again. Note that, fixing of lift ambiguity to these 4-simplicies sharing the triangle $f$ as in the Appendix \ref{app:deformation_phase} leads to $g_{ve} \tilde{G}_f(e) G_f(e) g_{ev} =1$. Then we have
\begin{equation}
	\sum_{v \in \partial f} ( \tilde{\phi}_{e'vef}+\phi_{e'vef}) = 0 \mod 2 \pi
\end{equation}
where the $\pi$ ambiguity is fixed.
Combine with (\ref{eq:delta_phi}), we have
\begin{equation}
	\begin{split}
		\sum_{v \in \partial f} \phi_{e'vef} &= - \sum_{v \in \partial f} \tilde{\phi}_{e'vef} \\
		&= \frac{(2-m_f)\pi-\varepsilon_f}{2} \mod \pi
	\end{split}
\end{equation}
As a result, the total phase for bulk triangles is
\begin{equation}
	\exp(S_f) = \exp\Big\{\ui r \sum_{f\ \mathrm{bulk}} A_f \left[(2-m_f)\pi-\varepsilon_f\right] \Big\}
\end{equation}
Again, the exponent is a Regge action when all bulk $m_f$ are even, i.e. every internal $f$ has even number of vertices.

Note that, the above derivation assumes a uniform orientation $\sgn(V )$, but the asymptotic formula of the spinfoam amplitude is given by summing over all possible configurations of orientations. As suggested by \cite{Han:2011re}, at a critical solution, one can make a partition of $\mathcal{K}$ into subregions such that each region has a uniform orientation, so that the above derivation can be applied.


\subsection{split signature solutions}

In this subsection, we focus on a single 4-simplex. We consider a pair of the degenerate solutions ${g_{ve}^{\pm}}$ which can be reformulated as non-degenerate solutions in the flipped signature space $(-++-)$ here. When degenerate solutions are gauge equivalent, there exists only a single critical point, then there is a single phase depending on boundary coherent states.

Since (\ref{eq:g_diff_boundary}) and (\ref{eq:g_diff_internal}) hold for all $\text{SL}(2,\C)$ elements which solve critical equations, they also hold for degenerate solutions $g_{ve}^{\pm}$. Thus from (\ref{eq:g_diff_one}), we have
\begin{equation}
	\begin{split}
		&g^{\pm}_{ev} g^{\mp}_{ev} g^{\mp}_{ve'} g^{\pm}_{e'v} = \ue^{\mp 2 \Delta \theta_{e'vef} X_f^{\pm} \pm 2 \ui \Delta \phi_{e'vef} X_f^{\pm}}\\
		& \qquad =  \ue^{\mp 2 \Delta \theta_{e'vef} X_f^{\pm}}
	\end{split}
\end{equation}
Notice that since all $g^{\pm}_{ve} \in \text{SU}(1,1) \subset\text{SL}(2,\C) $, we have $2 \Delta \phi_{e'vef} = 0 \; \mod \; 2 \pi$ ($*X_f^{\pm}$ generates rotations in $v^g$-$u$ plane). 

From (\ref{eq:E90}), we have
\begin{widetext}
\begin{equation}
	\Phi^{\pm}(g_{ev} \tilde{g}_{ev} \tilde{g}_{ve'} g_{e'v}) = \Phi^{\pm}(g_{ev}) \Phi^{\pm} (\tilde{g}_{ev}) \Phi^{\pm}(\tilde{g}_{ve'})\Phi^{\pm} (g_{e'v}) = g^{\pm}_{ev} g^{\mp}_{ev} g^{\mp}_{ve'} g^{\pm}_{e'v}
\end{equation}
\end{widetext}
Since $\tilde{G}_{ve} = R_u G_{ve} R_u$, we have
\begin{equation}
	\Phi^{\pm}(R_{N_e} R_{N_{e'}}) = G^{\pm}_{ev} G^{\mp}_{ev} G^{\mp}_{ve'} G^{\pm}_{e'v}
\end{equation}

For $X_f$ in flipped signature space $M'$, from the definition of $\Phi^{\pm}$ in (\ref{def_phi_flip}), we have
\begin{equation}
	\Phi^{\pm}(*' X_f) = \pm \Phi^{\pm}( X_f) = \pm v_{ef}^{g\pm} =\pm \Phi^{\pm}(X_f^{\pm})
\end{equation}
where we know $X_f^{\pm}=v_{ef}^{g\pm} \wedge u$ in degenerate case, and $X_f^{\pm}$ can be regarded as bivectors in $\text{so}(V) \sim \wedge^2V$. Then we have
\begin{equation}
	\Phi^{\pm}(\ue^{2 \Delta \theta_{e'vef} *' X_f}) = \ue^{\mp 2 \Delta \theta_{e'vef} X_f^{\pm}}
\end{equation}
where we identify the $\text{SO}(1,2)$ acting on $V$ to the one acting on $M'$. 

Therefore, $\Delta \theta$ contribution to the phase difference in degenerate solutions $\{ g^{\pm} \}$ is identified to the $\Delta \theta$ written in flipped signature solutions $\{g \}$ satisfying $\Phi^{\pm}(g)=g^{\pm}$. $\Delta \theta$ is given by
	\begin{equation}
		R_{N_e} R_{N_{e'}} = \ue^{2 \Delta \theta_{e'vef} *' X_f}
		\end{equation}
	where $X_f$ is the bivector from flipped signature solutions
		\begin{equation}
			X_f = \frac{B_f}{n_f}= - r \frac{*' (N_{e'}^{\Delta} \wedge N_e^{\Delta})}{|*‘ (N_{e'}^{\Delta} \wedge N_e^{\Delta})|}
		\end{equation}

From the fact that geometrically,
\begin{equation}
	R_{N_e} R_{N_{e'}}=R_{N_e^{\Delta}} R_{N_{e'}^{\Delta}} = \ue^{2 \theta_f \frac{N_{e}^{\Delta} \wedge N_{e'}^{\Delta}}{ |N_{e}^{\Delta} \wedge N_{e'}^{\Delta}|}} \; ,
\end{equation}
where  $\theta_{f}\in\mathbb{R}$ is a boost dihedral angle. We have
\begin{equation}
	- r \Delta \theta_{e'vef} = \theta_f, \qquad 2 \Delta \phi_{e'vef} = 0  \mod \;\; 2 \pi
\end{equation}
the phase difference is
\begin{equation}
	\Delta S_f^{\Delta} = 2 \ui r s_f \theta_{f} = 2 \ui r \frac{1}{\gamma} A_f \theta_f  \mod \; \pi \ui
\end{equation}
We can again fix the $\pi i$ ambiguity by using the method in Appendix \ref{app:deformation_phase}. There is no ambiguity in $\theta_f$ since it is a boost angle. As a result,
\begin{equation}
	\exp(\Delta S_f) = \exp\left(2 \ui r \frac{1}{\gamma}  A_f \theta_f  \right) 
\end{equation}

The generalization to simplicial complex is similar to the non-degenerate case, by substituting every $g$ and $\tilde{g}$ there with $g^{\pm}$.

\section{Conclusion and Discussion}
The present work studies the large-$j$ asymptotics limit of spin foam amplitude with timelike triangles in a most general configuration on a $4$d simplicial manifold with many $4$-simplicies. It turns out the asymptotics of spin foam amplitude is determined by critical configurations of the corresponding spinfoam action on the simplicial manifold. The critical configurations have geometrical interpretations as different types of geometries in separated subregions: Lorentzian $(-+++)$ $4$-simplicies, split $(--++)$ $4$-simplicies or degenerate vector geometries. The configurations come in pairs which corresponding to opposite global orientations in each subregion. In each sub-complex with globally oriented $4$-simplicies coming with the same signature, the asymptotic contribution to the spinfoam amplitude is an exponential of Regge action, up to a boundary term which does not affect the Regge equation of motion.

An important remark is that, for a vertex amplitude containing at least one timelike and one spacelike tetrahedron, critical configurations only give Lorentzian $4$-simplicies, while Euclidean and degenerate vector geometries do not appear. In all known examples of Lorentzian Regge calculus, the geometries are corresponding to such configuration, for example, the Sorkin triangulation \cite{Sorkin:1975ah} where each $4$-simplex containing $4$ timelike tetrahedra and $1$ spacelike tetrahedron. Since such configuration only gives Regge-like critical configurations which is supposed to be the result of simplicity constraint in spin foam models \cite{Perez:2012wv}, the result could open a new and promising way towards a better understanding of the imposition of simplicity constraint. Furthermore, Such configuration also naturally inherits the causal structure to spin foam models, which may open the possibility to build the connection between spin foam models and causal sets theory \cite{Sorkin:2003bx} or causal dynamical triangulation theories \cite{Ambjorn:2001cv,Ambjorn:2013tki}.

With this work, the asymptotics of Conrady-Hnybida spin foam model, with arbitrary timelike or spacelike non-degenerate boundaries, is now complete. In the present work we mainly concentrate on the case where each tetrahedron contains both timelike and spacelike triangles, which is the case in all Regge calculus geometry examples. The geometrical interpretation of the case where tetrahedron containing only timelike triangles is much more complicated and we only identify its critical configurations on special cases with the boundary data satisfies length matching condition and orientation matching condition. Further investigation is needed for all possible critical configurations in such case.

Moreover, in the present analysis we do not give the explicit form of measure factors of the asymptotics formula, which is important for the evaluation of the spin foam propagator and amplitude. The measure factor in EPRL model is related to the Hessian matrix at the critical configuration \cite{Alesci:2007tg,Bianchi:2011hp}. However, the measure factor for the triangulation with timelike triangles is a much more complicated function of second derivatives of the action, due to the appearance of singularities. A further study of such kind multidimensional stationary phase approximation, in particular, the derivation of the measure factor would be interesting.

The present work opens the possibility to have Regge geometries in Lorentzian Regge calculus emerges as critical configurations from spin foam model, which may leads to a semi-classical effective description of spin foam model. Especially, this may lead to a effective equation of motion for symmetry reduced models, e.g., FLRW cosmology or black holes, from the semi-classical limit of spin foam models. 

\begin{acknowledgments}
HL thanks the hospitality of the Department of Physics at Florida Atlantic University, where some of the research related to this work was carried out. MH acknowledges support from the US National Science Foundation through grant PHY-1602867, and Start-up Grant at Florida Atlantic University, USA. 
\end{acknowledgments}

\clearpage 
\appendix
\addtocontents{toc}{\protect\setcounter{tocdepth}{1}}

\begin{widetext}

	\section{Derivation of representation matrix}\label{app_f}
	This appendix shows the wigner matrix of continuous series in unitary irreps of $\text{SU(1,1)}$ group in the large $s$ approximation. 
	We begin with the introduction of the wigner matrix of continuous series given in \cite{Lindblad:1970tv}. Then by transformations of hypergeometric functions and saddle point approximation we obtain the representation matrix in large $s$ limit.
	\subsection{Wigner matrix}
	First let us introduce the parametrization of the $\text{SU}(1,1)$ group element $v$:
	\begin{equation}
	\begin{split}
		v(z)=&\ue^{\ui \phi J^3} \ue^{\ui t K^2} \ue^{\ui u K^1}=\left(
	\begin{array}{cc}
	 v_1 &v_2 \\ \bar{v}_2 & \bar{v}_1
	\end{array}
	\right)
	\end{split}
	\end{equation}
	where
	\begin{align}
		v_1=\ue^{\frac{\ui \phi }{2}} \left(\cosh \left(\frac{t}{2}\right) \cosh \left(\frac{u}{2}\right)-\ui \sinh \left(\frac{t}{2}\right) \sinh \left(\frac{u}{2}\right)\right)\\
		v_2=e^{\frac{\ui \phi}{2}} \left(\ui \cosh \left(\frac{u}{2}\right) \sinh \left(\frac{t}{2}\right)-\cosh \left(\frac{t}{2}\right) \sinh \left(\frac{u}{2}\right)\right)
	\end{align}
	Note that the generators defined here is complex version of what we used in the main part. In this parametrization, the wigner matrix which defined as
	\begin{equation}
		D^{j}_{m\lambda \sigma}(v)= \bra{j,m} v \ket{j \lambda \sigma}
	\end{equation}
	can be expressed by \cite{Lindblad:1970tv}
	\begin{equation}\label{eq:d_b}
		D^{j}_{m\lambda\sigma}=\ue^{\ui m \phi}d^{j}_{m\lambda\sigma}\ue^{\ui \lambda u}=\ue^{\ui m \phi}S^{j}_{m \lambda \sigma} \big{(} T^{j}_{m\lambda}F^{j}_{m, \ui \lambda}(\beta )-(-1)^{\sigma} T^{j}_{-m \lambda}F^{j}_{-m, \ui \lambda}(\bar{\beta}) \big{)} \ue^{\ui \lambda u}
	\end{equation}
	where
	\begin{align}
		&F^{j}_{m,\ui \lambda}(\beta)=(1-\beta)^{(m-\ui \lambda)/2} \beta^{(m+\ui \lambda)/2} {{}_2F_1} \left( -j+m,j+m+1; m+\ui \lambda +1;\beta \right)\\
		&T^{j}_{m \lambda}= \frac{1}{\Gamma(-m-j) \Gamma(m+1+\ui \lambda)}
	\end{align}
{Here ${{}_2F_1}(a,b,c,z)$ refers to Gaussian hypergeometric function, and $\Gamma(z)$ is the Gamma function.}
	Normalization factor $S^j_{m \lambda \sigma}$ reads
	\begin{equation}\label{eq:s_j_m}
		\begin{split}
		S^j_{m \lambda \sigma}=\sqrt{\frac{\Gamma(m-j)}{\Gamma(m+j+1)}}\frac{2^{j-1} \Gamma(-j+\ui \lambda)}{\ui ^{\sigma} \sin(\pi/2(-j-\ui \lambda+\sigma)))}
		\end{split}
	\end{equation}
	with $\beta=(1-\ui \sinh(t))/2$.
	
	Above equation (\ref{eq:d_b}) can be written in terms of normalized spinors $v=(v_1,v_2)$ in $\text{SU}(1,1)$ inner product $\langle v,v \rangle=1$. According to the parametrization, we have
	\begin{align}
		&v_1+v_2=\ue^{-\frac{u}{2}+\frac{\ui \phi }{2}} \left(\cosh \left(\frac{t}{2}\right)+\ui \sinh \left(\frac{t}{2}\right)\right),
		&v_1-v_2=\ue^{\frac{u}{2}+\frac{\ui \phi }{2}} \left(\cosh \left(\frac{t}{2}\right)-\ui \sinh \left(\frac{t}{2}\right)\right)
	\end{align}
	Wigner matrix $D$ can be written in terms of $v$ and $\bar{v}$
	\begin{equation}\label{eq:a10}
		D^{j}_{m\lambda\sigma}=S^{j}_{m\lambda \sigma} \big{(} T^{j}_{m\lambda}F^{j}_{m, \ui \lambda}(v)-(-1)^{\sigma} T^{j}_{-m\lambda}F^{j}_{-m, \ui \lambda}(\bar{v}) \big{)}
	\end{equation}
	with
	\begin{equation}\label{eq:f_v_m}
	\begin{split}
	F^{j}_{m,\ui \lambda}(v)=&2^{-m}(v_1+v_2)^{(m-\ui \lambda)} (v_1-v_2)^{(m+\ui \lambda)} \cross\\
	& {{}_2F_1}\left( -j+m,j+m+1; m+\ui \lambda +1; (\bar{v}_1+\bar{v}_2)(v_1-v_2)/2 \right)
	\end{split}
	\end{equation}
	\subsection{Asymptotics of Gauss hypergeometric function}
	According to (\ref{eq:d_b}), we need to evaluate the hypergeometric function  
	\begin{equation}\label{eq:hyp_org}
	\begin{split}
	{{}_2F_1} \left( -j+m,j+m+1; m+\ui \lambda +1;\beta \right),\;\;{{}_2F_1} \left( -j-m,j-m+1; -m+\ui \lambda +1; 1-\beta \right)
	\end{split}
	\end{equation}
	The function itself is complicated. However, we only need the asymptotics behavior with $j \sim m \sim \lambda \gg 1$ in our case. According to (\ref{eq:su11_f}), $m$ is chosen to be $n/2$ which related to $j=-1/2+\ui s$ by simplicity constraint (\ref{eq:simplicity}). Correspondingly, $\lambda$ is also chosen to be related to $s$.
	
	\subsubsection{Transformation of original function}
	First we would like to transform the original function to a more convenient form. According to the transformation properties of hypergeometric function, we have
	\begin{align}
	&{{}_2F_1} \left( -j+m,j+m+1; m+\ui \lambda +1; \beta \right)=(1-\beta)^{-m+\ui \lambda}{{}_2F_1} \left( j +\ui \lambda+1,-j+\ui \lambda; m+\ui \lambda +1; \beta \right) \label{eq:hgf_b1_1}\\
	 &{{}_2F_1} \left( -j-m,j-m+1; -m+\ui \lambda +1; 1-\beta \right)=(\beta)^{m+\ui \lambda}{{}_2F_1} \left( j +\ui \lambda+1,-j+\ui \lambda; -m+\ui \lambda +1; 1-\beta \right) \label{eq:hgf_b2_1}\\
	 &\frac{\sin(\pi(-m+\ui \lambda))}{\pi \Gamma(m+\ui \lambda + 1)}{{}_2F_1} \left( -j+m,j+m+1; m+\ui \lambda +1; \beta \right) \notag\\
	 =&\beta^{-m -\ui \lambda} \frac{{{}_2F_1} \left( j-\ui \lambda +1,-j-\ui \lambda; m-\ui \lambda +1; 1-\beta \right) }{\Gamma(m-\ui \lambda +1)\Gamma(j+\ui \lambda +1)\Gamma(\ui \lambda -j)}  \label{eq:hgf_b1_2}\\
	 &\;\;\;\;-(1-\beta)^{-m+\ui \lambda}\frac{ {{}_2F_1} \left( j +\ui \lambda+1,-j+\ui \lambda; -m+\ui \lambda +1; 1-\beta \right) }{\Gamma(-m+\ui \lambda +1)\Gamma(-j+m)\Gamma(j+m+1)} \notag\\
	 &\frac{\sin(\pi(m+\ui \lambda))}{\pi \Gamma(-m+\ui \lambda + 1)}{{}_2F_1}\left( -j-m,j-m+1; -m+\ui \lambda +1; 1-\beta \right) \notag\\
	 =&(1-\beta)^{m -\ui \lambda} \frac{{{}_2F_1} \left( j-\ui \lambda +1,-j-\ui \lambda; -m-\ui \lambda +1; \beta \right)}{\Gamma(-m-\ui \lambda +1)\Gamma(j+\ui \lambda +1)\Gamma(\ui \lambda -j)} \label{eq:hgf_b2_2}\\
	 &\;\;\;\;\;\;-(\beta)^{m+\ui \lambda} \frac{{{}_2F_1} \left( j +\ui \lambda+1,-j+\ui \lambda; m+\ui \lambda +1; \beta \right)}{\Gamma(m+\ui \lambda +1)\Gamma(-j-m)\Gamma(j-m+1)} \notag
	\end{align}
	From (\ref{eq:hgf_b2_1}) and (\ref{eq:hgf_b1_2}), we have
	\begin{equation}\label{eq:hgf_-m}
		\begin{split}
			 &{{}_2F_1} \left( -j-m,j-m+1; -m+\ui \lambda +1; 1-\beta \right)=\Gamma(-m+\ui \lambda +1)\Gamma(-j+m)\Gamma(j+m+1) \cross \\
			  &\left( -\frac{(\beta)^{m+\ui \lambda}\sin(\pi(-m+\ui \lambda))}{\pi \Gamma(m+\ui \lambda + 1)}{{}_2F_1} \left( j +\ui \lambda+1,-j+\ui \lambda; m+\ui \lambda +1; \beta \right) \right.\\
			 &\left. +\frac{(1-\beta)^{m -\ui \lambda} }{\Gamma(m-\ui \lambda +1)\Gamma(j+\ui \lambda +1)\Gamma(\ui \lambda -j)}{{}_2F_1} \left( j-\ui \lambda +1,-j-\ui \lambda; m-\ui \lambda +1; 1-\beta \right) \right)\\
		\end{split}
	\end{equation}
	Similarly, from (\ref{eq:hgf_b1_1}) and (\ref{eq:hgf_b2_2}), we have
	\begin{equation}\label{eq:hgf_m}
		\begin{split}
			 &{{}_2F_1} \left( -j+m,j+m+1; m+\ui \lambda +1; \beta \right)=\Gamma(m+\ui \lambda +1)\Gamma(-j-m)\Gamma(j-m+1) \cross \\
	 &\left( -\frac{(1-\beta)^{-m+\ui \lambda}\sin(\pi(m+\ui \lambda))}{\pi \Gamma(-m+\ui \lambda + 1)}{{}_2F_1} \left( j +\ui \lambda+1,-j+\ui \lambda; -m+\ui \lambda +1; 1-\beta \right) \right.\\
	 &\left. +\frac{\beta^{-m -\ui \lambda} }{\Gamma(-m-\ui \lambda +1)\Gamma(j+\ui \lambda +1)\Gamma(\ui \lambda -j)}{{}_2F_1} \left( j-\ui \lambda +1,-j-\ui \lambda; -m-\ui \lambda +1; \beta \right) \right)\\
		\end{split}
	\end{equation}
	Then in terms of (\ref{eq:hgf_b1_1}) and (\ref{eq:hgf_-m}), the function $d^{j}_{m\lambda\sigma}$ can be written as
	\begin{equation}\label{eq:d_1}
	\begin{split}
		d^{j}_{m\lambda\sigma}(\beta)=&S^j_{m \lambda \sigma} \bigg[  \left( 1+(-1)^{\sigma} \tan(\pi(-m+\ui \lambda)) \right)  \\
		& \cross \frac{(1-\beta)^{(-m+\ui \lambda)/2} \beta^{(m+\ui \lambda)/2}{{}_2F_1} \left( j +\ui \lambda+1,-j+\ui \lambda; m+\ui \lambda +1; \beta \right)}{\Gamma(-m-j)\Gamma(m+\ui \lambda + 1)} \\
		 &-(-1)^{\sigma}\frac{\beta^{(-m-\ui \lambda)/2} (1-\beta)^{(m-\ui \lambda)/2}  {{}_2F_1} \left( j-\ui \lambda +1,-j-\ui \lambda; m-\ui \lambda +1; 1-\beta \right) }{\Gamma(m-\ui \lambda +1)\Gamma(j+\ui \lambda +1)\Gamma(\ui \lambda -j)  \Gamma^{-1}(j+m+1)} \bigg]
	\end{split}
	\end{equation}
	Now we only need to evaluate the hypergeometric function ${{}_2F_1} \left( j +\ui \lambda+1,-j+\ui \lambda; m+\ui \lambda +1; \beta \right)$, since ${{}_2F_1} \left( j-\ui \lambda +1,-j-\ui \lambda; m-\ui \lambda +1; 1-\beta \right)$ is nothing else but the complex conjugation of the previous one. 
	Similar, start from (\ref{eq:hgf_b2_1}) and (\ref{eq:hgf_m}), we have
	\begin{equation}
	\begin{split}
		d^{j}_{m\lambda\sigma}(\beta)=&S^j_{m \lambda \sigma}\bigg[ \left( -\tan(\pi(m+\ui \lambda)) -(-1)^{\sigma} \right) \\
		& \cross \frac{(1-\beta)^{(-m+\ui \lambda)/2} \beta^{(m+\ui \lambda)/2}{{}_2F_1} \left( j +\ui \lambda+1,-j+\ui \lambda; -m+\ui \lambda +1; 1-\beta \right)}{\Gamma(m-j)\Gamma(-m+\ui \lambda + 1)}\\
		 &+\frac{\beta^{(-m-\ui \lambda)/2} (1-\beta)^{(m-\ui \lambda)/2}  {{}_2F_1} \left( j-\ui \lambda +1,-j-\ui \lambda; -m-\ui \lambda +1; \beta \right) }{\Gamma(-m-\ui \lambda +1)\Gamma(j+\ui \lambda +1)\Gamma(\ui \lambda -j)  \Gamma^{-1}(j-m+1)} \bigg]
	\end{split}
	\end{equation}
	Clearly the two expression obey the relation $d^j_{m \lambda \sigma}({\beta})=-(-1)^{\sigma}d^j_{-m \lambda \sigma}(\bar{\beta})$
	
	\subsubsection{Saddle point approximation}
	From (\ref{eq:d_1}), we need the large $s$ approximation of the hypergeometric function
	 ${{}_2F_1} \left( j +\ui \lambda+1,-j+\ui \lambda; m+\ui \lambda +1; \beta \right)$. Here we will only concentrate on the the parameters such that $m=n/2=\gamma s$ and $\lambda \sim s$ are satisfied. In this choice, all the parameters will scale together with $s$. A choice of $\lambda$ is $\lambda = -s$. The generalization to parameters where $m$ and $\lambda$ scales with $\Lambda$ but takes different value is straight forward. Noted the smearing of $\lambda$ requires to calculate $\lambda = -s_0 + \epsilon$ where $\epsilon \ll \lambda$.
	 
	 For simplicity, we will transform the original function as
	\begin{equation}\label{eq:hgf_trans_1}
	\begin{split}
		&{{}_2F_1} \left( j +\ui \lambda+1,-j+\ui \lambda; m+\ui \lambda +1; \beta \right)\\
		=&(1-\beta)^{-1/2}{{}_2F_1} \left( j +\ui \lambda+1,j+m+1; m+\ui \lambda +1; \frac{\beta}{\beta-1} \right) \;\;\;\;\; \text{with} \; \lambda=-s, m=\gamma s, \gamma>0\\
		=&(1-\beta)^{-1/2} {{}_2F_1} \left( \frac{1}{2}, \frac{1}{2}+(\gamma+\ui)s; (\gamma-\ui) s +1; \frac{\beta}{\beta-1} \right)
	\end{split}
	\end{equation}
	We will use the integral representation for Hypergeometric functions \cite{olver2010nist}:
	\begin{align}
		\label{eq:hgf_int_1}&{}_2F_1 (a,b;c; z)=\frac{\Gamma(1+b-c)\Gamma(c)}{2 \pi \ui \Gamma(b)} \int_{0}^{1+} \frac{t^{b-1}(t-1)^{c-b-1}}{(1-z t)^a} dt, \;\;\; \text{if} \; c-b \notin N \; \& \; \Re (b)>0
	\end{align}
	The validity region for these equations is $|\arg(1-z)| < \pi$. In (\ref{eq:hgf_int_1}), the integration path is the anti-clockwise loop that starts and ends at $t=0$, encircles the point $t=1$, and excludes the point $t= 1/z$. 
	In our case, we have $\Re(c-b) = 1/2$ and $\Re(b)=1/2 + m=1/2+\gamma s$ which satisfy the requirement.
	Thus with (\ref{eq:hgf_int_1}) we rewrite the original hypergeometric function as
	\begin{equation}
		{{}_2F_1} \left(\frac{1}{2}, \frac{1}{2}+(\gamma+\ui)s; (\gamma-\ui) s +1; \frac{\beta}{\beta-1} \right)=\frac{G(s)}{2 \pi \ui} \int_{0}^{1+} dt f(t,\beta) \ue^{s \Psi(t)}
	\end{equation}
	where $\Psi(t)$ and $f(t,\beta)$ are
	\begin{equation}
		\Psi(t)=(\gamma+\ui) \ln t - 2 \ui \ln (t-1), \;\; f(t,\beta)=\left(t(t-1)(1-\frac{\beta t}{\beta-1}) \right)^{-\frac{1}{2}} 
	\end{equation}
	and $G(s)$ is 
	\begin{equation}
	\begin{split}
		G(s)=\frac{\Gamma(\frac{1}{2}+ 2 \ui s) \Gamma((\gamma-\ui) s +1)}{\Gamma(\frac{1}{2}+(\gamma+\ui)s)}\sim \frac{\sqrt{2 \pi (\gamma-\ui) s} ((\gamma-\ui))^{(\gamma-\ui)s} (2\ui)^{2 \ui s}}{((\gamma+\ui))^{(\gamma+\ui)s}}
	\end{split}
	\end{equation}
	Here we use the asymptotic formula of $\Gamma$ functions
	\begin{equation}\label{eq:asym_gamma}
	\Gamma(z) \sim \sqrt{2 \pi} z^{z-1/2} \ue^{-z}
	\end{equation}
	{{Note that $| G(s)| \sim \sqrt{s} \exp(-\pi s) $. We will see later the contribution form $\exp(-\pi s)$ will cancel the contribution form $|\exp(s \Psi(t))|$ at the saddle point $t_0$. }}

	Clearly when $\beta/(\beta-1) \neq 1$,  we have three branch points $t=0$, $t=1$ and $t=(\beta-1)/\beta$ for $f(t,z)$ and two branch points $t=0$ and $t=1$ for $\Psi(t)$. The branch cuts for $\Psi(t)$ on the real axis are given by $(- \infty, 0]$ and $(0,1]$, which can be seen in Fig. \ref{fig:m>0}. We need to exclude the point $t_{\beta} = (\beta-1)/\beta$ from the path. 
	\begin{figure}
	  \centering
		\includegraphics[width=0.47\textwidth]{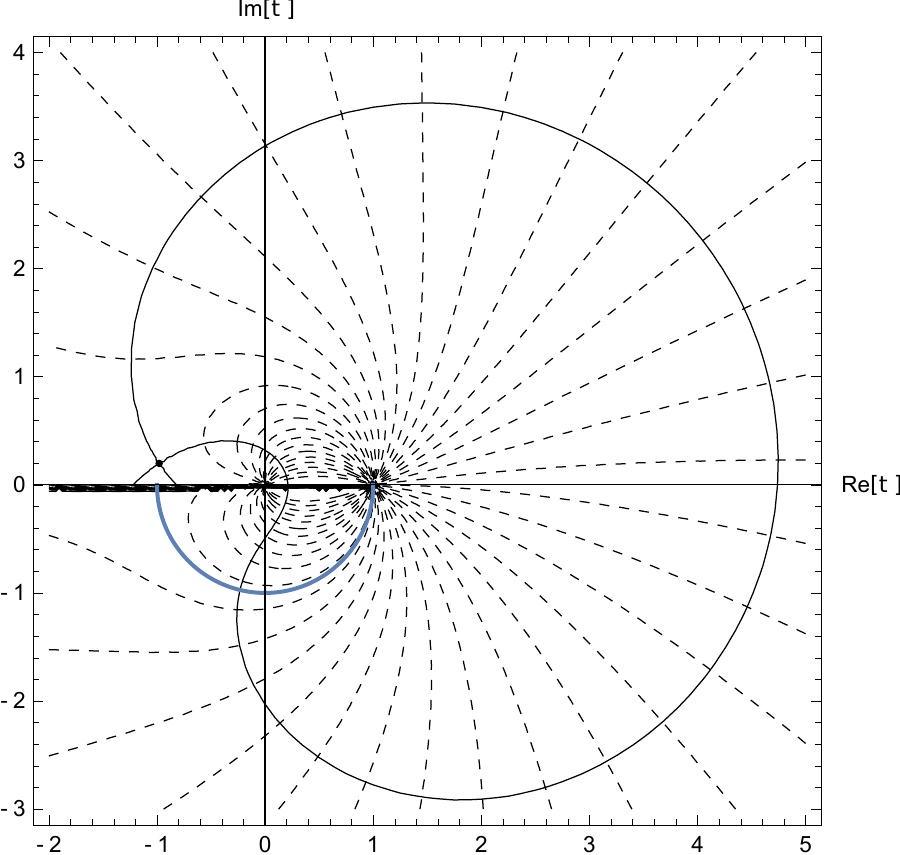}
		\caption{The value of $\Re(\Phi(t))$ (dash line) and the steepest decent and ascend path (black line) over the $t$-complex plane for $\gamma=0.1$. The blue line shows the position of possible poles $t_{\beta}$ of $f$}
		\label{fig:m>0}
	\end{figure}
	
	There is one saddle point $t_0$ given by the solution of the equation $\Psi'(t)=0$
	\begin{equation}
		 t_0=\frac{\gamma +\ui}{\gamma-\ui}
	 \end{equation} 
	consequently, at the saddle point $\Re(\Psi(t_0))=\pi$. The steepest decent and ascend curves are shown in Fig. \ref{fig:m>0}. The original integration path then can be deformed as the steepest decent curve and two equal real part curve of $\Psi(t)$.
	
	The corresponding value at the saddle point $t_0$ reads
	 \begin{align}
		 \ue^{s \Psi(t_0)}=\left(\frac{\gamma+\ui}{\gamma-\ui} \right)^{(\gamma+\ui) s} \left( \frac{2 \ui}{\ui+\gamma} \right)^{-2 \ui s}, \;\;\; f(t_0,\beta)=\frac{(2\ui)^{\ui\epsilon}}{\sqrt{2\ui}}\left(\frac{ \gamma - \ui (1-2 \beta)}{1-\beta} \right)^{-\frac{1}{2}-\ui \epsilon} \left(\frac{\gamma +\ui}{(\gamma-\ui)^3} \right)^{-\frac{1}{2}}
	 \end{align}
	 and 
	 \begin{equation}
		 \Phi''(t_0)=\frac{-\ui (\gamma -\ui)^3}{2 (\gamma +\ui)},\;\; \alpha=arg(n \Psi''(t_0))=\frac{\pi }{2}-\arg \left(\frac{\text{sgn}(\gamma +\ui)}{\text{sgn}(\gamma -\ui)^3}\right),\;\; \theta=\frac{\pi-\alpha}{2}
	 \end{equation}
	Then by the saddle point approximation we have
	\begin{equation}\label{eq:hgf_1_f}
	\begin{split}
		I=\frac{G(n)}{2 \pi \ui} \int_{C} dt f(t) \ue^{s \Psi(t)} & \sim \frac{\ue^{s \Psi(t_0)+\ui \theta}}{\sqrt{n}} \left( f(t_0) \sqrt{\frac{2\pi}{|\Phi''(t_0)|}} +\mathcal{O}(s^{-1})\right), \;\; as \;\; s \to \infty\\
		& \sim {\sqrt{\gamma-\ui}} \left(\frac{ \gamma - \ui  (1-2 \beta)}{1-\beta} \right)^{-1/2} + \mathcal{O}(s^{-1/2})
	\end{split}
	\end{equation}

	Note that the generalization to $\lambda= -s_0 + \delta$ or $s= s_0 + \delta$ leads to a modification with $ \left(\frac{ \gamma - \ui  (1-2 \beta)}{1-\beta} \right)^{- \ui \delta} $.
	
	We also need to consider the branch point $t_{\beta}=(\beta-1)/\beta$. When it lives outside the contour $C$, the integration over contour $C$ is exactly the path required by (\ref{eq:hgf_int_1}). Thus in this case we get the asymptotics of the hypergeometric function with usual saddle point method as (\ref{eq:hgf_1_f}). However, when $(\beta-1)/\beta$ inside the contour, we need to deform the contour to exclude the branch point and the branch cut due to $(\beta-1)/\beta$. A possible way is we choose the branch cut along one of the steepest decent path start at $(1-\beta)/\beta$, and deform the contour $C$ exclude the branch point and branch cut, which may gives a non-trivial contribution to the asymptotic expansion. Since $t_{\beta}=(\beta-1)/\beta$ is a $1/2$ order branch point, according to \cite{miller2006applied}, in this case, the contribution comes from branch point is given by
	\begin{equation}\label{asym_branch}
		\begin{split}
			I_1 &\sim 2\sqrt{\pi} \frac{G(n)}{2 \pi \ui} \ue^{s \Psi(t_{\beta})}f(t_{\beta},\beta) \left(t_{\beta}- \frac{\beta-1}{\beta} \right)^{\frac{1}{2}} \left( \frac{1}{s |\Psi'(t_{\beta})|} \right)^{\frac{1}{2}} + \mathcal{O}(s^{-1/2})\\
			& \sim \left( {1-\beta} \right)^{(\gamma+ \ui )s } \beta^{ (- \gamma + \ui)s } \frac{\sqrt{2 (\gamma-\ui)} (-1)^{\gamma s} 2^{2 \ui s} ((\gamma-\ui))^{(\gamma-\ui)s}}{((\gamma+\ui))^{(\gamma+\ui)s}} \sqrt{\frac{1-\beta}{|-\ui(1-2\beta)+ \gamma|}} +\mathcal{O}(s^{-1/2})
		\end{split}
	\end{equation}
	Since the asymptotics contribution contains power of $s$ in terms of $\ue^{s \Psi(t)}$, the full asymptotics of the function will comes from the largest $Re(\Psi(t))$ of $t_0$ and $t_{\beta}$. In our case, $t_{\beta}$ is in the negative imaginary half plane
	\begin{equation}
		t_{\beta}=\frac{\beta-1}{\beta}=\frac{\bar{\beta}}{\beta}
	\end{equation}
	And it is easy to show
	\begin{equation}
	\Re(\Psi(t_{\beta}))=\left\{ \begin{array}{ll} -\pi, & t< 0 \\ \pi & t>0 \end{array}  \right.
	\end{equation}
	When $t > 0$, the contribution from $t_{\beta}$ is lower than $t_0$ in arbitrary order after multiply by power $s$, and the final result is given by (\ref{eq:hgf_1_f}).
	The contribution form the branch point only exist when $\sinh(t) + \gamma  < \epsilon_0<0 $ and the contribution reads
	\begin{equation}\label{eq:a34}
		I =I_0-I_1
	\end{equation}
	And in this case the final asymptotics is given by the sum of (\ref{eq:hgf_1_f}) and (\ref{asym_branch}). A special case is when the branch point locates near the critical point $|t_0 - t_\beta| \leq \epsilon_{0 }$ , where the result is
	\begin{equation}\label{eq:a35}
		\begin{split}
			I &\sim \frac{G(n)}{2 \pi \ui} \left( \frac{\pi \ue^{\ui \pi (-1/4+ \theta/2)}}{\Gamma(1/4)} f(t_0) \left(t_{0}- \frac{\beta-1}{\beta} \right)^{\frac{1}{2}} \left( \frac{2}{s |\Psi''(t_0)} \right)^{-\frac{1}{4}} \ue^{s \Psi(t_0)} +\mathcal{O}(s^{-3/4}) \right) \\
			&\sim \frac{2\sqrt{\pi}{s}^{1/4} }{\Gamma(1/4)} (-\ui (\gamma-\ui)(\gamma+\ui))^{1/4} +\mathcal{O}(s^{-1/4})
		\end{split}
	\end{equation}
	Note that, for the continuos of the approximation on $\beta$, we have $\epsilon_{0} \sim s^{-1/2}$.
Fig (\ref{fig:error}) shows the error level of above asymptotics result when $s=100$.
	\begin{figure}
	  \centering
		\includegraphics[width=1\textwidth]{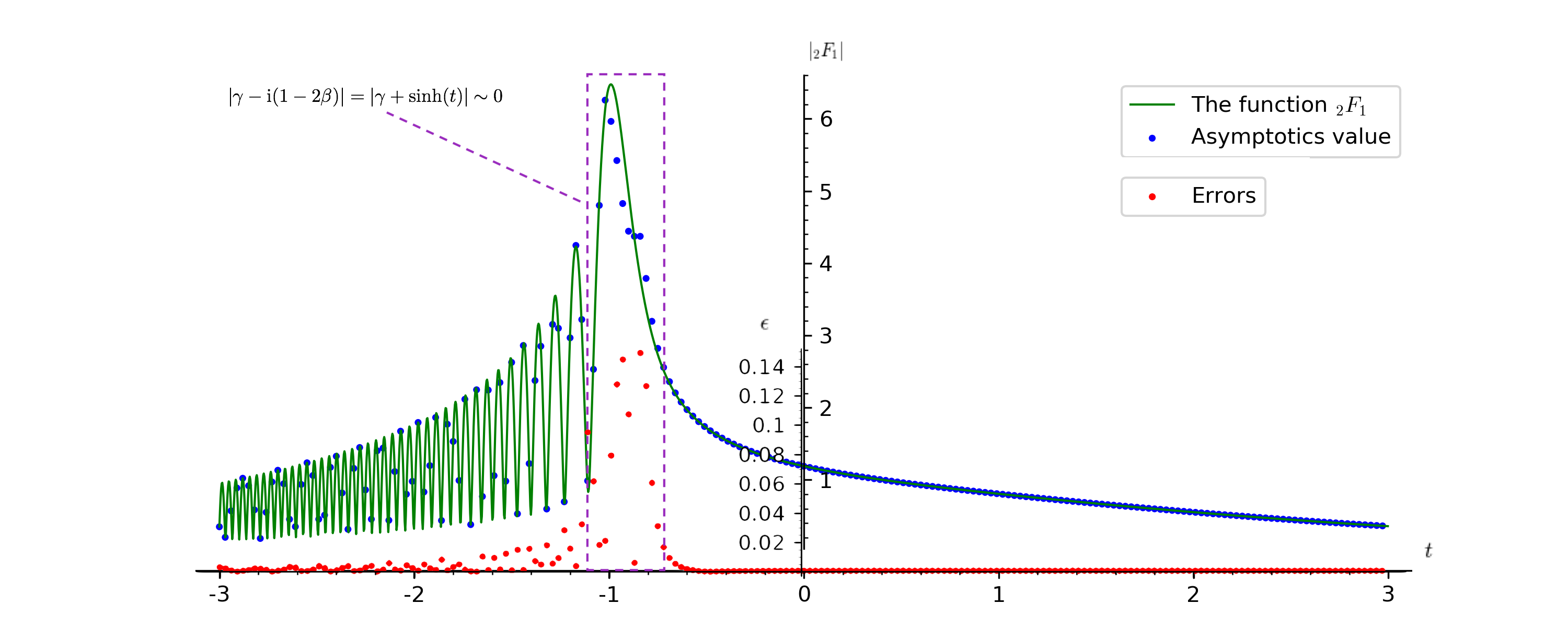}
		\caption{The function ${{}_2F_1} \left( j +\ui \lambda+1,-j+\ui \lambda; m+\ui \lambda +1; \beta \right)$ as shown in (\ref{eq:hgf_trans_1} ) and it's asymptotics result $I$ given as (\ref{eq:hgf_1_f}), (\ref{eq:a34}) and (\ref{eq:a35}) respectively, with $t \in [-3,3]$, $s=100$, $\gamma=1$. The absolute error is defined as $\epsilon=|(|I|-|{}_2F_1|)|/|{}_2F_1|$. }
		\label{fig:error}
	\end{figure}

	\subsubsection{Result}
	Now we can write out the final result, according to (\ref{eq:hgf_trans_1}), we have
	 \begin{equation}
		 {{}_2F_1} \left( j -\ui s+1,-j-\ui s; n/2-\ui s +1; \beta(t) \right) \sim \frac{\sqrt{\gamma-\ui}(1+\ui)}{\sqrt{2 ({\ui \gamma +  (1-2 \beta)})}} + \mathcal{O}(s^{-1/2})
	 \end{equation}

	From (\ref{eq:d_1}), for $\sinh(t) > - \gamma$ we have
	\begin{equation}\label{eq:d_1_1}
	\begin{split}
		{d_0}^j_{n/2,-\ui \lambda,\sigma} \sim & S^j_{m \lambda \sigma}\left( \frac{1}{ \sqrt{ s ({ \gamma - \ui  (1-2 \beta)})}} + \mathcal{O}(s^{-1}) \right) \left( \frac{\left( 1-(-1)^{\sigma} \ui \right) (1-\beta)^{(-\frac{n}{2}+\ui \lambda)/2} \beta^{(\frac{n}{2}+\ui \lambda)/2}}{\Gamma(-\frac{n}{2}-j)\Gamma(\frac{n}{2}+\ui \lambda + 1/2)} \right.\\
		 &\;\;\;\;\;\left. -(-1)^{\sigma}\frac{\beta^{(-\frac{n}{2}-\ui \lambda)/2} (1-\beta)^{(\frac{n}{2}-\ui \lambda)/2} }{\Gamma(j+\ui \lambda +1)\Gamma(\ui \lambda -j)} \right)
	\end{split}
	\end{equation}
	where we use the approximation
	\begin{equation}
		\Gamma(-\frac{n}{2}-j)\Gamma(\frac{n}{2}+\ui \lambda + 1) \sim 2 \pi \sqrt{(\gamma-\ui)s} (-(\gamma+\ui)s)^{-(\gamma+\ui)s} ((\gamma-\ui)s)^{(\gamma-\ui)s} \ue^{2\ui s}
	\end{equation}
	\begin{equation}
		\Gamma(\frac{n}{2}-\ui \lambda +1)\Gamma(j+\ui \lambda +1)\Gamma(\ui \lambda -j)  \Gamma^{-1}(j+m+1) \sim \sqrt{2} \pi 
	\sqrt{(\gamma + \ui)s} (-2 \ui s)^{-2 \ui s} \ue^{2 \ui s}
	\end{equation}
	for $\sinh(t) < -\gamma$, the contribution from the extra branch point reads
	\begin{equation}
		\begin{split}
			{d_1}^j_{n/2,-\ui s,\sigma} \sim &  S^j_{m \lambda \sigma} \left( \frac{\sqrt{2}}{\sqrt{s |\gamma - \ui (1-2\beta)|}} +\mathcal{O}(s^{-1}) \right) \left( \frac{ \left( 1-(-1)^{\sigma} \ui \right) (1-\beta)^{(\frac{n}{2}-\ui \lambda)/2} \beta^{(-\frac{n}{2}-\ui \lambda)/2}}{\sqrt{2}\Gamma(j+\ui \lambda +1)\Gamma(\ui \lambda -j)} \right.\\
			 &\;\;\;\;\;\left. -(-1)^{\sigma}\frac{\sqrt{2} \beta^{(\frac{n}{2}+\ui \lambda)/2} (1-\beta)^{(-\frac{n}{2}+\ui \lambda)/2} }{\Gamma(-\frac{n}{2}-j)\Gamma(\frac{n}{2}+\ui \lambda + 1/2)} \right)
		\end{split}
	\end{equation}
	One check the final result is approximately
	\begin{equation}
		\begin{split}
			d^j_{n/2,-\ui s,\sigma} &= {d_0}^j_{n/2,-\ui s,\sigma} - {d_1}^j_{n/2,-\ui s,\sigma} \sim   S^j_{m \lambda \sigma} \left(\frac{1}{\sqrt{s |\gamma - \ui (1-2\beta)|}} + +\mathcal{O}(s^{-1})\right) \cross\\
			 &\left( \frac{\left( 1-(-1)^{\sigma} \ui \right) (1-\beta)^{(-\frac{n}{2}+\ui \lambda)/2} \beta^{(\frac{n}{2}+\ui \lambda)/2}}{\Gamma(-\frac{n}{2}-j)\Gamma(\frac{n}{2}+\ui \lambda + 1/2)} -(-1)^{\sigma}\frac{\beta^{(-\frac{n}{2}-\ui \lambda)/2} (1-\beta)^{(\frac{n}{2}-\ui \lambda)/2} }{\Gamma(j+\ui \lambda +1)\Gamma(\ui \lambda -j)} \right)
		\end{split}
	\end{equation}
	When $|\gamma - \ui (1-2\beta)| < \epsilon$, which means the branch point near the saddle point, we have
	\begin{equation}
		\begin{split}
			d^j_{n/2,-\ui s,\sigma}  \sim & S^j_{m \lambda \sigma} \left( \frac{2 \sqrt{\pi} (-\ui(1+\gamma^2))^{1/4} s^{1/4}}{ \Gamma(1/4) \sqrt{s }} +\mathcal{O}(s^{-3/4})\right)\left( \frac{  \left( 1-(-1)^{\sigma} \ui \right) (1-\beta)^{(-\frac{n}{2}+\ui \lambda)/2} \beta^{(\frac{n}{2}+\ui \lambda)/2}}{\Gamma(-\frac{n}{2}-j)\Gamma(\frac{n}{2}+\ui \lambda + 1/2)} \right.\\
			&\;\;\;\;\;\left. -(-1)^{\sigma}\frac{ \beta^{(-\frac{n}{2}-\ui \lambda)/2} (1-\beta)^{(\frac{n}{2}-\ui \lambda)/2} }{\Gamma(j+\ui \lambda +1)\Gamma(\ui \lambda -j)} \right)
		\end{split}
	\end{equation}

	\subsection{full representation matrix}
	According to (\ref{eq:a10}), now we can write out $D$ matrix in terms of group elements $v$:
	\begin{equation}\label{eq:final_D_matrix}
		\begin{split}
		D_{m, \lambda}(z)=&\frac{{S}^j_{m, \lambda ,\sigma}}{\sqrt{s_0}} \left( \frac{H( |{\gamma + \Im(\bar{v}_1 v_2))}|- \epsilon)}{\sqrt{|{\gamma + \Im(\bar{v}_1 v_2))}|}} + H(\epsilon -  |{\gamma + \Im(\bar{v}_1 v_2))}| ) \frac{2 \sqrt{\pi} (1+\gamma^2)^{1/4} s_0^{1/4}}{\sqrt{\pi} \Gamma(1/4)} \right)\\
		  &\left( T^{j}_{+\sigma} \Big(\frac{v_1-v_2}{\sqrt{2}}\Big)^{m+ \ui \lambda} \Big(\frac{\overline{v_1-v_2}}{\sqrt{2}}\Big)^{-m+ \ui \lambda} -T^{j}_{- \sigma} \Big(\frac{v_1+v_2}{\sqrt{2}}\Big)^{m - \ui \lambda} \Big(\frac{\overline{v_1+v_2}}{\sqrt{2}}\Big)^{-m - \ui \lambda}\right) +\mathcal{O}(s^{-3/4})
		\end{split}
	\end{equation}
	where $H$ is the Heaviside step function
	\begin{equation}
	H(x) \left\{ \begin{array}{ll} 0 & x \leq 0\\ 1 & x >0 \end{array} \right.
	\end{equation}
	$\epsilon$ is defined as
	\begin{equation}
		\epsilon = \frac{\Gamma(1/4)^2}{4 \pi \sqrt{(1+\gamma^2) s} }
	\end{equation}
	such that $D$ is continuous for $v$. Note that the contribution from $|{\gamma + \Im(\bar{v}_1 v_2))}| < \epsilon$ is actually a regulator of the $1/2$ order singular points because of $|{\gamma + \Im(\bar{v}_1 v_2))}|$. In the inner product this regulator naturally arises as the asymptotics with $1/2$ order singular points. In this sense, we can ignore the regulator since we are only interested in the inner product in the amplitude. 
	The constant is given by
	\begin{align}\label{eq:def_t}
	 &T^j_{+ \sigma}=\frac{1-(-1)^{\sigma} \ui}{\Gamma(-m-j )\Gamma(m-j )}\\
	 &T^j_{- \sigma}=\frac{(-1)^{\sigma}}{\Gamma(j+\ui \lambda +1)\Gamma(\ui \lambda -j) }
	\end{align}
	with $S$ given in (\ref{eq:s_j_m}).
	In the asymptotics limit, we have
	\begin{align}
	 &{S}^j_{m\lambda\sigma}\overline{{S}} {}^j_{m\lambda\sigma} \sim \frac{\pi}{2\cosh (2 \pi s)}, \\
	 &T^j_1\overline{T} {}^j_1 \sim \frac{2 \cos(\pi(-m-\ui s))\cos(\pi(m-\ui s))}{\pi^2} \sim \frac{\cosh(2 \pi s)}{\pi^2}, \quad \text{when} \quad s \gg 1  \\
	 &T^j_2\overline{T} {}^j_2 \sim \frac{\cosh(2 \pi s)}{\pi^2}.
	\end{align}
	where we use the asymptotic approximation of Gamma function
	\begin{equation}
		\lim_{z \to \infty} \frac{\Gamma(z+\epsilon)}{\Gamma(z) z^{\epsilon}} = 1
	\end{equation}
	
    Form the parity property of representation matrix, we have
	\begin{equation}
		D^{\sigma j}_{-m, \lambda}(v)=-(-1)^{\sigma}\ue^{-\ui \pi m}D^{\sigma j}_{m, \lambda}(\bar{v})
	\end{equation}
	
	\section{Analysis of singularities and corresponding stationary phase approximation}\label{ana_singu}
	In this appendix we concentrate on the analysis of singularities appears in the denominator of the integrand of vertex amplitude.
	\subsection{Analysis of singularities}
	For simplicity, we consider one vertex case for some $v$ mainly. As we show, the amplitude enrolls the integration in the form
	\begin{equation}
		I = \int \prod_{e} dg_{ve} \int \prod_{f} \Omega_{vf}  \prod_{f} \frac{1}{h_{vef}h_{ve'f}} e^{S_{vf}}
	\end{equation}
	where $h$ is a real valued function
	\begin{equation}
		h_{vef}={|\langle Z_{vef},Z_{vef} \rangle|} \sqrt{\left|\gamma - \ui( 1- \frac{2 {\langle l^{-}_{ef}, Z_{vef} \rangle \langle Z_{vef}, l^{+}_{ef} \rangle}}{{\langle Z_{vef},Z_{vef} \rangle}} )\right|}
	\end{equation}
	Here each dual face is determined by two edges $f=(e,e')$.
	Note that the square root part inside $h_{vef}$ is the spinor representation for the square root term inside the wigner $d$ matrix:
	\begin{equation}\label{eq:B3}
		\left |{\gamma - \ui( 1- \frac{2 {\langle l^{\-}_{ef}, Z_{vef} \rangle \langle Z_{vef}, l^{+}_{ef} \rangle}}{{\langle Z_{vef},Z_{vef} \rangle}} )} \right| =| {\gamma + \Im(v_1 \bar{v}_2) } | 
	\end{equation}
	The zero sets of $h$ is given by $\langle Z_{vef},Z_{vef} \rangle = 0$ or $| {\gamma + \Im(v_1 \bar{v}_2) }|=0$.
	
	We can rewrite the original $\langle Z_{vef},Z_{vef} \rangle$ as
	\begin{equation}
			  \langle Z_{vef},Z_{vef} \rangle = 2 \Re({\langle l^{-}_{ef}, Z_{vef} \rangle \langle Z_{vef}, l^{+}_{ef} \rangle} ) = \Re(f)
	\end{equation}
	where we define $f$ as
	\begin{equation}
		f :=2 {\langle l^{-}_{ef}, Z_{vef} \rangle \langle Z_{vef}, l^{+}_{ef} \rangle}
	\end{equation}
	In this notation $h_{vef}$ becomes
	\begin{equation}
		h_{vef} = |\Re(f)| \sqrt{|\gamma + \frac{\Im(f)}{\Re(f)}|} = |f| |\cos(\phi_f)| \sqrt{ |\gamma + \tan(\phi_f)|}
	\end{equation}
	
	Suppose the function $f$ are linearly independent to each other. This requirement is the same as require the boundary tetrahedron $l^{\pm}_{ef}$ is non degenerate. In this case, we can define a coordinate transformation among the set of the original coordinates $(z, g) \to (\Re(f), \Im(f), z' ,g')$. The coordinate transformation only transfer among the number of $f$ variables and leaves the left invariant, e.g. we only transfer $40$ variables in one vertex case and leave the other $4$ invariant. The elements of Jacobian matrix of the transformation $J(f)$ is given by
	\begin{align}
		& \frac{\partial (\Re(f_{vef}))}{\partial z} = \overline{\frac{\partial (\Re(f_{vef}))}{\partial \bar{z}}}= \delta_z \langle Z_{vef},Z_{vef} \rangle = (g_{ve} \eta Z_{vef} )^T\\
		& \frac{\partial (\Im(f_{vef}))}{\partial z} =  \ui (\delta_z \langle Z_{vef},Z_{vef} \rangle - 2 \delta_z {\langle l^{-}_{ef}, Z_{vef} \rangle \langle Z_{vef}, l^{+}_{ef} \rangle} = \ui((g_{ve} \eta Z_{vef} - 2 g \eta l^{+}_{ef} \langle l^{-}_{ef}, Z_{vef} \rangle)^T\\
		& \frac{\partial (\Re(f_{vef}))}{\partial g} = \delta_g \langle Z_{vef},Z_{vef} \rangle = \langle L^{\dagger} Z_{vef},Z_{vef} \rangle + \langle Z_{vef}, L^{\dagger}Z_{vef} \rangle \label{eq:B10}\\
		& \frac{\partial (\Im(f_{vef}))}{\partial g} = \ui ( \langle L^{\dagger} Z_{vef},Z_{vef} \rangle + \langle Z_{vef}, L^{\dagger}Z_{vef} \rangle \\
		& \qquad \qquad \qquad \qquad - 2 {\langle l^{-}_{ef}, Z_{vef} \rangle \langle L^{\dagger} Z_{vef}, l^{+}_{ef} \rangle} - 2 {\langle l^{-}_{ef}, L^{\dagger} Z_{vef} \rangle \langle Z_{vef}, l^{+}_{ef} \rangle} ) \notag
	\end{align}
	where $L$ represents generators of $\text{SL}(2,\C)$. Note that $\delta_g \langle Z_{vef},Z_{vef} \rangle$ is zero when $L$ are $\text{SU}(1,1)$ generators. However, the Jacobian is non zero in general, e.g. in one vertex case of vertex $v$, we have the non-trivial contribution from terms like 
	\begin{equation}
	\begin{split}
		&\partial_{g_1} (13,14,15), \quad \partial_{g_2} (21,24,25),  \quad \partial_{g_3} (31,32,35), \quad \partial_{g_4} (42,43,45),\\
		 &\quad \partial_z (12, 23, 34, 41, 51,52,53,54)
	\end{split} 
	\end{equation}
	where $12$ is the representation of $ef$ label in terms of numbers labelling edges and corresponding faces $(e_1,e_2)$.
	Apart from those $0$ in (\ref{eq:B10}), other zeros of matrix elements only possible when $Z= \zeta l^{\pm}$. The Jacobian matrix in this case is given by ($Z= \zeta l^{+}$ as an example),
	\begin{align}
		& \frac{\partial (\Re(f_{vef}))}{\partial z} = \overline{\frac{\partial (\Re(f_{vef}))}{\partial \bar{z}}}= (g_{ve} \eta Z_{vef} )^T\\
		& \frac{\partial (\Im(f_{vef}))}{\partial z} =\overline{\frac{\partial (\Im(f_{vef}))}{\partial \bar{z}}}= - \ui(g_{ve} \eta Z_{vef})^T \\
		& \frac{\partial (\Re(f_{vef}))}{\partial g} =\langle L^{\dagger} Z_{vef},Z_{vef} \rangle + \langle Z_{vef}, L^{\dagger}Z_{vef} \rangle =\left\{ \begin{array}{ll} 0, & L =F \\ 2 \langle Z_{vef}, L^{\dagger}Z_{vef} \rangle, & L = \ui {F} \end{array} \right.\\
		& \frac{\partial (\Im(f_{vef}))}{\partial g} = \ui ( \langle Z_{vef}, L^{\dagger}Z_{vef} \rangle-{ \langle L^{\dagger} Z_{vef}, Z_{vef} \rangle})=\left\{ \begin{array}{ll}  2 \ui \langle Z_{vef}, L^{\dagger}Z_{vef} \rangle, & L = {F}\\0, & L = \ui F  \end{array} \right.
	\end{align}
	Clearly the Jacobian matrix is still well defined and leads to non zero Jacobian. 
	
	After this coordinate transformation, the original integration becomes
	\begin{equation}
		I = \prod_{v} \int \frac{\Omega'}{J(f)}\prod_{e, f} \ud \Re(f_{vef}) \ud \Im(f_{vef})  \prod_{f} \frac{e^{S_{vf}}}{|\Re(f_{vef})|| \Re(f_{ve'f})| \sqrt{|\gamma + \frac{\Im(f_{vef})}{\Re(f_{vef})}|} \sqrt{|\gamma + \frac{\Im(f_{ve'f})}{\Re(f_{ve'f})}|}}
	\end{equation}
	With a further polar coordinate transformation
	\begin{equation}
		\rho_{vef}=\sqrt{\Re(f_{vef})^2 + \Im(f_{vef})^2}, \qquad \phi_{vef} = \arg(f_{vef}) \; \in \; [0, \pi/2)
	\end{equation}
	whose Jacobian is given by
	\begin{equation}
		J^1_{vef}= \frac{1}{\rho_{vef}}
	\end{equation}
	The Jacobian is well defined except on the points where $|f|=0$. After the coordinates transformation, we have
	\begin{equation}\label{eq:B21}
		I = \int {\Omega'} \prod_{e, f} \int \ud \rho_{vef} \int_{0}^{\pi/2} \ud \phi_{vef} \frac{1}{J(\rho, \phi)}  \prod_{f} \frac{e^{S_{vf}}}{|\cos(\phi_{vef})| |\cos(\phi_{ve'f})| \sqrt{|\gamma + \tan(\phi_{vef})||\gamma + \tan(\phi_{ve'f})|}}
	\end{equation}
	Clearly all possible singular points are $1/2$ order. The singular points due to $|\gamma + \tan(\phi_{ve'f})|$ and due to $|\cos(\phi_{ve'f})|$ are separated. The integration respects to $\rho$ does not have singularities.
	
	\subsection{Multidimensional Stationary phase approximation}
	In appendix \ref{app_f}, we already use the saddle point approximation when there is a branch point appearing in the non-scaled function $g(x)$. When adapting to the stationary phase approximation, for the $1/2$ order singular point locates exactly at the critical point, the result is the following:
	\begin{equation}
		I = \int \frac{g(x)}{\sqrt{x}} \ue^{\Lambda S(x)} \sim g(x_c) \frac{\pi \ue^{\ui \pi (\mu-2)/8}}{\Gamma(3/4)}\left(\frac{2}{\Lambda|S''(x_c)|}\right)^{1/4} \ue^{\Lambda S(x_c)} 
	\end{equation}
	where $\Lambda \sim \infty$ and $S$ is pure imaginary. Note that the dominate part here is given by the $-1/4$ order of $\Lambda$ instead of $-1/2$ as in the asymptotic formula without singularities. The regulator appears in (\ref{eq:final_D_matrix}) is exactly this $1/4$ order difference.

	However, this asymptotic formula only hold for single variable integral. We will generalize this single variable approximation to multi variables case. Recall Fubini's theorem:
	\begin{theorem}
		Let $w = f(x_1,x_2, \dots, x_n)$ be a $n$ variable valued complex function. If the integral of $f$ on the domain $B=\prod_i^n I_n$ where $I_n$ are intervals in $\R$ is absolutely convergent:
		\begin{equation}
			\int_{B} |f(x_1,x_2,\dots,x_n)| d (x_1,x_2,\cdots, x_2) < \infty,
		\end{equation}
		then the multiple integral will give the same result as the iterated integral,
		\begin{equation}
			\int_{A \cross B} |f(x,y)| d(x,y) = \int_{A}(\int_{B} f(x,y) dy) dx =  \int_{B}(\int_{A} f(x,y) dx) dy
		\end{equation}
		The result is independent of the iterate order.
	\end{theorem}
	Here from (\ref{eq:B21}) we have the integral in the form  
		\begin{equation}
			I = \int d^n x \prod_{i=1}^{j} (x_i)^{-1/2} g(x) \ue^{\ui S(x)}
		\end{equation}
		where $S(x) \in \R$, $x \in \R^n$, $j < n$ and $g(x)$ is analytic. $j<n$ illustrates the fact that only in a subspace of the total variables space will have singularities.
		Then in a closed region $M$ where the stationary phase points (solutions of $\delta S =0$) exists, we have
		\begin{equation}
			\int_{M} d^n x |\prod_i (x_i)^{-1/2} g(x) \ue^{\ui S(x)}| \sim \int_{M} d^n x |\prod_i (x_i)^{-1/2} \tilde{g}(x) | < \infty
		\end{equation}
		From Fubini's theorem, we then can write the multi-dimensional integral as iterated integral. For the original variables, since the singularities exist only in a subspace of the total variables space, we can always perform a coordinate transformation, such that variables with singularities are separated from those do not have, as we show in (\ref{eq:B21}). Then the final result is given by performing the stationary phase approximation iteratively. Each step one may use the usual stationary phase approximation or the one with singularities. The lowest order of the total integration is given by picking lowest order approximation of each single integration.
	
	However, due to technical reason, we would like to derive the saddle point equations directly from $S(x)$ instead of evaluate it iteratively. According to the approximation, each single valued integral is dominated by the phase $S(x_0)$ where $x_0$ is the solution of saddle point equation $\delta_x S(x)=0$. Then iteratively, the saddle points are given by 
	\begin{equation}\label{eq:B25}
		\begin{split}
		\delta_{x_1} S(x_1,x_2, \dots, x_n) &= 0, \\
		 \delta_{x_2} S(x_1^0, x_2, \dots, x_n)&=\left(\delta_{x_1} S(x)\frac{\partial x_1^0}{\partial x_2}  + \delta_{x_2} S(x)\right) \Big|_{x_1=x_1^0} = \delta_{x_2} S(x)|_{x_1=x_1^0} =0,\\
		& \vdots \\
		 \delta_{x_n} S(x_1^0, x_2^0, \dots, x_n) &= \delta_{x_n} S(x)|_{x_1=x_1^0,x_2=x_2^0, \dots, x_{n-1}=x_{n-1}^0} =0 \\
		\end{split}
	\end{equation}
	where $x_i^0(x_{i+1}, \dots, x_n)$ is the solution of the corresponding equation of motion $\delta_{x_i}(x_1^0, \dots, x_{i-1}^0, x_i, x_{i+1},\dots, x_n)$ respect to $x_i$. As one can see from (\ref{eq:B25}), the above equation of motion is nothing else but we solve the original equation of motion $\{E_n = \delta S(x)\}$ iteratively. Thus they have the same solutions. The saddle points given by the two method will coincide to each other. {Note that, for variables whose saddle points are near the singularities, the induced measure which contains second derivatives of the action will be given in the order $1/4$ in contrast to $1/2$ for those do not have singularities. As a result, there is no general Hessian term in contrast to the previous EPRL approximation, and the measure is more involved as some special functions of second derivatives of the action. As a result finally we have order $I \sim g(\Lambda) \Lambda^{-a/2-b/4} $ for $b$ variables have singular points.}
	
	\section{Analysis of critical points in bivector representation}\label{app:bivector_critical}
	In this appendix we will analysis and reformulate the critical point equations we get in Sec. \ref{sec3} in bivector representation. The analysis is done for all possible actions appearing in the amplitude (\ref{eq:amp}).
	
	\subsubsection{$S_{vf+}$ case}
	From (\ref{eq:sol_parall}) and (\ref{eq:va_z_s+}) in $S_{vf+}$ case,
	\begin{align}\label{eq:s+_parallel}
		&g_{ve} \eta l^{+}_{ef} =\frac{\bar{\zeta}_{vef}}{\bar{\zeta}_{ve'f}} g_{ve'} \eta  l^{+}_{e'f}
		&g_{ve} \; J \; \tilde{Z}_{vef} = \frac{\bar{\zeta}_{ve'f}}{\bar{\zeta}_{vef}} g_{ve'} \; J \; \tilde{Z}_{vef}
	\end{align}
	we have
	\begin{equation}
		g_{ve} \eta l^{+}_{ef} \otimes (l^{-}_{ef} + \alpha_{vef} l^{+}_{ef})^{\dagger} g_{ev} = g_{ve'} \eta l^{+}_{ef} \otimes (l^{-}_{e'f} + \alpha_{ve'f} l^{+}_{e'f})^{\dagger} g_{e'v}
	\end{equation}
	with the fact that $\langle l^{+}, l^{+} \rangle = 0$ and $\langle l^{-}, l^{+} \rangle =1$. With (\ref{eq:bivector_tensorp}), the above equation can be written as
	\begin{equation}
		g_{ve} ( V_{ef} + \ui \bar{\alpha}_{vef} W^{+}_{ef}) g_{ev} =  g_{ve'} ( V_{e'f} + \ui \bar{\alpha}_{ve'f} W^{+}_{e'f}) g_{e'v}
	\end{equation}
	In spin-1 representation, this equation reads
	\begin{equation}\label{eq:s+_pa}
		g_{ve} (V_{ef} + (\Im(\alpha_{vef}) + \Re({\alpha}_{vef}) * ) W^{+}_{ef}) g_{ev} =  g_{ve'} (V_{e'f} + (\Im(\alpha_{ve'f})+ \Re({\alpha}_{ve'f}) * ) W^{+}_{e'f}) g_{e'v}
	\end{equation}
	We can define a bivector $X_{vef}$
	\begin{equation}
		 X_{vef} = V_{ef} + (\Im(\alpha_{vef})+ \Re({\alpha}_{vef}) * ) W^{+}_{ef}
	 \end{equation} 
	Easy to check $X$ is a simple bivector which can be expressed as
	\begin{equation}\label{eq:xvef+}
		X = * (v + \Im(\alpha) w^{+} ) \wedge (u - \Re(\alpha) w^{+}) = * ( \tilde{v} \wedge \tilde{u} )
	\end{equation}
	Here by the definition of $v$ and $w$, we have
	\begin{equation}\label{eq:tilde_v}
		 \tilde{v}^I= (\tilde{v}^0, -\tilde{v}^2, \tilde{v}^1,0 )\;, \qquad \tilde{w}^I= ({w^+}^0, -{w^+}^2, {w^+}^1,0 )\;,
	 \end{equation}
	 where 
	 \begin{equation}
		\tilde{v}^i= - 2 \langle l^{-} + \ui \Im(\alpha) l^{+}, F^{i} l^{+} \rangle, {w^{+}}^i=  2 \ui \langle l^{+}, F^{i} l^{+} \rangle
	 \end{equation}
	 One can check $\tilde{v}^I \tilde{v}_I = \tilde{u}^I \tilde{u}_I = 1$, thus $X$ is timelike.
	(\ref{eq:s+_pa}) implies
	\begin{equation}
		(G_{ve} \tilde{v}_{vef}) \wedge (G_{ve} \tilde{u}_{vef}) = (G_{ve'} \tilde{v}_{ve'f}) \wedge (G_{ve'} \tilde{u}_{ve'f}).
	\end{equation} 
	which reminds us define
	\begin{equation}\label{eq:x_vf_+}
		X_{f}(v) : = G_{ve} X_{vef} G_{ev} = G_{ve'} X_{ve'f} G_{e'v}
	\end{equation}
	Noted that, from this equation, we have
	\begin{equation}
		(G_{ve} u)_I X_{f}^{IJ}(v) = - \Re(\alpha_{vef}) (G_{ve} w_{ef}^{+})^{J}
	\end{equation}
	which is $0$ only when $\Re(\alpha_{vef}) = 0$.
	
	Go back to equations we get from the variation respecting to $g$, clearly (\ref{eq:va_e}) and (\ref{va_e1_+}) can be written as
	\begin{align}
		&\sum_{f} \epsilon_{ef}(v) \langle l^{-} + \ui \Im(\alpha) l^{+}, F^{i} l^{+} \rangle = 0 \\
		&\sum_{f} \epsilon_{ef}(v) \Re(\alpha) \langle l^{+}, F^{i} l^{+} \rangle = 0
	\end{align}
	In terms of 4 vectors $\tilde{v}$ and $w$, these equation reads
	\begin{align}\label{eq:clo_+}
		&\sum_{f}  \epsilon_{ef}(v) G_{ve}\tilde{v}_{vef} = 0 
		&\sum_{f} \epsilon_{ef}(v) \Re(\alpha_{vef}) G_{ve} w^{+}_{ef}= 0
	\end{align}
	where $\tilde{v}$ is defined by (\ref{eq:tilde_v}).
	Then we can write (\ref{eq:clo_+}) as
	\begin{equation}
		\sum_f \epsilon_{ef}(v) X_{f}(v) =0
	\end{equation}
	which is a closure condition to bivectors.
	
	\subsubsection{$S_{vf-}$ case}
	In this case, from (\ref{eq:sol_parall}) and (\ref{eq:va_z_s-}) we have
	\begin{align}
		&g_{ve}\; \eta n_{vef} =\frac{\bar{\zeta}_{vef} \Re(\alpha_{vef})}{\bar{\zeta}_{ve'f} \Re(\alpha_{ve'f}) } g_{ve'} \; \eta  n_{ve'f} \\
		&g_{ve} \; J \; \tilde{Z}_{vef} = \frac{\bar{\zeta}_{ve'f}}{\bar{\zeta}_{vef}} g_{ve'} \; J \; \tilde{Z}_{vef}
	\end{align}
	where $n_{ef} := l^{+}_{ef} + \ui (\gamma \Re(\alpha_{vef}) + \Im(\alpha_{vef})) l^{-}_{ef}$. Note with equation (\ref{eq:trans_v_x-}), we see $n$ does not change for different vertex $v$: $n_{ef}(v)=n_{ef}(v')$. $n$ defined here satisfies the relation in Lemma \ref{lemma:l+o}, thus according to Lemma \ref{lemma:basis}, $\{n, l^{-} \}$ forms a null basis. 
	With $n$ and $l^{-}$, $\tilde{Z}$ can be rewritten as
	\begin{equation}
		Z= l^{+} + \alpha l^{-} = n + (1- \ui \gamma) \Re(\alpha) l^{-}
	\end{equation}
	This leads to the tensor product equation
	\begin{equation}\label{eq:c19}
		g_{ve} \frac{\eta n_{ef }}{\Re(\alpha_{ef})} \otimes (n_{ef} + (1- \ui \gamma ) \Re(\alpha_{vef}) l^{-}_{ef})^{\dagger} g_{ev} = (e \to e')
	\end{equation}
	The right part of above equation means exchange all the $e$ in left part to $e'$. 
	
	In terms of bivector variables, according to (\ref{eq:bivector_tensorp}), we have
	\begin{equation}
		g_{ve} (V_{ef} + \frac{(\ui - \gamma)W^{+}_{ef}}{( 1+\gamma^2 )\Re(\alpha_{vef})}) g_{ev} = (e \to e')
	\end{equation}
	Noted now $V$ is the space-like bivector generated by $n$ with $l^{-}$ and $W^{+}$ is null bivector generated by $n$ with itself. 
	Again bivector $X_{vef} := V_{ef} - (\gamma - *) W_{ef}/((1+\gamma^2)\Re(\alpha) )$ is a simple bivector. $X_{vef}$ can be written as
	\begin{equation}\label{eq:xvef-}
		X_{vef} = * (( v_{ef} - \frac{\gamma}{(1+\gamma^2)\Re(\alpha_{vef})}w_{ef}^{+} ) \wedge (u - \frac{\gamma}{(1+\gamma^2)\Re(\alpha_{vef})} w^{+}_{ef})) = *(\tilde{v}_{vef} \wedge \tilde{u}_{vef})
	\end{equation}
	where
	\begin{equation}
		 \tilde{v}^I= (\tilde{v}^0, -\tilde{v}^2, \tilde{v}^1,0 ), \qquad {w^{+}}^i=  2 \ui \langle n, F^{i} n \rangle
	 \end{equation} 
	Here
	\begin{equation}
		\tilde{v}^i= 2 \langle n, F^i (l^{-} - \frac{\ui \gamma n}{(1+\gamma^2)\Re(\alpha)} ) \rangle, {w^{+}}^i=2 \ui \langle n, F^i n \rangle
	\end{equation} 
	$\tilde{v}^I \tilde{v}_I = \tilde{u}^I \tilde{u}_I = 1$ implies $X$ is timelike.
	
	Then (\ref{eq:c19}) leads to
	\begin{equation}
		X_{f}(v):= G_{ve} X_{vef} G_{ev} = G_{ve'} X_{ve'f} G_{e'v}
	\end{equation}
	which is the parallel transport of $X$ between edge $e$ and $e'$. With (\ref{eq:xvef-}), we can write $X_{f}(v)$ as
	\begin{equation}
		X_{f}(v) = G_{ve} \tilde{v}_{vef}) \wedge (G_{ve} \tilde{u}_{vef}
	\end{equation}
	Note here again we have
	\begin{equation}
		(G_{ve} u)_I  X_{vf}^{IJ} = -  \frac{1}{(1+\gamma^2)\Re(\alpha)} (G_{ve} w_{ef}^{+})^{J}
	\end{equation}
	which is some null vector and can not be 0.
	
	Form (\ref{eq:va_e}) and (\ref{va_e1_-}), we have the following equations of motion from variation respecting to $g$
	\begin{align}
		&\sum_{f} \epsilon_{ef}(v) \langle n, F^{\dagger} (l^{-} - \frac{\ui \gamma n}{(1+\gamma^2)\Re(\alpha)} ) \rangle =0
		&\sum_{f} \epsilon_{ef}(v) \frac{\langle n, F^{\dagger} n \rangle}{\Re(\alpha)} =0
	\end{align}
	In terms of 4-vectors, 
	\begin{align}\label{eq:clo_-}
		&\sum_{f} \epsilon_{ef}(v) G_{ve} {v}_{ef} =0
		&\sum_{f} \epsilon_{ef}(v) \frac{G_{ve}  w^{+}_{ef}}{\Re(\alpha)} =0
	\end{align}
	which leads to
	\begin{equation}
		\sum_f \epsilon_{ef}(v) X_{f}(v) =0
	\end{equation}
	
	\subsubsection{$S_{vfx}$ case}
	We will use $S_{vfx-}$ as an example, the $S_{vfx+}$ will be exactly the same but switch $e$ and $e'$ here. From the critical point equations (\ref{eq:sol_parall}) and (\ref{eq:va_z_sx}),
	\begin{equation}
		\begin{split}
			&(\gamma - \ui)s_f \frac{g_{ve} \eta l^{+}_{vef}}{\bar{\zeta}_{vef}} = - \ui s_f \frac{ g_{ve'} \eta n_{ve'f} }{\bar{\zeta}_{ve'f} \Re(\alpha_{ve'f}) }\;,\\
			& \quad  g_{ve} \bar{\zeta}_{vef} J \; (l^{-}_{ef} + \alpha_{vef} l^{+}_{ef} )=g_{ve'} \bar{\zeta}_{ve'f} J \; (l^{+}_{e'f} + \alpha_{ve'f} l^{-}_{e'f} )
		\end{split}
	\end{equation}
	With the equation (\ref{eq:trans_v_x-}) from the variation respecting to $\text{SU}(1,1)$ group elements $v_{ef}$, in this case $n = l^{+}$, and $\tilde{Z}_{ve'f}$ can be written as $\tilde{Z}_{ve'f}= l^{+}_{e'f} + (1-\ui \gamma ) \Re(\alpha_{ve'f}) l^{-}_{e'f}$.
	
	The tensor product between the two equations leads to
	\begin{equation}
	\begin{split}
		&(\ui \gamma +1) g_{ve} ( \eta l^{+}_{ef} \otimes (l^{-}_{ef})^{\dagger}+ \bar{\alpha}_{vef} \eta l^{+}_{ef} \otimes (l^{+}_{ef})^{\dagger}) g_{ev} =  g_{ve'} \; \eta n_{ve'f} \otimes (\frac{n_{ve'f}}{\Re(\alpha_{ve'f})} + (1-\ui \gamma) l^{-}_{e'f} )^{\dagger} \; g_{e'v}\\
		&= g_{ve'} \;  (\frac{\eta n_{ve'f} \otimes n_{ve'f}}{\Re(\alpha_{ve'f})} + (1+ \ui \gamma) \eta n_{ve'f} \otimes  (l^{-}_{e'f} )^{\dagger} ) \; g_{e'v}
	\end{split}
	\end{equation}
	In bivector representation
	\begin{equation}
	\begin{split}
		&g_{ve} ( V_{ef} + \ui \bar{\alpha}_{vef} W^{+}_{ef}) g_{ev} = g_{ve'} \;  (V_{e'f} + \frac{(\ui - \gamma) W^{+}_{ve'f}}{\Re(\alpha_{e'f})(1+\gamma^2) }  ) \; g_{e'v}
	\end{split}
	\end{equation}
	Easily to see one recovers the corresponding bivectors in $S_{vf\pm}$ case respectively. Thus the equation implies
	\begin{equation}
		X_{f}(v) := g_{ve} X_{vef} g_{ev} = g_{ve'} X_{ve'f} g_{e'v}
	\end{equation}
	with $X_{vef}$ defined by (\ref{eq:xvef+}) and $X_{ve'f}$ defined by (\ref{eq:xvef-}).
	The closure constraint, in these case, are the combination of corresponding equation in (\ref{eq:clo_+}) or (\ref{eq:clo_-}) according to their representations in $S_+$ or $S_-$. Then we still have
	\begin{equation}
		\sum_f \epsilon_{ef}(v) X_{f}(v) =0
	\end{equation}
	
	\section{Brief review of critical point equations with spacelike triangles in timelike tetrahedra}\label{app:sp_face}
	In this appendix we briefly summarize the critical point equations for spacelike triangles in a timelike tetrahedron. The result was derived in \cite{Kaminski:2017eew}.
	As we described before, spacelike faces corresponding to the discrete series representation of $\text{SU}(1,1)$ group. In this case, the simplicity constraint implies
	\begin{equation}
		\rho_f = \gamma j_f, \qquad n_f/2 = j_f
	\end{equation}
	with areas spectrum asymptotically given by $A_f =\gamma \sqrt{j_f(j_f+1)} \sim \rho_f =\gamma j_f$.
	
	The embedded coherent state reads
	\begin{equation}
		\begin{split}
			f^{j \alpha}_{\xi}=(\alpha \langle z,z \rangle)^{\ui \rho/2 -1-j} (\alpha \langle \xi^{\alpha}, \bar{z} \rangle)^{- 2 j}
		\end{split}
	\end{equation}
	where $\alpha= \pm = \langle z,z \rangle$ for spinors $z$. $\xi$ are spinors defined as
	\begin{equation}
		 \xi^{\alpha}=v^{-1 \dagger} \xi_0^{\alpha}, \qquad \text{with} \quad \left\{ \begin{array}{l} \xi_0^{+} = (1,0)^{T}\\ \xi_0^{-}=(0,1)^{T} \end{array}\right., \quad v \in \text{SU}(1,1)
	 \end{equation} 
	With these coherent states, it's immediately to see the action reads
	\begin{equation}
		\begin{split}
			S_{vf}^{\pm}
			&=\ui \gamma j_f \ln \frac{\langle Z_{vef},Z_{vef} \rangle}{\langle Z_{ve'f},Z_{ve'f} \rangle} - j_f \ln \frac{{\langle \xi_{e'f}^{\pm},Z_{ve'f} \rangle}^2 {\langle Z_{vef},\xi_{ef}^{\pm} \rangle}^2}{{\langle Z_{vef},Z_{vef} \rangle}{\langle Z_{ve'f},Z_{ve'f} \rangle} }
		\end{split}
	\end{equation}
	Here we use the simplicity constraint $\rho_f = 2 \gamma j_f$. $Z_{vef}$ is again defined by $Z_{vef}= g_{ve}^{\dagger} \bar{z}_{vf}$
	The real parts of the action reads
	\begin{equation}
		\begin{split}
		 \Re S= -j_f \Re \ln \frac{{\langle \xi_{e'f},Z_{ve'f} \rangle}^2 {\langle Z_{vef},\xi_{ef} \rangle}^2}{{\langle Z_{vef},Z_{vef} \rangle}{\langle Z_{ve'f},Z_{ve'f} \rangle} } \leq 0
		\end{split}
	\end{equation}
	From $\Re S_0=0$, we have
	\begin{equation}\label{eq:re_c_spa}
		\begin{split}
		Z_{vef} = \zeta_{vef} \xi_{ef}^{\pm}
		\end{split}
	\end{equation}
	Due to $Z_{vef}= g_{ve}^{\dagger} \bar{z}_{vf}$, this equation leads to
	\begin{equation}\label{eq:pa_jxi}
		g_{ve} J \xi_{ef}^{\pm} = \frac{\bar{\zeta}_{ve'f}}{\bar{\zeta}_{vef}} g_{ve'} J \xi_{e'f}^{\pm}
	\end{equation}
	The variation of the action reads
	\begin{equation}
		\begin{split}
		 \delta S_{vf}= j_f (1+\ui \gamma) \frac{\delta \langle Z_{vef},Z_{vef} \rangle}{\langle Z_{vef},Z_{vef} \rangle} +j_f (1-\ui \gamma)  \frac{\delta \langle Z_{ve'f},Z_{ve'f} \rangle}{\langle Z_{ve'f},Z_{ve'f} \rangle}-2j_f \left( \frac{\delta \langle \xi_{ef}^{\pm},Z_{vef} \rangle}{\langle \xi_{ef}^{\pm},Z_{vef} \rangle}+\frac{\delta \langle Z_{ve'f},\xi_{ef}^{\pm} \rangle}{\langle Z_{ve'f},\xi_{e'f}^{\pm} \rangle} \right)
		\end{split}	
	\end{equation}
	
	\subsubsection{critical point equation}
	Note that the variation takes the same properties as in timelike triangle case, where the variation respects to $z$ leads to
		\begin{equation}
			\begin{split}
				&\delta_z S =j_f (1+\ui \gamma) \frac{(g_{ve} \eta Z_{vef})^{T}}{\langle Z_{vef},Z_{vef} \rangle} \\
				&\qquad + j_f (1-\ui \gamma)  \frac{(g_{ve'} \eta Z_{ve'f})^{T}}{\langle Z_{ve'f},Z_{ve'f} \rangle}-2j_f  \frac{(g_{ve'} \eta \xi^{\pm}_{e'f})}{\langle Z_{ve'f},\xi^{\pm}_{e'f} \rangle}
			\end{split}
		\end{equation}
	After inserting (\ref{eq:re_c_spa}), we have
	\begin{equation}\label{eq:pa_etaxi}
		g_{ve} \eta \xi^{\pm}_{ef} = \frac{\bar{\zeta}_{vef}}{\bar{\zeta}_{ve'f}} g_{ve'} \eta \xi^{\pm}_{e'f}
	\end{equation}
	One can check that the variation respects to $\text{SU}(1,1)$ group elements $v_{ef}$ is trivial. The variation respects to $\text{SL}(2,\C)$ group elements $g_{ve}$ leads to
	\begin{equation}
	\begin{split}
		\delta S &= \sum_{f+} j_{f} (1+\ui \gamma) \frac{\langle L^{\dagger} Z_{vef},Z_{vef} \rangle+\langle  Z_{vef}, L^{\dagger} Z_{vef} \rangle}{\langle Z_{vef},Z_{vef} \rangle} -2j_{f} \frac{\langle \xi^{\pm}_{ef}, L^{\dagger} Z_{vef} \rangle}{\langle \xi^{\pm}_{ef},Z_{vef} \rangle}\\
		 &\sum_{f-} j_f (1-\ui \gamma) \frac{\langle L^{\dagger} Z_{vef},Z_{vef} \rangle+\langle  Z_{vef}, L^{\dagger} Z_{vef} \rangle}{\langle Z_{vef},Z_{vef} \rangle} -2j_f \frac{ \langle L^{\dagger} Z_{vef},\xi^{\pm}_{ef} \rangle}{\langle Z_{vef},\xi^{\pm}_{ef} \rangle}
	\end{split}
	\end{equation}
	Applying (\ref{eq:re_c_spa}), we have
	\begin{equation}
	\begin{split}
		\delta S &= \sum_{f+} j_f (1+\ui \gamma) \left( {\langle L^{\dagger} \xi^{\pm}_{vef},\xi^{\pm}_{vef} \rangle+\langle  \xi^{\pm}_{vef}, L^{\dagger} \xi^{\pm}_{vef} \rangle} \right)-2j_f {\langle \xi^{\pm}_{ef}, L^{\dagger} \xi^{\pm}_{ef} \rangle}+\\
		 &\sum_{f-} j_f (1-\ui \gamma) \left( {\langle L^{\dagger} \xi^{\pm}_{vef},\xi^{\pm}_{vef} \rangle+\langle  \xi^{\pm}_{vef}, L^{\dagger} \xi^{\pm}_{vef} \rangle} \right)-2j_f {\langle L^{\dagger} \xi^{\pm}_{ef},  \xi^{\pm}_{ef} \rangle}\\
	\end{split}
	\end{equation}
	where $f \pm$ means face $f$ is incoming or outing edge $e$ correspondingly.
	This leads to six equations with the generators of $\text{SL}(2, \C)$ group, which reads
	\begin{align}
		\label{eq:sp_cl_1}&\delta S =  -2 \sum_{f} \epsilon_{ef}(v) j_f{\langle \xi^{\pm}_{ef}, F^{\dagger} \xi^{\pm}_{ef} \rangle}\\
		\label{eq:sp_cl_2}&\delta S = 2 \ui \gamma \sum_{f} \epsilon_{ef}(v) j_f{\langle \xi^{\pm}_{ef}, \tilde{F}^{\dagger} \xi^{\pm}_{ef} \rangle}
	\end{align}
	Again $\epsilon_{ef}(v)$ here is the signature determined up to a global sign by
	\begin{equation}
		\epsilon_{ef}(v) = -\epsilon_{e'f}(v), \epsilon_{ef}(v) = -\epsilon_{ef}(v')
	\end{equation}
	for the triangle $f$ shared by the tetrahedra $t_e$ and $t_{e'}$.
	
	\subsubsection{geometrical interpretation}
	We can define a vector from $\xi_{ef}$
	\begin{equation}
		n^{i}_{ef}=-2 \ui \langle \xi_{ef}^{\pm}, F^{i} \xi_{ef}^{\pm} \rangle
	\end{equation}
	which is the $\text{SU}(1,1)$ action on the unit time-like vector $n_0= -2 \ui \langle \xi_{0}^{\pm}, F^{i} \xi_{0}^{\pm} \rangle=\{\pm 1,0,0\}$.
	The encoding of this vector in four dimensional Minkowski space is given by
	\begin{equation}
		n^{I}_{ef}=\{ n^3_{ef}, -n^2_{ef}, n^1_{ef} ,0 \}= \bra{\xi^{\pm}_{ef}} \sigma^{I} \ket{\xi^{\pm}_{ef}}-\bra{\xi^{\pm}_{ef}}\ket{\xi^{\pm}_{ef}}
	\end{equation}
	Clearly $n^{I}_{ef}$ is timelike vector and future directed with $\zeta^{+}_{ef}$ while past directed with $\zeta^{-}_{ef}$.
	
	Then there is a nature $\text{SL}(2,\C)$ bivector defined by
	\begin{equation}\label{eq:bi_sp}
		X_{ef}=- 2 \ui \langle \xi^{\pm}_{ef}, F^{i} \xi^{\pm}_{ef} \rangle E_i=- \ui( \eta \xi^{\pm}_{ef} \otimes (\xi^{\pm}_{ef})^{\dagger} - \frac{1}{2} I_2 )
	\end{equation}
	which in spin-1 representation reads
	\begin{equation}
			X_{ef}^{IJ}=\left( \begin{array}{llll} 0 & n^1 & n^2 & 0\\ -n^1 & 0 & n^3 & 0\\ -n^2 & -n^3 & 0 & 0\\0 & 0 & 0 & 0 \end{array} \right) = * (n^{I}_{ef} \wedge u^{I})
	\end{equation}
	Clearly from (\ref{eq:pa_jxi}) and (\ref{eq:pa_etaxi}), $X_{ef}$ satisfy the parallel transport equation
	\begin{equation}\label{eq:para_trans_sp}
		X_{f}(v) = g_{ve} X_{ef} g_{ev}=g_{ve'} X_{e'f} g_{e'v}
	\end{equation}
	and satisfies
	\begin{equation}\label{eq:sim_sp}
		(G_{ve} u) . X_{f}(v) =0
	\end{equation}
	The bivector is then again scaled as $B_f(v) = 2 A_f X_f(v) = 2 \gamma j_f X_f(v)$, where $|B_f| = 2 A_f $.
	The equation (\ref{eq:sp_cl_1}) and (\ref{eq:sp_cl_2}) then can be written as equations of $B_f$:
	\begin{align}
		&\delta_{g} S =  \frac{\ui}{2\gamma} \sum_{f} \epsilon_{ef}(v) B_f(v) =0\\
		&\delta_{\tilde{g}} S = \frac{1}{2} \ui \sum_{f} \epsilon_{ef}(v) B_f(v)=0
	\end{align}
	
	\section{Geometric interpretation and reconstruction}\label{app:geo}
	In this appendix we summarize the geometric reconstruction theorems for tetrahedron with spacelike triangles only in {\cite{Barrett:2009mw, Barrett:2009gg,Han:2011re, Han:2011rf,Kaminski:2017eew}}, and extend them to general tetrahedron may contains also timelike triangles. We start with a single simplex $\sigma_v$ corresponding to a vertex $v$, and then generalize the result to general simplicial manifold with many simplices. For simplicity, we introduce a short hand notation for a single simplex $\sigma_v$:
	\begin{equation}\label{eq:e1}
		\begin{split}
			N_i := N_{e_i}(v) \qquad B_{ij}^{G} = -B_{ji}^{G} = \epsilon_{e_i e_j}(v) B_{e_i e_j}(v) \qquad B_{ij}^{G} = * (v_{ij}^{G} \wedge N_i)
		\end{split}
	\end{equation}
	where $e_ie_j$ represents the face determined by the dual edge $e_i$ and $e_j$, and $i=0,1,...,4$, and $v_{ij}$ here is the trianlges normal scaled with the area : $v_{ij}^2 = \pm 4 A_{ij}^2$. 
	
	Note that here we will assume our boundary data to be a geometric boundary data, which means they satisfy length matching condition and orientation matching condition. The detailed meaning of these conditions will become clear later. The geometric boundary data is necessary to get a Regge like geometric solution. For non-geometric boundary data, there will be at most one solution up to gauge equivalence, which is an analogy to the result in EPRL model \cite{Barrett:2009gg,Barrett:2009mw}. 
	
	\subsection{non-degenerate condtion and classification of the solution}
	To begin with, we would like to introduce the non-degenerate condition. We will first consider non-degenerate simplicies and then move to degenerate case. For the boundary data, non-degenerate means for a boundary tetrahedron any $3$ out of $4$ face normal vectors $n_{ef}$ span a 3-dimensional space. With non-degenerate boundary data, for any $3$ different edges $i,j,k$ in a $4$ simplex one of the following holds
	\begin{itemize}
	\item $N_{ei}= \pm N_{ej}$ and $N_{ej}= \pm N_{ek}$ ,
	\item $N_{ei} \neq N_{ej}$
	\end{itemize}
	The first case can be further proved that leads to all $N_i$ are parallel by using the closure constraint of $B_{ij}$. This result was first proved in \cite{Barrett:2009gg} and later by \cite{Kaminski:2017eew}.
	
	The only non-degenerate case is then specify by the following non-degeneracy condition
	\begin{equation}
		\prod_{e1,e2,e3,e4=0}^5 \det(N_{e1},N_{e2},N_{e3},N_{e4}) \neq 0
	\end{equation}
	which means any $4$ out of $5$ normals are linear independent and span a $4$ dimensional Minkowski space. Since $N_{e}(v)=g_{ve} N^0$, it is easy to see the non-degenerate condition is actually a constraint on $\{ g_{ve} \}$.
	
	\subsection{Nondegenerate geometry on a 4-simplex}
	For simplicity, we start with one 4-simplex $\sigma_v$ in $4$ dimensional Minkowski space $M=R^4$ here.
	For each 4-simplex $\sigma_{v}$ dual to the vertex $v$, we associate it with a reference frame. In this reference frame, the $5$ vertices of the 4-simplex $[p_0, p_1, p_2, p_3, p_4 ]$ have the coordinates
		$p_i:(x_i^I) = (x_i^0, x_i^1,x_i^2,x_i^3)$.
	Based on these coordinates, we introduce vectors $y_i$, $a$ as well as covector $A$ in an auxiliary space $R^5$,
	\begin{equation}
		y_i = (x_i^I, 1)^{T}, \qquad \text{and} \quad a=(0, ... ,0 , 1)^{T}, \quad A=a^{T}
	\end{equation}
	We define the $k+1$ vector in $R^5$
	\begin{equation}
		\tilde{V}_{\alpha_0, ..., \alpha_k} = y_{\alpha_0} \wedge ... \wedge y_{\alpha_k}
	\end{equation}
	where $\alpha_i \in \{ 0, \cdots, 5\}$.
	With covector $A$, for $k$-vectors $\Omega$ in $R^5$ satisfying $A  \llcorner \Omega =0$, we can identify it with a $k$-vector in $M$. For example, since $A \llcorner A \llcorner \tilde{V}_{\alpha_0, ..., \alpha_5} =0$, we then induce a $4$-vector in $M$ from $\tilde{V}_{\alpha_0, ..., \alpha_5}$,
	\begin{equation}
		{V}_{\alpha_0, ..., \alpha_5}= A \llcorner \tilde{V}_{\alpha_0, ..., \alpha_k}=(y_{\alpha_1}-y_{\alpha_0}) \wedge ... \wedge (y_{\alpha_5}-y_{\alpha_0})
	\end{equation}
	This vector is actually  $4!$ times the volume $4$-vector of $4$-simplex:
	\begin{equation}\label{eq:_vol_4v}
		 {V}_{\alpha_0, ..., \alpha_4}=(x_{\alpha_1}-x_{\alpha_0}) \wedge ... \wedge (x_{\alpha_4}-x_{\alpha_0}) = E_{\alpha_1 \alpha_0} \wedge ... \wedge E_{\alpha_5 \alpha_0}
	\end{equation} 
	$E_{\alpha_i \alpha_0}^{I}=x_{\alpha_i}^I-x_{\alpha_0}^I$ is the edge vector related to the oriented edge $l_{\alpha_i \alpha_0}=[p_{\alpha_i},p_{\alpha_0} ]$. Notice that the volume $4$-vector comes with a sign respecting to the order of points.

	We further define $3$-vector and bivector by skipping some points
	\begin{eqnarray}
		&V_{i} = (-1)^i V_{0... \hat{i} ... 4}\\
		&B_{ij}= A \llcorner \tilde{V}_{0... \hat{i} ... n} = \left\{ \begin{array}{l} (-1)^{i+j+1} V_{0... \hat{i} ... \hat{j} ... 4} \quad i <j \\(-1)^{i+j} V_{0... \hat{j} ... \hat{i} ... 4} \quad i > j \end{array} \right.
	\end{eqnarray}
	where $\hat{i}$ means omitting $i_{th}$ elements.
	We have the following properties for $V_i$ and $B_{ij}$
	\begin{align}
		&\sum_{i} V_i =0, \\
		&B_{ij}=-B_{ij}m \qquad \forall_{i} \sum_{j \neq i} B_{ij}=0,
	\end{align}
	One can further check that $B_{ij}$ can be written as
	\begin{equation}\label{eq:b_ee}
		B_{ij} = \frac{1}{2} (-1)^{\text{sgn}(\sigma)} \epsilon^{ijkmn} E_{mk} \wedge E_{nk}
	\end{equation}
	And one has $B_{ij}^2 = \pm 4 A_{ij}^2$ with $A_{ij}$ is the area of the corresponding spacelike or timelike triangles in non-degenerate case.
	
	Suppose the volume $4$-vector of $4$-simplex $V_{0,...,4}$ is non-degenerate. In this case any 4 out of 5 $y_{i}$ are linearly independent. One can introduce a dual basis $\hat{y}_i$ and $\tilde{y}_i$ defined by
	\begin{equation}
		\hat{y}_i \lrcorner y_j = \delta_{ij}, \qquad \hat{y}_i=\tilde{y}_i + \mu_i A, \quad \tilde{y}_i \lrcorner a=0
	\end{equation}
	with properties
	\begin{equation}
		\sum_{i} \hat{y}_i =A, \qquad \sum_{i} \tilde{y}_i =0
	\end{equation}
	$\tilde{y}_i$ here can be regarded as covectors belong to $M$. With $\tilde{y}_i$, we have
	\begin{equation}
		V_i=-\tilde{y}_i \lrcorner V_{0...4}, \quad B_{ij}= \tilde{y}_j \lrcorner \tilde{y}_i \lrcorner V_{0...4}
	\end{equation}
	Thus covectors $\tilde{y}_i$ are conormal to subsimplices $V_i$. And by using Hodge star, we have
	\begin{equation}
		V_i= -Vol * \tilde{y}_i, \qquad B_{ij} = -Vol * (\tilde{y}_j \wedge \tilde{y}_i )
	\end{equation}
	where the volume $Vol >0$ is the absolute value of the oriented $4$-volume
	\begin{equation}
		V_4 : = \det(V_{0,...,4}) = \text{sgn}(V_4) Vol
	\end{equation}
	It can be shown that
	\begin{equation}
		\frac{1}{V_4} = \epsilon^{ijkl} \det(\tilde{y}_i, \tilde{y}_j,\tilde{y}_k,\tilde{y}_l ) 
	\end{equation}
	and the co-frame vector $E_{ij}$ is given by
	\begin{equation}
			E_{ij} = V_4 \epsilon_{ijklm}(v) * (\tilde{y}^k \wedge \tilde{y}^l \wedge \tilde{y}^m)
	\end{equation}
	
	If the subsimplices $V_i$ are non-degenerate, by introducing normalized vectors $N_i$, we can write $\tilde{y}_i$ as
	\begin{equation}
		\tilde{y}_i = \frac{1}{Vol} W_i N_i,  \quad N_i \cdot N_i = t_i, \; W_i>0
	\end{equation}
	where $t_i=\pm 1$ distinguish spacelike or timelike normals respectively. This leads to
	\begin{equation}\label{eq:e22}
		B_{ij}= - \frac{1}{Vol} W_i W_j *( N_j \wedge N_i), \qquad \sum_i W_i N_i =0
	\end{equation}
	In order to make the normal out-pointing, we redefine the normalized normal vectors $N_i$ by
	\begin{equation}
		N^{\Delta}_i = -t_i N_i, \quad W^{\Delta}_i = - t_i W_i \qquad \sum_i W_i^{\Delta} N_i^{\Delta} =0
	\end{equation}
	such that $N^{\Delta}_i$ are out-pointing.
	
	\subsection{Reconstruct geometry from non-degenerate critical points}
	We begin with the reconstruction of normals. Recall in critical point equations (\ref{eq:cri_eq_final_mix}), normals $N_e$ satisfying
	\begin{equation}
		 \forall_{f \in t_e } \eta_{IJ} {N_e}^{I} B_{f}(v)^{JK} = 0 
	\end{equation} 
	If there is another normal vector $N$ satisfy the same condition for some edge $e$, easy to see we have
	\begin{equation}
		\forall_{f \in t_e } \;\; B_{f}(v) \sim * (N \wedge N_{e} )
	\end{equation}
	which means for an edge $e$, $B_{ef}$ are proportional to each other. This clearly contrary to the fact that we have a non degenerate solution. Thus, for given bivectors which are the solution of the critical point equation, if we require a vector $N$ satisfies
	\begin{equation}\label{eq:e27}
		\forall_{f \in t_e } \eta_{IJ} {N}^{I} B_{f}(v)^{JK} = 0 
	\end{equation}
	for a edge tetrahedron $t_e$, we then have $N=\pm N_e$ after normalization. The condition (\ref{eq:e27}) is sufficient and necessary.
	
	Considering a $4$-simplex $\sigma_v$ at some vertex $v$, the critical point equation (\ref{eq:cri_eq_final_mix}) can be written in short hand notation we introducing in (\ref{eq:e1}) as
	\begin{equation}\label{eq:e28}
		B_{f}(v) = B_{ij}^{\{G\}} = - B_{ji}^{G}, \quad N_i \llcorner B_{ij}^{\{G\}}= 0, \qquad \sum_{j} B^{\{G\}}_{ij} = 0
	\end{equation}
	Now we give normalized vectors $N_i$ satisfying non-degenerate condition. If we require the bivectors satisfy (\ref{eq:e28}), they are uniquely determined up to a constant  $\lambda \in R$
	\begin{align}\label{eq:e29}
		{B'}_{i j}= \lambda W_i W_j * (N_j \wedge N_i) 
	\end{align}
	Here $W_i \in R$ are non zero and determined by
	\begin{equation}
		\sum_{i} W_i N_i=0
	\end{equation}
	The proof is stated first in \cite{Han:2011re} and later \cite{Kaminski:2017eew}.
	Note that the bivector $B_{ij}$ is independent of the choice of signature of normal vectors $N$ since the sign of $W$ and $N$ will change simultaneously. $\lambda$ can be fixed up to a sign by the normalization of ${B'}_{ij}$
	\begin{equation}
		|B_f|^2 = - 4 \gamma^2 s_f^2 = - 4 A_f^2
	\end{equation}
	
	Then it can be proved that non-degenerate geometric solution determines $4$ simplex specified by bivectors $B^{\Delta}$ uniquely up to shift and inversion such that
	\begin{equation}
		B^{\Delta}_{ij} = r B_{ij}^{\{G\}}
	\end{equation}
	where $r=\pm 1$ is the geometric Plebanski orientation. The construction can be done as follows. With given $5$ normals $N_i$, we take any $5$ planes orthogonal to $N_i$. With the non-degeneracy condition, they cut out a $4$ simplex $\Delta'$ which is uniquely determined up to shifts and scaling. According to (\ref{eq:e22}) and (\ref{eq:e29}), bivectors of the reconstructed $4$ simplex $B^{\Delta'}_{ij}$ related to $B_{ij}$ as
	\begin{equation}
		 B^{\Delta'}_{ij} = \lambda B_{ij}^{\{G\}}
	\end{equation} 
	Then the identity of the normalization will determines the scaling up to a sign
	\begin{equation}
		B_{ij}^{\{G\}}= r B^{\Delta'}_{ij} = -\frac{1}{Vol} r W_i^{\Delta} W_j^{\Delta} * (N_j^{\Delta} \wedge N_i^{\Delta} )
	\end{equation}
	where $Vol$ is the $4!$ volume of the 4-simplex. 
	
	Let us move to the boundary tetrahedron. Since $G_{e}$ is a $\text{SO}(1, 3)$ rotation, it action then keeps the shape of tetrahedrons. Thus the tetrahedron with bivectors $B_{ij} = *(v_{ij} \wedge u_i)$ has the same shape with the tetrahedron with face bivectors $B_{ij}^{\{G\}} = G_i * (v_{ij} \wedge u_i )$. For given $v_{ij} $,
	when the boundary data is non-degenerate, we can cut out a tetrahedron with planes perpendicular to $v_{ij}$ in the 3 dimensional Minkowski space orthogonal to $u$. Clearly, the face bivectors of this tetrahedron satisfy
	\begin{equation}
		B_{ij} = \lambda'_{ij} * (v_{ij} \wedge u)
	\end{equation}
	with $\lambda'_{ij}$ arbitrary real number. However, from the closure constraint, we have
	\begin{equation}
		\sum_{j:j \neq i} B'_{ij} = * ( \sum_{j: j \neq i } \lambda'_{ij} v_{ij} ) \wedge u =0
	\end{equation}
	Since $\forall_j \; v_{ij}.u =0$, the above closure equation implies
	\begin{equation}
		 \sum_{j: j \neq i } \lambda'_{ij} v_{ij} =0 
	 \end{equation}
	 which according closure with $v_{ij}$ leads to
	 \begin{equation}
		\exists_{\lambda} : \lambda_{ij}^{'} = \lambda
	 \end{equation}
	Thus, for every edge $e_i$, there exists a tetrahedron determined uniquely up to inversion and translation with face bivectors 
		\begin{equation}
			B_{ij} = r_{i} (v_{ij} \wedge u) 
		\end{equation}
		in the subspace perpendicular to $N_i$ with $r_{i} = \pm 1$.

	 The edge lengths of the tetrahedron is then determined uniquely by $v_{ij}$. We denote ${l^{i}_{jk}}^2$ the signed square lengths of the edge between faces $ij$ and $ik$. The length matching condition can be expressed as
	 \begin{equation}
		l_{(ijk)}^2:= {l^{i}_{jk}}^2 = {l^{j}_{ik}}^2 ={l^{k}_{ij}}^2
	 \end{equation}
	 The non-degenerate solution exists if and only if the lengths satisfy length matching condition. In case when length matching condition is satisfied, we can write $l_{(ijk)}^2$ using the missing indices different from $i,j,k$ as $l_{(ml)}^2$, with this notation, one introduce lengths Gram matrix of the $4$ simplex
	 \begin{equation}
	 G^l = \left( \begin{array}{lllll} 0& 1& 1 & \cdots & 1 \\ 1 & 0 & l_{01}^2 & \cdots & l_{04}^2 \\ 1 &  l_{10}^2 & 0 & \cdots & l_{24}^2 \\ \vdots & \vdots & \vdots & \ddots & \vdots \\ 1 & l_{40}^2 & l_{41}^2 & \cdots & 0 \\   \end{array} \right)
	 \end{equation}
	 The signature of $G^l$ corresponds to the signature of reconstructed $4$ simplex. We denote the signature as $(p,q)$.
	Based on $G^l$ is degenerate or not, we have 
	\begin{itemize}
		\item If $G^l$ is non degenerate, then there exist a unique up to rotation, shift and reflection non degenerate $4$ simplex with signature $(p,q)$. There are two non-equivalent $4$ simplex up to rotations and shift. The normals of two reconstructed $4$ simplicies $\{N_i\}$ and $\{N'_i\}$ are related by
		\begin{equation}
			N'_i = (-1)^{s_i} G N_i = G I^{s_i} N_i
		\end{equation}
		\item If $G^l$ is degenerate, then there exist a unique up to rotation and shift degenerate $4$ simplex with signature $(p,q)$. The $4$ volume in this case is $0$.
	\end{itemize}

	The signature here is related to the signature of boundary tetrahedron. For all boundary tetrahedra being timelike, the possible signatures are Lorentzian $(- +++)$, split $(-++-)$ or degenerate $(-++0)$. For all boundary tetrahedra being spacelike, the possible signatures are Lorentzian $(- +++)$, Euclidean $(++++)$ or degenerate $(0+++)$. For boundary data contains both spacelike and timelike tetrahedra, the only possible reconstructed $4$ simplex is in Lorentzian signature $(-+++)$.
	
	\subsection{Gauge equivalent class of solutions}
	Suppose we have a non-degenerate geometric boundary data and the $4$ volume is non-degenerate, then we can reconstruct geometric non-degenerate $4$-simplex up to orthogonal transformations. Suppose we have this reconstructed $4$-simplex with geometric bivectors $B^{\Delta}_{ij}$ with normals $N^{\Delta}_i$. From these normals, we can introduce
	\begin{equation}\label{eq:v_delta}
		v_{ij}^{\Delta} =  - \frac{1}{Vol} \left( W_i^{\Delta} W_j^{\Delta} N_j^{\Delta} - \frac{ W_i^{\Delta} W_j^{\Delta} N_i^{\Delta} \cdot N_j^{\Delta} }{(N_i^{\Delta})^2} N_i^{\Delta} \right)
	\end{equation}
	Easy to check that $v_{ij}^{\Delta} \cdot N_i^{\Delta} = 0 $ and $B_{ij}^{\Delta} = * (v_{ij}^{\Delta} \wedge N_i^{\Delta}) $. Thus these are nothing else but normals of faces of the $i$th tetrahedron recovered from bivectors $B_{ij}^{\Delta}$. Easy to check that we have
	\begin{equation}
		v_{ij}^{\Delta} \cdot v_{ik}^{\Delta} = v_{ij} \cdot v_{ik}
	\end{equation}
	by the fact that $B_{ij}^{\Delta} \cdot B_{ik}^{\Delta}= B_{ij} \cdot B_{ik}$. We can introduce group elements $G^{\Delta}_i \in O$ for each $i$ satisfy 
	\begin{equation}
		G_i^{\Delta} u = N_i^{\Delta}, \qquad \forall_{j: j \neq i} \; G_i^{\Delta} v_{ij} = v_{ij}^{\Delta}
	\end{equation}
	Note that there are only 4 independent conditions out of 5.
	
	We would like compare these group elements $G_i^{\Delta}$ obtained from $B_{ij}^{\Delta}$ with $G_i$ from critical point solution. From reconstruction of bivectors and normals, we know that
	\begin{equation}\label{eq:E38}
		B_{ij}^{\Delta} = (-1)^s B_{ij}^{\{G\}}, \qquad N_i = (-1)^{s_i} N_i^{\Delta}
	\end{equation}
	where $(-1)^s \; \text{with} \; s \in \{0,1 \} $ and $s_i \in \{0,1 \}$.
	The condition leads to
	\begin{equation}
		\begin{split}
			&* ( G_i v_{ij} \wedge N_i) = B_{ij}^{\{G\}} = (-1)^s B_{ij}^{\Delta}\\
			& = (-1)^s  * ( v_{ij}^{\Delta} \wedge N_i^{\Delta} ) = * ((-1)^{s+s_i} v_{ij}^{\Delta} \wedge N_i )
		\end{split}
	\end{equation}
	Since $N_i \cdot v_{ij}^{\Delta} = N_i \cdot G_i v_{ij} =0 $, we have
	\begin{equation}
		G_i v_{ij} = (-1)^{s+s_i} v_{ij}^{\Delta}, \qquad G_i N = (-1)^{s_i} N_i^{\Delta}
	\end{equation}
	which implies
	\begin{equation}\label{eq:g_gdelta}
		G_i = G_i^{\Delta} I^{s_i} (I R_{N})^s
	\end{equation}
	
	For $G_i \in \text{SO}$, we have $\det G_i =1$, then from (\ref{eq:g_gdelta})
	\begin{equation}\label{eq:gdelta_det}
		\det G_i^{\Delta} = (-1)^s
	\end{equation}
	Since there is only one reconstructed 4 simplex up to rotations from $O$, thus two $G^{\Delta}$ solutions are related by
	\begin{equation}
		 G_{i}^{\Delta'} = G G_{i}^{\Delta}, \quad G \in O
	\end{equation}
	which means
	\begin{equation}
		\forall_i \frac{\det G_{i}^{\Delta'}}{\det G_{i}^{\Delta}} = \det G
	\end{equation}
	This condition reminds us to introduce an orientation matching condition for boundary data where the reconstructed 4 simplex have
	\begin{equation}
		\forall_{i} \; \det G_i^{\Delta} = r \qquad r \in \{-1, 1 \}
	\end{equation}
	We call the boundary data as the geometric boundary data if it satisfy the length matching condition and orientation matching condition.

	After we choose reconstructed 4 simplex, we have fixed the value of $s$ by
	\begin{equation}
		r = (-1)^s
	\end{equation}
	and it is Plebanski orientation. However $s_i$ is still arbitrary.
	
	With (\ref{eq:g_gdelta}) and (\ref{eq:gdelta_det}), we can identify the geometric solution and reconstructed 4-simplicies. Up to SO rotations, there are two reconstructed 4 simplices. The two classes of simplicies solutions are related by reflection respect to any normalization 4 vector $e_{\alpha}$
	\begin{equation}
		B_{ij}^{\tilde{G}} = R_{e_{\alpha}}(B_{ij}^{\{G\}}), \qquad s' =s +1
	\end{equation}
	which means
	\begin{equation}\label{eq:E53}
		\tilde{G}_i = R_{e_{\alpha}} G_i (I R_{u}) \in \text{SO}(1,3)
	\end{equation}
	With the gauge choice that $G_i \in \text{SO}_{+}(1,3)$, we can rewrite (\ref{eq:E53}) as
	\begin{equation}
		\tilde{G}_i = R_{e_{0}} I^{r_i} G_i R_{u} 
	\end{equation}
	such that $\tilde{G}_i \in \text{SO}_{+}(1,3)$. It is direct to see $r_i = 0$ for $u$ timelike and $r_i=1$ for $u$ spacelike.
	
	\subsection{Simplicial manifold with many simplicies}
	The above interpretation and reconstruction are with in single $4$-simplex case. Now we will generalize the result to simplicial manifold with many simplicies. We will consider two neighboring $4$ simplicies where there corresponding center $v$ and $v'$ are connected by a dual edge $e=(v,v')$. For a short hand notation, we will use prime to represent the parallel transported bivector and normals from simplex with center $v'$ to $v$, e.g. $N'_i = G_{vv'}N_i(v')$. We denote the edge $e=(v,v')$ as $e_0$.
	
	Since $N_e(v) = G_{ve}u$ and $N_{e}(v') = G_{v'e} u$, we have $N_e(v) = G_{vv'} N_{e}(v')$ for $G=(v,v')$. From the reconstruction theorem, with (\ref{eq:E38}), we have
	\begin{equation}\label{eq:f55}
		N_{0}^{\Delta} = (-1)^{s_0 + {s'}_0} {N'}_{0}^{\Delta}
	\end{equation}
	From the parallel transport equation $X_f(v) = g_{vv'} X_{f}(v') g_{v'v}$, with the fact $\epsilon_{ef}(v)=-\epsilon_{ef}(v')$,
	we have
	\begin{equation}\label{eq:e50}
		 B_{0i}^{\{ G \}} = - r(v) \frac{1}{Vol} W_i^{\Delta} W_0^{\Delta} *  ( N_{i}^{\Delta} \wedge N_{0}^{\Delta}) =  r(v') \frac{1}{Vol'} {W'}_i^{\Delta} {W'}_0^{\Delta} *  ( {N'}_{i}^{\Delta} \wedge {N'}_{0}^{\Delta})
	\end{equation}
	where $B_{0i}^{\Delta}$ is the geometric bivector corresponding to the triangle $f$ dual to face determined by $e,e_i,e'_i$.
	Now similar to (\ref{eq:v_delta}), we can define
	\begin{equation}
		{v}_{0i}^{\Delta}(v) =  - \frac{1}{Vol} \left( W_0^{\Delta}(v) W_i^{\Delta}(v) N_i^{\Delta}(v) - \frac{ W_0^{\Delta}(v) W_i^{\Delta}(v) N_0^{\Delta}(v) \cdot N_i^{\Delta}(v) }{(N_0^{\Delta}(v))^2} N_0^{\Delta}(v) \right)
	\end{equation}
	which satisfies $v_{0i}^{\Delta}(v) \cdot N_{0}^{\Delta}(v) =0$.
	The geometrical group elements $\Omega_{vv'}^{\Delta} \in O(1,3)$ is defined from
	\begin{equation}\label{eq:f58}
		v^{\Delta}_{0i}(v) = \Omega_{v v'}^{\Delta} v^{\Delta}_{0i}(v'), \quad N^{\Delta}_{0}(v) = \Omega_{v v'}^{\Delta} N^{\Delta}_{0}(v')
	\end{equation}
	(\ref{eq:e50}) now reads 
	\begin{equation}\label{eq:f59}
		B_{0i}^{\{ G \}} = r(v) * (v^{\Delta}_{0i}(v) \wedge N^{\Delta}_{0}(v)) = -r(v') *( G_{vv'} v^{\Delta}_{0i}(v') \wedge G_{vv'} N^{\Delta}_{0}(v'))
	\end{equation}
	From (\ref{eq:f55}) and (\ref{eq:f59}), with the fact that, $v_{0i}^{\Delta}(v) \cdot N_{0}^{\Delta}(v) =  G_{vv'} v_{0i}^{\Delta}(v') \cdot G_{vv'} N_{0}^{\Delta}(v') = 0$, we have
	\begin{equation}
		v^{\Delta}_{0i}(v) = - (-1)^{s_0+s'_0} r(v) r(v') G_{v v'} v_{0i}^{\Delta}(v'), \quad N_{0}^{\Delta}(v) = (-1)^{s_0+s'_0} G_{vv'} N_{0}^{\Delta}(v')
	\end{equation}
	Compare with (\ref{eq:f58}),
	\begin{equation}\label{eq:f61}
		\Omega_{v v'}^{\Delta} = G_{vv'} I I^{s_0+s'_0} (I R_{N_0(v')})^{s+s'}, \qquad \det \Omega_{v v'}^{\Delta} = (-1)^{s+s'}
	\end{equation}
	where $s$ and $s'$ is determined by $(-1)^{s} = r(v)$ and $(-1)^{s'} = r(v')$.
	Note that, from the fact $N_0(v') = G_0(v') u = I^{s'_0} N_{0}^{\Delta}(v')$, and $R_{N} = G R_{u} G^{-1}$, we have $R_{N_0^{\Delta}}  = R_{N_0}$. One can check that the (\ref{eq:f61}) can be written as
	\begin{equation}
		\Omega_{v v'}^{\Delta} = I I^{s_0+s'_0} I^{s+s'} G_{ve} R_{u}^{s+s'} G_{ev'} = I G_{ve}^{\Delta} G_{ev'}^{\Delta}
	\end{equation}
	which coincide with the geometric solution for single simplex. Note that, after fixing a pair of compatible values of $s$ and $s'$, another pair of compatible values are given by $s+1$ and $s'+1$ due to the common tetrahedron $t_e$ shared by two $4$ simplices. This is nothing else but reflecting simtounesly every $4$ simplex connects with each other. Then according to (\ref{eq:E53}), these two possible non gauge equivalent solutions are related by
	\begin{equation}\label{eq:g_f_tilde_o_1}
		\tilde{G}_f = \left\{\begin{array}{l} R_{u_e} G_f(e) R_{u_e} \quad \text{internal faces} \\ I^{r_{e1}+r_{e0}} R_{u_{e1}} G_f(e_1,e_0) R_{u_{e0}} \quad \text{boundary faces} \end{array} \right.
	\end{equation} 
	where $G_f=  \prod_{v \subset \partial f} G_{e'v}G_{ve}$ is the face holonomy.
	
	For a simplicial manifold, we will introduce the consistent orientation. For two $4$ simplex  $\sigma_v$ and $\sigma_{v'}$  share a same tetrahedron $t_e$, we say they are consistently oriented if their orientation satisfies $[p_0,p_1,p_2,p_3,p_4]$ and $-[p_0,p_1,p_2,p_3,p_4]$. Therefore we have $\epsilon^{01234}(v)= - \epsilon^{01234}(v')$ for the orientation in (\ref{eq:b_ee}). The orientated volume then contains a minus sign in $V'$. 
	
	From (\ref{eq:f55}) and (\ref{eq:e50}), we have
	\begin{equation}\label{eq:n'_n-n0}
		{N'}_{i}^{\Delta} = - (-1)^{s_0 + {s'}_0} r(v) r(v') \frac{W_i^{\Delta} W_0^{\Delta} Vol'}{{W'}_i^{\Delta} {W'}_0^{\Delta} Vol} N_{i}^{\Delta} + a_i N_{0}^{\Delta}
	\end{equation}
	where $a_i$ are some coefficients s.t. $\sum_i {W'}_i^{\Delta} {N'}_{i}^{\Delta} = -{W'}_0^{\Delta} {N'}_{0}^{\Delta}$. We introduce $\tilde{y}$ where $\tilde{y}_i = \frac{1}{Vol}W_{i}^{\Delta} N_{i}^{\Delta}$, then
	\begin{equation}
		B_{0i}^{G} = -r(v) Vol * (\tilde{y}_i \wedge \tilde{y}_0), \quad 	{\tilde{y}'}_{i} = - (-1)^{s_0 + {s'}_0} r(v) r(v') \frac{W_0^{\Delta} }{ {W'}_0^{\Delta}} \tilde{y}_{i} + \tilde{a}_i \tilde{y}_{0}
	\end{equation}
	where $\tilde{a}_i$ are coefficients s.t. $\sum_i \tilde{y}_i = - \tilde{y}_0$. We then have
	\begin{equation}
		-\frac{1}{V'}=\det({\tilde{y'}}_0,{\tilde{y'}}_1, {\tilde{y'}}_2 , {\tilde{y'}}_3) = (- r(v) r(v'))^3 \left( \frac{W_0^{\Delta}}{ {W'}_0^{\Delta}} \right)^2 \frac{Vol}{Vol'} \det({\tilde{y}}_0,{\tilde{y}}_1, {\tilde{y}}_2 , {\tilde{y}}_3) = -\tilde{r}(v) \tilde{r}(v') \left( \frac{W_0^{\Delta}}{ {W'}_0^{\Delta}} \right)^2 \frac{1}{V'}
	\end{equation}
	where we define $\tilde{r}(v) =r(v) \text{sgn}(V(v))$. The equation results in $\tilde{r}(v) =\tilde{r}(v') = \tilde{r}$. Therefore $\tilde{r}= \text{sgn} (V(v)) r(v)$ is a global sign on the entire triangulation after we choose compatible orientation. The equation also implies $|W_0^{\Delta}| = |{W'}_0^{\Delta}|$. With the fact that normal vector $N_0^{\Delta}$ and ${N'}_0^{\Delta}$ are in the same type (spacelike or timelike), we have $W_0^{\Delta} = {W'}_0^{\Delta}$ . Thus (\ref{eq:n'_n-n0}) leads to
	\begin{equation}\label{eq:f68}
		{N'}_{i}^{\Delta} = - (-1)^{s_0 + {s'}_0} \text{sgn}(V V') \frac{W_i^{\Delta} W_0^{\Delta} Vol'}{{W'}_i^{\Delta} {W'}_0^{\Delta} Vol} N_{i}^{\Delta} + a_i N_{0}^{\Delta} = \mu_e N_{i}^{\Delta} + a_i N_0^{\Delta}
	\end{equation}
	where we define a sign factor $\mu_e :=- (-1)^{s_0 + {s'}_0} \text{sgn}(V V')$. One can see that, for a edge $E_{lm}$ in the tetrahedron $t_e$ shared by $\sigma_v$ and $\sigma_{v'}$, we have
	\begin{equation}
		E'_{lm} = V' \epsilon_{lmjk}(v') * (\tilde{y'}^j \wedge \tilde{y'}^k \wedge \tilde{y'}^0)= \mu_e V \epsilon_{lmjk}(v) * (\tilde{y}^j \wedge \tilde{y}^k \wedge \tilde{y}^0) = \mu_e E_{lm}
	\end{equation}
	The equation thus implies the co-frame vectors on all edges of tetrahedron $t_e$ at neighboring vertices $v$ and $v'$ are related by
	\begin{equation}\label{eq:f70}
		E_{l}(v) = \mu_e  G_{vv'} E_{l}(v')
	\end{equation}
	Since $E_{l}(v') \perp N_{0}(v')$, the relation is a direct consequence of (\ref{eq:f61}) with the fact $\tilde{r}(v) =\tilde{r}(v') = \tilde{r}$.
	This relation shows that, the vectors $E$ in a tetrahedron shared by two $4$ simplicies $\sigma_{v}$ and $\sigma_{v'}$ satisfies
	\begin{equation}
		g_{l_1 l_2}:=\eta_{IJ} E^I_{l_1}(v) E^J_{l_2}(v) = \eta_{IJ} E^I_{l_1}(v') E^J_{l_2}(v')
	\end{equation}
	where $g_{l_1 l_2}$ is the induced metric on the tetrahedron and it is independent of $v$. If the oriented volume of these two neighboring $4$-simplices are come with the same signature, i.e. $\text{sgn}(V(v))= \text{sgn}(V(v'))$,
	We can associated a reference frame in each $4$ simplex $\sigma_v$ and the frame transformation is given by $\Omega_{vv'}= \mu_e G_{vv'} \in \text{SO}(1,3)$. The matrix $\Omega_{e=(v,v')}$ is a discrete spin connection compatible with the co-frame then. Note that, since $\tilde{r}(v) = r(v) \sgn(V(v))$ is a global sign, globally orienting $\sgn(V(v))$ will make $r=r(v)$ a global orientation on the dual face.
	
	Let us go back to the original geometric rotation $\Omega_{vv'}^{\Delta}$. Suppose we orient consistently all pairs of $4$ simplicies on the simplicial complex $\mathcal{K}$. We then choose a sub-complex with boundary such that, with in it the oriented volume $\sgn(V)$ is a constant. Then for the holonomy along edges of an internal face, we have
	\begin{equation}\label{eq:f72}
		\Omega_{f}^{\Delta}(v) = \Omega_{v_0 v_n}^{\Delta} \Omega_{v_{n} v_{n-1}}^{\Delta} \cdots \Omega_{v_1 v_0}^{\Delta} =I^n I^{s_{0n}+s_{n,n-1}+ \cdots + s_{10}} G_{v_0 v_n} G_{v_{n} v_{n-1}} \cdots G_{v_1 v_0} = \mu_e G_{f}(v)
	\end{equation}
	while for a boundary face,
	\begin{equation}\label{eq:f73}
		\Omega_{f}^{\Delta}(v_n,v_0) = \Omega_{v_n v_{n-1}} \cdots \Omega_{v_1 v_0} = I^n I^{s_{n,n-1}+\cdots + s_{10}} G_{v_0 v_n} G_{v_{n} v_{n-1}} \cdots G_{v_1 v_0} = \mu_e G_{f}(v_n,v_0) 
	\end{equation}
	where $n$ is the number of internal edges belong to the face $f$.
	Here $\mu_e= I^n \prod_{e \in f} I^{s_{e}} = \pm 1, s_{e=(v,v')}= s_{ve}+s_{ve'}$ is independent from orientation. 
	
	Suppose the edges of the triangle due to face $f$ is given by $E_{l1}(v)$ and $E_{l2}(v)$, then from (\ref{eq:f70}) and (\ref{eq:f72}-\ref{eq:f73}), we have
	\begin{equation}\label{eq:parallel_e}
		G_f(v) E_{l}(v) = \mu_e E_{l}(v), \text{or} \qquad G_f(v_n,v_0) E_{l}(v_0) = \mu_e E_{l}(v_n)
	\end{equation}
	For the normals $N_0(v)$ and $N_1(v)$ which othrognal to the triangle due to $f$, from (\ref{eq:f68}) and (\ref{eq:f72}-\ref{eq:f73}), we have
	\begin{equation}
		G_f(v) N_1(v)^{\Delta} = a N_{0}(v)^{\Delta} + b N_{1}(v)^{\Delta}, G_f N_1(v) \cdot E_{l1}(v) = G_f N_1(v) \cdot E_{l2}(v) = 0, 
	\end{equation}
	For boundary faces with boundary tetrahedron $t_{e_n}$ and $t_{e_0}$, similarly, we have
	\begin{equation}
		G_f(v_n,v_0) N_{e0}(v_0) \cdot E_{l1}(v_n) = G_f(v_n,v_0) N_{e0}(v_0) \cdot E_{l2}(v_n) = 0
	\end{equation}
	
	\subsection{Flipped signature solution and vector geometry}\label{app:flip_sig}
	Now let us consider degenerate case, where the $4$ volume is $0$ and $G_i$ can be gauge fixed to its subgroup $G_i \in SO(1,2)$ for timelike tetrahedron. In this case, the $4$-normals of boundary tetrahedra are then gauge fix to be $\forall_i\; N_i=u$. We can introduce a auxiliary space ${M^{4}}'$ with metric $g_{\mu \nu}'$ from $M^4$ by flipping the norm of $u$
	\begin{equation}
		g_{\mu \nu}'= g_{\mu \nu} - 2{ u_{\mu} u_{\nu}}
	\end{equation}
	where $g_{\mu \nu}$ is the metric in $M^4$. We will use prime to all the operations in ${M^4}'$. For the norm of $u$, we have
	\begin{equation}
		t = u \cdot u, \qquad t' = -t  = u \cdot' u
	\end{equation}
	Notice that for the subspace $V$ orthogonal to $u$, the restriction of both scalar product coincide. Thus for vectors in $V$ we can use both scalar product. The Hodge dual operation satisfies ${*'}^2=-*^2 = t = -t'$.
	
	For the subspace $V$, we can introduce maps $\Phi^{\pm}$
	\begin{equation}
		\Phi^{\pm}: \Lambda^2 {M^4}' \to V, \quad \Phi^{\pm}(B) = t' (\pm B - t' *' B) \cdot' u = (\mp B + *' B) \cdot' u
	\end{equation}
	where $B$ is a bivector in ${M^4}'$. Clear for a vector $v \in V$, we have
	\begin{equation}
		\Phi^{\pm}( *' (v \wedge u) ) = v
	\end{equation}
	The map $\Phi^{\pm}$ naturally induce a map from $G \in \text{SO}(2,2)$ to the subgroup $h \in \text{SO}(1,2)$, which defined by
	\begin{equation}
		\Phi^{\pm}(G B G^{-1}) = \Phi^{\pm}(G) \Phi^{\pm}(B)
	\end{equation}
	where
	\begin{equation}
		\Phi^{\pm}(G) \in O(V)
	\end{equation}
	Easy to see when $G=h \in \text{SO}(1,2)$, we have $\Phi^{\pm}(h) = h$. And one can further prove that the condition is sufficient and necessary as shown in \cite{Kaminski:2017eew}.
	
	Clearly for given bivectors $B_{ij}^{\{G\}}= G_i * (v_{ij} \wedge u)$ in $M'$, if $B_{ij}^{\{G\}}= -B_{ji}^{\{G\}}$, we have
	\begin{equation}
		 v^{\{G\} \pm}_{ij} = - v^{\{G\} \pm}_{ji}, \;\; v^{\{G\} \pm}_{ij}= \Phi^{\pm}(G) v_{ij}=\Phi^{\pm}(B_{ij}^{\{G\}}) 
	\end{equation}
	and the closure $\sum_i B_{ij}^g = 0$ leads to 
	\begin{equation}
		\sum_i v^{\{G\} \pm}_{ij} =0
	\end{equation}
	 One can prove the condition is necessary. In other words, if we have $g^{\pm}_i$ such that $v^{\{G\} \pm}_{ij} = - v^{\{G\} \pm}_{ji}$, we can always build unique $G_i \in SO(M')$ (up to $I^{s_i}$ ) which constitute a $SO(M')$ solution. 
	
	In summary we see that there is an 1-1 correspondence between 
		\begin{itemize}
			\item pair of two non-gauge equivalent vector geometries,
			\item geometric $SO(M')$ non-degenerate solution.
		\end{itemize}
		The two vector geometries are obtained from $SO(M')$ solutions $\{g_{ve} \}$ as $g^{\pm}_{ve}=\Phi^{\pm}(g_{ve})$.
	This is the flipped signature case for a Gram matrix with given geometric boundary data. For example, with all boundary tetrahedra timelike, the signature of reconstructed non-degenerate $4$ simplex is split $(-++-)$.
	
	From the reconstruction for non-degenerate solutions, we have the orientation matching condition for the geometric group elements $G^{\Delta \pm} \in O(V)$ where
	\begin{equation}
		G^{\Delta \pm}_i v_{ij} = v_{ij}^{\Delta \pm}, \qquad v_{ij}^{\Delta \pm}=\Phi^{\pm}(B_{ij}^{\Delta})
	\end{equation}	
	One can show that, in flipped signature case, this condition becomes
	\begin{equation}
		\det G_{ve}^{\Delta} = \det G_{ve}^{\Delta \pm} 
	\end{equation}
	
	Since the critical point solutions are in 1-1 correspondence with reconstructed 4 simplicies up to reflection and shift. As a direct result from (\ref{eq:E53}), for non-degenerate boundary data satisfying length matching condition and orientation matching condition, there are two gauge inequivalent solutions corresponding to reflected 4 simplicies which are related by
	\begin{equation}
		\tilde{G}=R_{u} G R_{u}
	\end{equation}
	where $\tilde{G}$ and $G$ represent two gauge equivalent series. 
		Two non-equivalent geometric $SO(M')$ non-degenerate solutions then satisfy
		\begin{equation}\label{eq:E90}
			\Phi^{\pm}(\tilde{G})= \Phi^{\pm}(R_{u} G R_{u}) = \Phi^{\mp}(g)
		\end{equation}

	Finally, when the $SO(M')$ solution is degenerate, we can assume $N_i=u$ by gauge transformations. In this case, we see $\Phi^{+}(G)=\Phi^{-}(G) = h$. Thus the vector geometries are gauge equivalent. The inverse is also true. When the vector geometries are gauge equivalent, we have $\Phi^{+}(G)=\Phi^{-}(G)$, which means there exists $G_i$ (uniquely up to gauge transformations) such that after gauge transformations $N_i=G_i u =u$. This corresponds to the degenerate reconstructed $4$ simplex with zero $4$-volume.
	
	\section{Derivation of rotation with dihedral angles}\label{app_rr_d}
	In this appendix, we prove the following equation
		\begin{equation}
			R_{N_i} R_{N_{j}}=\Omega_{ij}=\ue^{2 \theta_{ij} \frac{N_i \wedge N_j}{|N_i \wedge N_j|}}
		\end{equation}
	which is used in Sec. \ref{sec6}.
	For two normalized spacelike vector $N_{i}$, $N_{j}$, $N_{i}^I N_{iI}=N_{j}^J N_{jJ}=1$, compatible with (\ref{eq:def_dihedral_1}) and (\ref{eq:def_dihedral_2}), we have
	\begin{align}
	&	N_{i}^I N_{jI}=\cos \theta_{ij},\\
	& |N_j \wedge N_i|^2=-|* N_j \wedge N_i|^2 =\sin^2 (\theta_{ij})
	\end{align}
	For $N_{i}$, $N_{j}$ are timelike and the signature of plane span by $N_{i} \wedge N_{j}$ is mixed in flipped signature case, we have
	\begin{align}
		&	N_{i}^I N_{jI}=\cosh \theta_{ij},\\
		&  |N_j \wedge N_i|^2=|*' N_j \wedge N_i|^2 = - \sinh^2 (\theta_{ij})
	\end{align}
	
	Now from
	\begin{equation}
		(R_{N}) ^I_J=I - \frac{2 N^I N_J}{N \cdot N} =I - 2 t N^I N_J
	\end{equation}
	where we define $t := N^I N_I$. 
	Easy to see for a vector $v$ in $N_i \wedge N_j$ plane,
	\begin{equation}
	\begin{split}
		&R_{N_i} R_{N_{j}} v=(I-2 t N_i^K N_{iI})(I-2 t N_j^I N_{jJ}) v^J\\
		&= v -2 t (N_i \cdot v) N_i -2 t (N_j \cdot v) N_j+4 (N_i \cdot N_j)( N_j \cdot v) N_i
	\end{split}
	\end{equation}
	which leads to
	\begin{align}
		R_{N_i} R_{N_{j}} -R_{N_j} R_{N_{i}} =4 (N_i \cdot N_j) N_i \wedge N_j\\
		\Tr (R_{N_i} R_{N_{j}}) = 4 (N_i \cdot N_j)^2-2
	\end{align}
	Let us introduce spacetime rotations $\Omega \in SO_{\pm}(1,3)$. For connected components in Lorentzian group, two group elements $\Omega$ and $\Omega'$ are equal is they satisfy
	\begin{equation}
		\Omega - \Omega^{-1} = \Omega' - {\Omega'}^{-1}, \qquad \Tr(\Omega) = \Tr(\Omega')
	\end{equation}
	The space rotation can be written using bivectors as
	\begin{equation}
		\Omega_{ij}=\ue^{2 \theta_{ij} \frac{N_i \wedge N_j}{|N_i \wedge N_j|}}= \cos (2 \theta_{ij}) + \sin (2 \theta_{ij})\frac{N_i \wedge N_j}{|N_i \wedge N_j|}
	\end{equation}
	and for spacelike normal vectors we have
	\begin{align}
		\Omega_{ij} -\Omega_{ji}=2 \sin (2 \theta_{ij})\frac{N_i \wedge N_j}{|N_i \wedge N_j|}=4 (N_i \cdot N_j)(N_i \wedge N_j) \\
		\Tr (\Omega_{ij})= 2 \cos (2 \theta_{ij})=2 (2\cos^2(\theta_{ij})-1)=4 (N_i \cdot N_j)^2-2
	\end{align}
	while for timelike normal vectors span a mixed signature plane, $\Omega$ is a boost, 
	\begin{equation}
		\Omega_{ij}=\ue^{2 \theta_{ij} \frac{N_i \wedge N_j}{|N_i \wedge N_j|}}= \cosh (2 \theta_{ij}) + \sinh (2 \theta_{ij})\frac{N_i \wedge N_j}{|N_i \wedge N_j|}
	\end{equation}
	with
	\begin{align}
		\Omega_{ij} -\Omega_{ji}=2 \sinh (2 \theta_{ij})\frac{N_i \wedge N_j}{|N_i \wedge N_j|}=4 (N_i \cdot N_j)(N_i \wedge N_j) \\
		\Tr (\Omega_{ij})= 2 \cosh (2 \theta_{ij})=2 (2\cosh^2(\theta_{ij})-1)=4 (N_i \cdot N_j)^2-2
	\end{align}
	Notice that here $|N_i \wedge N_j|$ is defined as 
	\begin{equation}
		|N_i \wedge N_j| = \sqrt{| |N_i \wedge N_j|^2 | } ,
	\end{equation}
	
	Thus in both case we have
	\begin{equation}
		R_{N_i} R_{N_{j}}=\Omega_{ij}=\ue^{2 \theta_{ij} \frac{N_i \wedge N_j}{|N_i \wedge N_j|}}
	\end{equation}
	where $\theta_{ij}$ is angle between normals and related to the dihedral angle by (\ref{eq:def_dihedral_1}) and (\ref{eq:def_dihedral_2}).

	\section{Fix the ambiguity in the action}\label{app:deformation_phase}
	In this appendix we show how to choose the $\text{SL}(2,\C)$ lift to fix the ambiguity in the action. Note that here we only fix the ambiguity for single $4$-simplex $\sigma_v$ with boundary data, where the deficit angle $\Theta_f = \theta_f$ is the angle between normals. 
	The ambiguity (in one $4$ simplex $\sigma_v$ with boundary) which due to odd $n_f$ can be expressed as
	\begin{equation}
	\Delta S- \Delta S^{\Delta} = \ui r \sum_{f:n_f \; odd} \begin{array}{l} \Delta \phi - \Theta_f \qquad \text{non degenerate case}\\ \Delta \phi \qquad \text{split signature case} \end{array}
	\end{equation}
	The procedure we use here is an extension of the one used for spacelike triangles in \cite{Kaminski:2017eew}.
	\subsubsection{non-degenerate case}
	Suppose we have a non-degenerate solutions $\{ G_{ve}^0 \in SO(1,3) \}$ with normals $v^0_{ef}$ of triangles of non-degenerate boundary tetrahedra. The area of these triangles is given by spins $\gamma s_f^0 = \frac{n_f^0}{2}$. Define the following continuos path
	\begin{equation}
		G_{ve}(t), \quad v_{ef}(t), \quad u(t) = u = (0,0,0,1)^{T}, 
	\end{equation}
	where $\forall e G_{ve}^{0} = G_{ve}(0), v_{ef}^0 = v_{ef}(0)$. 
	Such that
	\begin{itemize}
		\item $\forall t \in [0,1]$, $\{G_{ve}(t)\}$ is a solution of critical point equations with boundary data where the normals of triangles of boundary tetrahedra are $v_{ef}(t)$,
		\item $\forall t \neq 1$ boundary data is non-degenerate, and $v_{ef}(1) \neq 0$,
		\item $\forall t \neq 1$ solution $\{G_{ve}(t)\}$ is non-degenerate,
		\item for $t=1$, pair of solutions $\{G_{ve}(t)\}$ and $\{\tilde{g}_{ve}(t) = R_{e_\alpha} g_{ve}(t) R_{u} \}$ are gauge equivalent.
	\end{itemize}
	In this path, the function
	\begin{equation}
		f(t) = \sum_{f:n_f \; odd} \; \Delta {\phi}_{eve'f}(t)  - r \Theta_f(t) \; \mod 2 \pi
	\end{equation}
	takes values in $\{0, \pi \}$ and changing continuously with the phase the difference from stationary points determined by $\{G_{ve}(t)\}$ and $\{\tilde{G}_{ve}(t) = R_{e_\alpha} G_{ve}(t) R_{u} \}$. Thus $f(t)$ is a constant. Since at $t=1$, we have two geometric solutions are gauge equivalent to each other, which means the lifts $g_{ve},\tilde{g}_{ve}$ of solutions satisfy
	\begin{equation}
		\forall_{e} \; \tilde{g}_{ve} = (-1)^{r_{ve}} g g_{ve}, \qquad r_{ve} = \{0,1\}
	\end{equation}
	From (\ref{eq:g_diff_one}),
	\begin{equation}
		\begin{split}
			&(-1)^{r_{ve}+ r_{ve'}} =  g_{ve} (\tilde{g}_{e'v} \tilde{g} _{ve})^{-1}g_{e'v} =\ue^{-2 \Delta \theta_{e'vef} X_f + 2 \ui \Delta \phi_{e'vef} X_f}
		\end{split}
	\end{equation}
	which leads to $ \Delta {\phi}_{eve'f}(1) = (r_{ve} + r_{ve'}) \pi \mod \; 2\pi$ since we have $(2 X_f)^2 =1$.
	We shall consider a subgraph of spin network which contains those odd $n$ links. The subgraph has even valence nodes. Thus we can decompose into Euler cycles. In those cycles every link of odd $n$ will appears exactly once. For a Euler cycle consisting edges with odd $n$, every edge will be counted twice, thus we have
	\begin{equation}
		\sum_{e \in cycle} \Delta {\phi}_{eve'f}(1)=\sum_{e \in cycle} 2 r_{ve} \pi = 0 \mod \; 2 \pi
	\end{equation}
	Also, from the fact that two geometrical solution is gauge equivalent $\forall_{e}\;\tilde{G}_{ve}=G G_{ve}$, we have $R_{N_{e}} R_{N_{e'}} = G_{ve} (\tilde{G}_{e'v} \tilde{G} _{ve})^{-1} G_{e'v} = 1$, thus
	\begin{equation}
		\Theta_f(1) = \tilde{r}_{f} \pi \mod 2 \pi, \qquad \tilde{r}_f = \tilde{r}_{ve} + \tilde{r}_{ve'} \in \{0,1\} \; .
	\end{equation}
	which can be fixed again using Euler cycles as for $\Delta \phi$.
	
	The path can be achieved by deforming solutions in the following way: First choose a timelike plane with simple normalized bivector $V$ at some vertex $v$ satisfies 
	\begin{equation}
		\forall_{f} V \wedge * B_{f} \neq 0 \; .
	\end{equation}
	The path is made by contracting the two directions in $*V$, and we donate the $t=1$ as the limit for contracting directions to $0$. From above condition we have $\lim_{t \to 1} B_f$ exist and keep nonzero. The dual action of the shrinking on geometric normal vectors $N^{\Delta}$ also have a limit which is their normalized components lying in $*V$ plane (after normalization). By suitable definition of boundary data, we can assume $G_{ve}(1) = \lim_{ \to 1} G_{ve}(t)$ exist. Now we end up with a highly degenerate 4-simplex which contained in a 2d plane and all bivectors are proportional to $V$.
	\subsubsection{split signature case}
	The treatment concerning degenerate solutions following the similar method. Start form the non-degenerate boundary data, where normals of triangles of boundary tetrahedra are given by $v_{ef}^0$ and area of these triangles are related to spins $n_f/2$. Suppose from these boundary data, we can reconstruct a non-degenerate $4$-simplex in flipped signature space $M'$. In this case, we have two non-gauge equivalent solutions $\{g_{ve}^{\pm}\}$. We define the following path
	\begin{equation}
		g_{ve}^{\pm}(t), \quad v_{ef}(t), \quad u(t) = u = (0,0,0,1)^{T},
	\end{equation}
	where $\forall e g_{ve}^{0\pm} = g_{ve}^{\pm}(0), v_{ef}^0 = v_{ef}(0)$. The path 
	satisfies
	\begin{itemize}
		\item $\forall t \in [0,1]$, $\{g_{ve}^{\pm}(t)\}$ are solutions of critical point equation with boundary data given by $v_{ef}(t)$,
		\item $\forall t \in [0, 1]$ boundary data is non-degenerate, e.g. the boundary tetrahedron is non-degenerate,
		\item $\forall t \neq 1$ solutions $\{g_{ve}^{\pm}\}$ are non-gauge equivalent thus we have a non-degenerate reconstructed $4$-simplex in $M'$
		\item for $t=1$, the reconstructed $4$-simplex is degenerate in $M'$.
	\end{itemize}
	Now the constant function $f(t) \in \{0,\pi \}$ reads
	\begin{equation}
		f(t) = \sum_{f:n_f \; odd} \; \Delta {\phi}_{eve'f}(t)   \; \mod 2 \pi
	\end{equation}
	Following the same argument in non-degenerate case, we have for the lifts
	\begin{equation}
		g^{+}_{ve}(1) = (-1)^{r_{ve}} g^{-}_{ve}(1)
	\end{equation}
	Based on the same consideration using Euler cycles, we have
	\begin{equation}
		f(1)=\sum_{f:n_f \; odd} \; \Delta {\phi}_{eve'f}(t) = 0 \mod 2 \pi
	\end{equation}
	Thus we have
	\begin{equation}
		\Delta S^0 - \Delta S^{\Delta 0} = 0 \mod 2 \pi
	\end{equation}
	
	The path is built by the following way: We choose a spacelike normal such that, in flipped signature space 
	\begin{equation}
		\forall_f N \wedge B_f \neq 0.
	\end{equation}
	The path is then made by contracting in the direction of $N$ in the flipped space $M'$. The contraction leads to a continuos path of non-degenerate solutions in $M'$ until $t=1$ where the $4$-simplex is degenerate.

	\end{widetext}
	\bibliographystyle{utphys}
	\bibliography{refs}
	
	\end{document}